\documentclass[11pt,a4paper]{article}
\usepackage{amsmath}
\usepackage{graphicx}
\usepackage{xcolor}
\usepackage[caption=false]{subfig}
\usepackage{mathrsfs,mathtools}
\usepackage{physics,amssymb}
\usepackage{siunitx}
\usepackage{bm}
\usepackage{booktabs}
\usepackage{braket}
\usepackage{listings}

\usepackage{cases}
\usepackage{comment}
\usepackage{soul}
\usepackage{cancel}
\usepackage{cases}
\usepackage[utf8]{inputenc}
\usepackage{url}
\usepackage{float}
\usepackage{longtable}
\usepackage[normalem]{ulem}
\usepackage{xspace}
\usepackage{xcolor}
\usepackage{aas_macros}
\setstcolor{red}
\usepackage{jcappub}
\hypersetup{colorlinks=true
,urlcolor=DARKBLUE
,anchorcolor=DARKBLUE
,citecolor=DARKBLUE
,filecolor=DARKBLUE
,linkcolor=DARKBLUE
,menucolor=DARKBLUE
,linktocpage=true
,pdfproducer=medialab
}

\newcommand{\Pz}{\mathcal P_{\zeta_G}}

\newcommand{\ddlnk}{\,\mathrm d\ln k}

\newcommand{\cC}{\mathfrak C}

\newcommand{\inner}[2]{\left(#1,#2\right)_{\mathcal P}}

\definecolor{MONZA}{HTML}{CF000F}
\definecolor{DARKBLUE}{HTML}{00008b}
\definecolor{DARKMAGENTA}{HTML}{8b008b}
\definecolor{DARKCYAN}{HTML}{008B8B}
\definecolor{DARKORANGE}{HTML}{FF8C00}
\definecolor{PURPLE}{HTML}{8800CC}

\begin{document}
\title{The statistics of curvature-profile dispersion in primordial black hole formation}
\author[a,b,c,d]{Albert Escriv\`a}

\affiliation[a]{Asia Pacific Center for Theoretical Physics, Pohang 37673, Republic of Korea}
\affiliation[b]{Department of Physics, Pohang University of Science and Technology, Pohang 37673, Republic of Korea}
\affiliation[c]{Institute for Advanced Research, Nagoya University, \\
Furo-cho Chikusa-ku, Nagoya 464-8601, Japan}
\affiliation[d]{Department of Physics, Nagoya University, \\
Furo-cho Chikusa-ku, Nagoya 464-8602, Japan}

\emailAdd{alberto.escriva@apctp.org}

\date{\today}
\abstract{In the standard curvature-perturbation scenario, PBHs form from the collapse of superhorizon curvature fluctuations after horizon re-entry. The predicted
abundance is exponentially sensitive to the collapse threshold and hence to the
shape of the primordial curvature profile. In this work we develop a finite-action
framework to describe curvature-profile dispersion around representative peak
profiles. Using a multipolar Fourier--Bessel decomposition, we separate the local
peak variables of the Gaussian field from residual radial and angular deformations,
normalized by their Gaussian action. We apply the formalism to spherical numerical-collapse examples in order to
isolate the effect of radial shape dispersion. For finite-width spectra, and in
the presence of logarithmic local non-Gaussianity, we compute the collapse
threshold as a function of a coherent shape variable and combine the result with
peak statistics. We find that the dominant contribution to the PBH abundance is
not necessarily the conditional-mean reference profile, nor simply the profile with the lowest threshold.
Instead, it is selected by a competition between the Gaussian cost of realizing a
coherent deformation and the exponential gain associated with lowering the
collapse threshold. Broad spectra and negative non-Gaussianity can make rare shape
deformations dominate the abundance. In the examples studied here, the dominant
branches can correspond to several-sigma coherent shape fluctuations while
enhancing the integrated abundance by orders of magnitude. Equivalently, including
shape dispersion can reduce the power-spectrum amplitude required to obtain a
fixed PBH abundance. Our results show that residual profile dispersion is a genuine statistical ingredient in PBH formation and can be quantitatively important for accurate abundance estimates.}

\maketitle
\flushbottom

\section{Introduction}
\label{sec:introduction}

Primordial black holes (PBHs) provide a direct connection between the physics of
the early Universe, gravitational collapse and present-day observations. Unlike
astrophysical black holes, PBHs may form from the collapse of sufficiently large
primordial inhomogeneities before the formation of stars and galaxies
\cite{Hawking:1971ei,Carr:1974nx}. Their possible abundance is therefore a
sensitive probe of the statistics of primordial fluctuations, of the thermal
history of the Universe and of the nonlinear threshold for gravitational collapse. This has motivated extensive work on PBHs as dark-matter candidates,
as possible progenitors of gravitational-wave events and as probes of inflation
and physics beyond the Standard Model
\cite{Sasaki:2018dmp,Carr:2020gox,Green:2020jor,Escriva:2022duf}. The formation of PBHs is a highly nonlinear problem that typically requires relativistic numerical simulations to investigate the phenomenology of their formation across different scenarios \cite{Shibata:1999zs,Niemeyer:1999ak,Hawke:2002rf,Musco:2004ak,Musco:2008hv,Nakama:2013ica,Bloomfield:2015ila,Deng:2016vzb,Deng:2017uwc,Escriva:2019sim,Yoo:2020lmg,Yoo:2021fxs,Escriva:2022pnz,Escriva:2022yaf,Escriva:2022bwe,Musco:2023dak,deJong:2023gsx,Escriva:2023nzn,Uehara:2024yyp,Yoo:2024lhp,Escriva:2024hsv,Padilla:2025bkv,Ning:2025ogq,Milligan:2025zbu,Kitajima:2025shn,Germani:2025hcu,Baumgarte:2025vvs,Cheng:2025eas,Baumgarte:2026igz,Ning:2026nfs,Yuwen:2026hxu}.

In the standard scenario (see \cite{Escriva:2022duf} for a list of other mechanisms), PBHs are produced when large-amplitude curvature
fluctuations re-enter the cosmological horizon during the radiation-dominated
epoch, which is the scenario we consider in this work. The final abundance depends exponentially on the collapse threshold
\cite{Carr:1974nx}. Consequently, even modest changes in the threshold, in the
definition of the perturbation amplitude, or in the assumed real-space profile
can lead to large changes in the predicted abundance. Numerical-relativity
studies (see Ref.~\cite{Escriva:2021aeh} for a review) have shown that the threshold is not a universal number when expressed
in terms of a local density contrast, but depends on the shape of the
perturbation (see for relevant works \cite{Musco:2018rwt,Escriva:2019nsa}). A robust
characterization is obtained in terms of the compaction function, first introduced in this context in Ref.~\cite{Shibata:1999zs}, and in particular in terms of its maximum value.

PBH formation is therefore a problem of both statistics and nonlinear dynamics.
The primordial power spectrum determines the probability distribution of
curvature profiles, while relativistic collapse determines which of those
profiles form black holes. A systematic treatment of this issue requires separating the statistical
description of the curvature field from the nonlinear collapse calculation. Peak
theory provides the natural language for this separation. In the BBKS (Bardeen-Bond-Kaiser-Szalay) peak theory construction \cite{Bardeen:1985tr}, a local maximum of a Gaussian field is characterized by its height, gradient and Hessian (see \cite{Green:2004wb} for the first work where peak theory was applied in the context of PBH abundance estimations). In the PBH context, the spherical part of the Hessian fixes
the curvature of the peak, while the traceless part describes the local ellipsoidal deformation. These variables, however, do not exhaust the possible shape information of a peak. Infinitely many coherent field configurations may share the same height, curvature and local ellipsoid while differing in their radial structure or in higher angular multipoles. Since the collapse threshold is, in general, a functional of the full curvature profile, these residual degrees of freedom can modify both the threshold and the abundance.

In several PBH abundance frameworks
(see for instance \cite{Germani:2018jgr,Yoo:2018kvb,Young:2019yug,Germani:2019zez,Yoo:2019pma,Young:2020xmk,Yoo:2020dkz,Germani:2023ojx,Pi:2024ert,Fumagalli:2024kgg}),
the connection between peak statistics and collapse dynamics is made tractable by reducing the stochastic ensemble of curvature configurations to a finite set of profile variables. In typical-profile implementations of peak theory, the collapse criterion is evaluated on the conditional mean profile determined by the chosen peak variables, while in other approaches part of the shape dependence is retained through a reduced threshold prescription, such as one depending on the compaction-shape parameter. These approaches capture important aspects of the profile dependence, but they do not explicitly integrate over finite-action residual deformations of the curvature profile around the chosen reference configuration. The purpose of the present work is to formulate this residual profile dispersion as an explicit statistical degree of freedom. We keep fixed the selected peak variables defining the reference
profile and introduce coherent, action-normalized deformations orthogonal to them. The
collapse threshold is then computed as a function of these deformations and combined with
their Gaussian action cost in the abundance estimate. The precise separation between the variables used to define the reference
configuration and the remaining profile-shape degrees of freedom depends
on the statistical prescription adopted. Different choices of peak
variables or conditioning criteria can therefore lead to different
parametrizations of the same underlying ensemble of curvature profiles.

The importance of accounting for dispersion around a central high-peak profile
has also been emphasized in previous studies. In Ref.~\cite{Atal:2019erb}, PBH formation was studied in a
single-field inflationary scenario in which the physical curvature perturbation
is nonlinearly related to an underlying Gaussian field. The statistical treatment
there was based on the high-peak profile of the Gaussian field: the median
configuration was written in terms of the normalized two-point function, and
shape dispersion was estimated by considering the conditional one-sigma envelope
around this median profile. These dispersed profiles were then mapped through
the perturbative or non-perturbative local non-Gaussian relation and evolved in
spherical symmetry to determine the collapse threshold. That analysis showed that profile dispersion can lead to a non-negligible spread in the
threshold amplitude of the curvature perturbation, especially at low
non-Gaussianity.

More recently, Ref.~\cite{Escriva:2025ftp} studied an ultra-slow-roll plateau model in which PBHs can arise both from adiabatic curvature perturbations and from relic vacuum bubbles. In that work the statistical ensemble of initial conditions was constructed directly from the power spectra of the field and momentum perturbations. The mean profiles for the scalar field were obtained as minimum-action configurations subject to central-amplitude constraints, while deviations from the mean were described by coherent finite-action deformations, parametrized by
standard Gaussian variables. The resulting profile ensemble was then propagated through nonlinear field evolution, bubble formation, the generalized non-Gaussian curvature map, and the PBH mass-function calculation. In the parameter region studied there, the mean profiles were found to dominate both the adiabatic and bubble-induced PBH contributions, with shape dispersion giving only subleading corrections.

These results and previous explorations motivate a model-independent and statistically controlled description of profile dispersion, as well as an assessment of whether, for a broader class of spectra, shape dispersion can give
a larger abundance contribution than the reference profile associated with the chosen conditioning. In this work we
formulate such a description in terms of finite-action coherent deformations of
the Gaussian curvature profile. We expand the curvature perturbation in a multipolar
Fourier--Bessel basis, with the power spectrum defining the Gaussian metric in
the space of profiles. In this language, the BBKS height, curvature and
ellipsoidal variables correspond to particular action-normalized directions,
while the remaining orthogonal directions describe genuine residual shape
dispersion beyond the BBKS local variables. This construction allows us to
compare the change in the collapse threshold induced by a coherent deformation
with the Gaussian cost required to realize it.

After developing the general formalism, we apply it to two spherical examples. The first is a sharply peaked finite-width
spectrum, where we study a coherent radial deformation at fixed central
amplitude and include the effect of local-type primordial non-Gaussianity. The
second is a finite-width scale-invariant spectrum, where the distinction between
height, curvature and residual radial dispersion can be made explicit. In both
cases we restrict the numerical collapse calculation to spherical symmetry, in
order to isolate the role of radial shape dispersion before addressing the full
non-spherical problem in future works.

The main message is that the PBH abundance is controlled by a competition
between the Gaussian cost of realizing a coherent shape deformation and the
change in the collapse threshold induced by that deformation. Therefore the dominant contribution need not come from the reference profile associated with the chosen conditioning. Instead, it can come from a rarer profile whose lower threshold compensates its statistical suppression.

The paper is organized as follows. In Sec.~\ref{sec:formalism} we introduce the
finite-action Fourier--Bessel decomposition and recover the BBKS spherical and
ellipsoidal sectors. In Sec.~\ref{sec:monochromatic_case_general} we specialize
the construction to the monochromatic limit and describe the generalization from the BBKS ellipsoid to
higher multipoles, while in
Sec.~\ref{sec:finite_width_nonspherical} we extend the discussion to
finite-width spectra. In Sec.~\ref{sec:numerical} we present the spherical
numerical-collapse examples and use the resulting threshold curves to estimate
the impact of shape dispersion on the PBH abundance and mass function. Finally,
in Sec.~\ref{sec:conclusions} we summarize our conclusions and discuss future
extensions.

\section{Theoretical framework}
\label{sec:formalism}
In this section, we present the theoretical development of our approach to account for the dispersion of curvature profiles.

A statistically homogeneous and isotropic Gaussian field contains infinitely many possible peak profiles. Even after fixing the height of a peak, its radial shape, ellipticity, prolateness, higher angular structure, and surrounding environment remain stochastic. For PBH formation, these differences matter because the collapse threshold is, in general, a nonlinear functional of the complete initial profile. The framework that we aim to develop in this work separates the problem into four logically distinct ingredients: i) the Gaussian statistics, specified by a power spectrum; ii) the peak variables held fixed in defining a reference profile; iii) a set of coherent, finite-action deformations around that reference profile; iv) the nonlinear collapse threshold determined by the relativistic numerical evolution.

The resulting profile family has the schematic form
\begin{equation}
  \zeta_G(\bm x)
  =
  \overline\zeta_G(\bm x\mid \cC)
  +
  \sum_\tau n_{\tau}\,\Delta_\tau(\bm x),
  \label{eq:intro-master}
\end{equation}
where $\bm x = (x_1,x_2,x_3)$ in three-dimensional space, $\cC$ is the chosen set of peak constraints, $\tau$ collectively labels angular and radial modes, and every $n_{\tau}$ is a standardized Gaussian coefficient if the modes are normalized in the Gaussian-action metric. Specifically, the coefficient $n_{\tau}$ is not a pointwise standard deviation. It is the number of Gaussian standard deviations along one complete coherent field configuration $\Delta_\tau(\bm x)$. If the modes are action orthonormal, the statistical cost of a configuration is $\Delta W=\sum_\tau n_\tau^2$.

\subsection{General multipolar Fourier--Bessel decomposition}
\label{subsec:general_fourier_bessel}

We start from the general multipolar Fourier--Bessel decomposition of the
Gaussian curvature perturbation. For a scalar field $\zeta_G({\bm x})$, we
write

\begin{equation}
\zeta_G({\bm x})
=
\sum_{\ell=0}^{\infty}
\sum_{m=-\ell}^{\ell}
\int d\ln k\,
B_{\ell m}(k)\,
j_\ell(kr)\,
Y_{\ell m}(\hat{\bm x}) .
\label{eq:general_fourier_bessel_expansion}
\end{equation}

Here $r=|{\bm x}|$, $\hat{\bm x}={\bm x}/r$, $j_{\ell}$ is the spherical
Bessel function and $Y_{\ell m}$ is a spherical harmonic, which is normalized with the angular-average measure \(d\Omega/(4\pi)\). We refer the reader to Appendix~\ref{appendix:harmonics} for the definitions used.

For a real field, if complex spherical harmonics are used, the coefficients obey the
usual reality condition $B_{\ell,-m}(k)=(-1)^m B_{\ell m}^{*}(k)$, where we use the convention $Y_{\ell,-m}(\hat{\boldsymbol{x}}) =
(-1)^m Y_{\ell m}^{*}(\hat{\boldsymbol{x}})$. Equivalently, one may work in an orthonormal real spherical-harmonic basis, in which
the independent expansion coefficients are real. In the rest of this work we keep the
compact complex notation \(Y_{\ell m}\), but whenever action-normalized amplitudes such
as \(n_{\ell m \alpha}\) are introduced, the index \(m\) is understood to label the \(2\ell+1\)
independent real harmonic components after imposing the reality condition. With this
convention, all amplitudes multiplying independent coherent modes are real Gaussian
variables, and the quadratic action is an ordinary sum of squares, with no double
counting of the \(m\) and \(-m\) modes.

Then, $B_{\ell m}(k)$ contains the radial spectral content of
the angular multipole $(\ell,m)$. The mode $\ell=0$ is the spherical component,
$\ell=2$ describes quadrupolar or ellipsoidal deformations, and $\ell\geq 3$
describes higher non-spherical distortions. The lowest radial direction in the $\ell=1$ sector corresponds to the
gradient of the field, or equivalently to a displacement of the peak position, and is set to zero when the peak is centered at the origin. Higher radial $\ell=1$ directions may remain as residual shape modes. The coefficients $B_{\ell m}$ can also be written through the inverse Fourier--Bessel transform

\begin{equation}
B_{\ell m}(k)
=
{2k^3\over \pi}
\int_0^\infty dr\,r^2\,
j_{\ell}(kr)
\int {d\Omega\over 4\pi}\,
\zeta_G(r,\hat{\bm x})\,
Y^{*}_{\ell m}(\hat{\bm x}) .
\label{eq:Blm_inverse}
\end{equation}

Motivated by the spherical-symmetry construction of Ref.~\cite{Escriva:2025ftp}, for a statistically homogeneous and isotropic Gaussian field with dimensionless
power spectrum ${\cal P}_{\zeta_G}(k)$, the Gaussian probability functional can
be written as
\begin{equation}
P[\zeta_G]
\propto
\exp\left[-{1\over 2}W[B]\right],
\end{equation}
where the Gaussian action\footnote{In this work we define peaks directly in the Gaussian curvature field $\zeta_G$ itself. Therefore no additional transfer or smoothing kernel is introduced in the peak variables; equivalently, in the notation of a filtered peak field $X(k)=S(k)\zeta_G(k)$, we set $S(k)=1$ throughout.} is
\begin{equation}
W[B]
=
\sum_{\ell,m}
\int d\ln k\,
{|B_{\ell m}(k)|^2 \over {\cal P}_{\zeta_G}(k)} .
\label{eq:gaussian_action_Blm}
\end{equation}
Thus the power spectrum defines the statistical metric in the space of profiles:
modes supported where ${\cal P}_{\zeta_G}(k)$ is large are statistically less
costly, while modes supported where the spectrum is small are exponentially
suppressed.

The constrained finite-action framework consists of decomposing the coefficients
into a reference part and coherent deviations,
\begin{equation}
B_{\ell m}(k)
=
\overline{B}_{\ell m}(k)
+
\Delta B_{\ell m}(k).
\label{eq:B_reference_plus_deviation}
\end{equation}
The reference coefficients $\overline{B}_{\ell m}(k)$ describe the conditional
mean profile, for example the spherical mean at fixed height or the BBKS mean at
fixed height, curvature and Hessian. The deviations can be expanded in
action-normalized spectral directions,
\begin{equation}
\Delta B_{\ell m}(k)
=
\sum_\alpha
n_{\ell m \alpha}\,
{\cal P}_{\zeta_G}(k)\,
q_{\ell \alpha}(k).
\label{eq:delta_B_lm_expansion}
\end{equation}

For each angular multipole \(\ell\), the residual fluctuations around the reference
profile span a radial spectral function space. We denote by \(q_{\ell \alpha}(k)\) an
orthonormal basis of this space, where the index \(\alpha=0,1,2,\ldots\) labels independent
radial, or spectral, deformation directions. These functions are not fixed uniquely;
they may be chosen in any convenient way, for example by Gram--Schmidt
orthonormalization of a set of radial templates or of the spectral directions associated
with the peak constraints. The only requirement is that they are orthonormal with
respect to the Gaussian covariance inner product defined below. Statistical isotropy
implies that the covariance is diagonal in the angular indices and degenerate in \(m\)
at fixed \(\ell\). Therefore the same radial orthonormal basis \(q_{\ell \alpha}(k)\) can be
used for all \(m=-\ell,\ldots,\ell\), while the amplitudes \(n_{\ell m \alpha}\) label the
independent real angular components. The radial spectral functions are then normalized by
\begin{equation}
\int d\ln k\,
{\cal P}_{\zeta_G}(k)\,
q_{\ell \alpha}(k)\,
q_{\ell \eta}(k)
=
\delta_{\alpha \eta}.
\label{eq:q_action_normalization}
\end{equation}
The associated real-space radial modes are
\begin{equation}
\mathcal{R}_{\ell \alpha}(r)
=
\int d\ln k\,
{\cal P}_{\zeta_G}(k)\,
q_{\ell \alpha}(k)\,
j_{\ell}(kr).
\label{eq:R_la_definition}
\end{equation}

The functions \(\mathcal{R}_{\ell \alpha}(r)\) are the real-space radial images of the orthonormal spectral directions \(q_{\ell \alpha}(k)\). While \(q_{\ell \alpha}(k)\) specifies the deformation in Fourier--Bessel space, \(\mathcal{R}_{\ell \alpha}(r)\) gives the corresponding coherent radial profile in real space for the angular multipole \(\ell\). Thus a single amplitude \(n_{\ell m \alpha}\) adds to the curvature field the deformation $ n_{\ell m \alpha}\mathcal{R}_{\ell \alpha}(r)Y_{\ell m}(\hat{\bm x}) $. The label \(\alpha\) should therefore be understood as a radial-shape label, not as an angular quantum number. Different choices of the orthonormal basis \(q_{\ell \alpha}\) correspond to different, but equivalent, choices of coherent radial deformation modes. The Gaussian cost is fixed by the spectral normalization of the \(q_{\ell \alpha}\), not by a pointwise normalization of \(\mathcal{R}_{\ell \alpha}(r)\) in real space.

Therefore, combining the previous expressions for $\Delta B_{\ell m}$ and $\mathcal{R}_{\ell \alpha}$, the full profile can be written as
\begin{equation}
\zeta_G({\bm x})
=
\overline{\zeta}_G({\bm x})
+
\sum_{\ell,m,\alpha}
n_{\ell m \alpha}\,
\mathcal{R}_{\ell \alpha}(r)\,
Y_{\ell m}(\hat{\bm x}) .
\label{eq:profile_mean_plus_modes}
\end{equation}
Because of the normalization in Eq.~\eqref{eq:q_action_normalization}, the
additional Gaussian cost is
\begin{equation}
\Delta W
=
\sum_{\ell,m,\alpha}
n_{\ell m \alpha}^{2}.
\label{eq:finite_action_cost}
\end{equation}
Hence \(n_{\ell m \alpha}\) is the standardized Gaussian amplitude of the coherent
deformation \(\mathcal{R}_{\ell \alpha}(r)Y_{\ell m}(\hat{\bm x})\). It measures a global
deviation along one orthonormal profile direction, rather than a pointwise
fluctuation of the curvature field.

The decomposition above gives a convenient coordinate system for profiles, but one still has to explain how the statistically preferred directions are selected. In a Gaussian theory, the natural notion of distance in profile space is the quadratic action \(W\). Therefore, the representative profile associated with a given physical condition is the profile of minimum action satisfying that condition. The following subsection illustrates this point in the simplest possible setting, with a single linear constraint. This warm-up also explains why coherent deformations naturally appear in the form \(\Delta B=n{\cal P}_{\zeta_G}q\), with \(n\) an action-normalized Gaussian coordinate.

\subsection{A useful warm-up: one linear constraint}
\label{subsec:one_constraint}

Before applying the formalism to BBKS peak variables \cite{Bardeen:1985tr}, it is useful to recall how
a constrained Gaussian profile is constructed. The basic question is the
following: among all profiles satisfying a prescribed linear condition, which
one is statistically preferred? Since the Gaussian probability is controlled by
the action \(W\), the answer is the profile that minimizes \(W\) subject to the
constraint. This is the conditional mean profile associated with that
constraint.

For this purpose, consider one fixed angular sector and suppress the indices
\((\ell,m)\). Equivalently, one may think of the monopole sector, where the
coefficient is \(B_{00}(k)\). We impose one linear constraint of the form
\begin{equation}
C[B]
=
\int d\ln k\,F(k)B(k)
=
c .
\label{eq:one_constraint}
\end{equation}
The kernel \(F(k)\) specifies which physical quantity is being fixed. For
example, in the spherical sector \(F(k)=1\) fixes the value of the field at the
origin, while \(F(k)=k^2\) fixes the Laplacian, or spherical curvature, at the
origin.

Among all profiles satisfying Eq.~\eqref{eq:one_constraint}, the representative Gaussian profile is the one that minimizes
\begin{equation}
W[B]
=
\int \ddlnk\,{|B(k)|^2\over \Pz(k)}.
\label{eq:one_constraint_action}
\end{equation}
The minimum-action profile is obtained by extremizing the Gaussian action
subject to this constraint. We introduce a Lagrange multiplier $\lambda$ and minimize
\begin{equation}
\mathcal S[B,\lambda]
=
\int \ddlnk\,{|B(k)|^2\over \Pz(k)}
-2\lambda\left[
\int \ddlnk\,F(k)B(k)-c
\right].
\label{eq:one_constraint_lagrange}
\end{equation}
Varying with respect to $B(k)$ gives
\begin{equation}
{B(k)\over \Pz(k)}
=
\lambda F(k),
\label{eq:one_constraint_variation}
\end{equation}
and therefore
\begin{equation}
\overline B(k)
=
\lambda \Pz(k)F(k).
\label{eq:one_constraint_solution_step1}
\end{equation}
The multiplier is fixed by the constraint:
\begin{equation}
c
=
\lambda
\int \ddlnk\,\Pz(k)F^2(k).
\label{eq:one_constraint_lambda_step}
\end{equation}
Defining
\begin{equation}
\sigma_F^2
=
\int \ddlnk\,\Pz(k)F^2(k),
\label{eq:sigma_F_definition}
\end{equation}
we obtain
\begin{equation}
\overline B(k)
=
{c\over \sigma_F^2}\Pz(k)F(k).
\label{eq:one_constraint_Bbar}
\end{equation}
The action of this minimum-action profile is $W[\overline B]=c^2/ \sigma_F^2$. This motivates the normalized variable $n=c/ \sigma_F$, and the normalized spectral direction $q(k)=F(k)/ \sigma_F$. Then the minimum-action profile becomes $\overline B(k)=n\Pz(k)q(k)$.

The corresponding real-space coherent profile is 
\begin{equation} 
\overline{\zeta}_G(r)
=
n\,\mathcal R(r),
\qquad
\mathcal R(r)
=
\int d\ln k\,
\mathcal P_{\zeta_G}(k)q(k)j_0(kr).
\end{equation} 
in the monopole case. Thus \(q(k)\) is the action-normalized spectral direction, while \(\mathcal{R}(r)\) is its real-space radial image. The action of this profile is simply $W[\overline B]=n^2$.

This calculation explains the origin of the finite-action coordinates used throughout this work. A coherent deformation is written as \(\Delta B(k)=n{\cal P}_{\zeta_G}(k)q(k)\) because this is the minimum-action way of realizing the corresponding linear condition, and the normalization of \(q(k)\) makes \(n\) a standard Gaussian variable. However, a peak is not specified by a single linear condition. In BBKS peak theory \cite{Bardeen:1985tr} one fixes several local quantities at the same point, such as the height, the gradient and the Hessian of the field. These quantities are in general statistically correlated. Therefore, before identifying independent shape directions, one must first account for their covariance and orthogonalize the corresponding kernels. We now describe this multi-constraint construction.

\subsection{Several constraints and orthogonalized directions}
\label{subsec:several_constraints}

We now generalize the previous one-constraint example to the case in which several peak variables are fixed simultaneously. This is the situation relevant for BBKS peak theory, where the height, gradient and Hessian of the field are all local linear functionals of the Fourier--Bessel coefficients. These quantities are generally correlated in the Gaussian ensemble, and therefore the physical variables themselves are not automatically independent Gaussian directions. The role of the construction below is to separate these correlated physical constraints into orthogonal, action-normalized directions. In this way, the Gaussian cost of a coherent deformation can be written as a simple sum of squares. Let the constraints be

\begin{equation} C_{I}[B] = \int d\ln k\,F_I(k)B(k) = c_I 
, \end{equation} 

where \(I\) labels the imposed variables. The kernels \(F_I(k)\) specify which physical quantities are being fixed. For example, in the spherical BBKS sector the peak height corresponds to \(F_\nu(k)=1\), while the spherical curvature corresponds to \(F_x(k)=k^2\).

Repeating the Lagrange multiplier argument gives
\begin{equation}
\overline B(k)
=
\Pz(k)\sum_{I}\lambda_{I}F_{I}(k).
\label{eq:multi_constraint_solution}
\end{equation}
The multipliers satisfy
\begin{equation}
c_{I}
=
\sum_{J}M_{I J}\lambda_{J},
\label{eq:M_lambda_relation}
\end{equation}
where
\begin{equation}
M_{I J}
=
\int \ddlnk\,\Pz(k)F_{I}(k)F_{J}(k).
\label{eq:constraint_covariance}
\end{equation}
The matrix $M_{ I J}$ is simply the covariance matrix of the constrained variables. Thus
\begin{equation}
\lambda_{I}
=
\sum_{J}(M^{-1})_{I J}c_{J},
\label{eq:lambda_multi_solution}
\end{equation}
and
\begin{equation}
\overline B(k)
=
\Pz(k)
\sum_{I,J}
F_{I}(k)(M^{-1})_{I J}c_{J}.
\label{eq:Bbar_multi_constraint}
\end{equation}
The corresponding minimum action is
\begin{equation}
W[\overline B]
=
\sum_{I,J}
c_{I}(M^{-1})_{I J}c_{J}.
\label{eq:multi_constraint_action}
\end{equation}

For the present work, the most useful interpretation of these equations is the following. The kernels $F_{I}$ tell us which physical peak variables are fixed. The covariance matrix $M_{I J}$ tells us how correlated those variables are. Orthogonalizing the kernels with respect to the inner product
\begin{equation}
\inner{f}{g}
=
\int \ddlnk\,\Pz(k)f(k)g(k)
\label{eq:P_inner_product_general}
\end{equation}
produces independent action-normalized directions $q_\tau(k)$. In that basis a general coherent deformation takes the form
\begin{equation}
\Delta B(k)
=
\Pz(k)\sum_\tau n_\tau q_\tau(k),
\label{eq:DeltaB_action_basis_general}
\end{equation}
with
\begin{equation}
\inner{q_A}{q_B}
=
\delta_{AB},
\label{eq:qA_orthonormal_general}
\end{equation}
and therefore
\begin{equation}
\Delta W
=
\sum_\tau n_\tau^2.
\label{eq:DeltaW_action_basis_general}
\end{equation}

This is the finite-action coordinate system used throughout the rest of the work. The original variables \(c_I\) are physical peak variables, while the orthogonalized variables \(n_{\tau}\) are statistically independent Gaussian coordinates. In the next subsection we apply this construction to the BBKS spherical sector. The height and curvature of the peak will appear as two correlated constraints, and the Gram--Schmidt procedure above will produce the independent height direction \(q_\nu\) and curvature direction \(q_x\).

\subsection{The BBKS sector in action-normalized variables}
\label{subsec:BBKS_ellipsoidal_sector}

We now apply the multi-constraint construction of
Sec.~\ref{subsec:several_constraints} to the local peak variables used in BBKS.
The local data of a peak are encoded in the height, the gradient and the
Hessian of the Gaussian curvature field. Since we work at a peak centered at
the origin, the dipole sector is set to zero. The remaining local BBKS variables
are the height, the trace of the Hessian, and its traceless part.

We first consider the monopole sector, \(\ell=0\). At the origin \(j_0(0)=1\),
and therefore
\begin{equation}
\zeta_G(0)=\int \ddlnk\,B_{00}(k).
\label{eq:zeta0_B00}
\end{equation}
Thus the height corresponds to the kernel
\begin{equation}
F_\nu(k)=1,
\qquad
\sigma_0^2=\int\ddlnk\,\Pz(k),
\qquad
\nu=\frac{\zeta_G(0)}{\sigma_0}.
\label{eq:nu_definition_detailed}
\end{equation}
The associated action-normalized direction is
\begin{equation}
q_\nu(k)=\frac{1}{\sigma_0},
\qquad
\mathcal{R}_\nu(r)=\int\ddlnk\,\Pz(k)q_\nu(k)j_0(kr).
\label{eq:q_nu_detailed}
\end{equation}

The second monopole variable is the spherical curvature of the peak. Since
\(\nabla^2 j_0(kr)=-k^2j_0(kr)\), one has
\begin{equation}
-\nabla^2\zeta_G(0)
=
\int\ddlnk\,k^2B_{00}(k).
\label{eq:laplacian_B00}
\end{equation}
We define the spectral moments and normalized curvature by
\begin{equation}
\sigma_j^2=\int\ddlnk\,k^{2j}\Pz(k),
\qquad
x_{\rm B}=\frac{-\nabla^2\zeta_G(0)}{\sigma_2}.
\label{eq:x_definition_detailed}
\end{equation}
The normalized height and curvature are correlated, with
\begin{equation}
\gamma_{\rm BBKS}
\equiv
\langle \nu x_{\rm B}\rangle
=
\frac{\sigma_1^2}{\sigma_0\sigma_2}.
\label{eq:height_curvature_overlap}
\end{equation}
Therefore the curvature direction must be orthogonalized against the height
direction. The Gram--Schmidt construction gives
\begin{equation}
q_x(k)
=
\frac{k^2/\sigma_2-\gamma_{\rm BBKS}/\sigma_0}{\sqrt{1-\gamma_{\rm BBKS}^2}},
\qquad
n_x
=
\frac{x_{\rm B}-\gamma_{\rm BBKS}\nu}{\sqrt{1-\gamma_{\rm BBKS}^2}},
\label{eq:q_x_detailed}
\end{equation}
with
\begin{equation}
\inner{q_\nu}{q_x}=0,
\qquad
\inner{q_x}{q_x}=1.
\label{eq:qx_orthonormality_check}
\end{equation}
The corresponding radial function is
\begin{equation}
\mathcal{R}_x(r)
=
\int\ddlnk\,\Pz(k)q_x(k)j_0(kr).
\label{eq:R_x_detailed}
\end{equation}
Thus the spherical BBKS conditional mean profile at fixed height and curvature
is
\begin{equation}
\overline{\zeta}_{G,\ell=0}^{\rm BBKS}(r)
=
\nu \mathcal{R}_\nu(r)+n_x \mathcal{R}_x(r).
\label{eq:spherical_BBKS_profile_detailed}
\end{equation}
This form makes explicit that \(q_\nu\) fixes the central amplitude, while
\(q_x\) fixes only the curvature component statistically independent of the
height. Indeed,
\begin{equation}
\mathcal{R}_\nu(0)=\sigma_0,
\qquad
\mathcal{R}_x(0)=0,
\end{equation}
and
\begin{equation}
-\nabla^2 \mathcal{R}_\nu(0)=\gamma_{\rm BBKS} \sigma_2,
\qquad
-\nabla^2 \mathcal{R}_x(0)=\sigma_2\sqrt{1-\gamma_{\rm BBKS}^2},
\end{equation}
so that
\begin{equation}
x_{\rm B}=\gamma_{\rm BBKS}\nu+\sqrt{1-\gamma_{\rm BBKS}^2}\,n_x.
\label{eq:xB_from_nx}
\end{equation}

We now include the traceless part of the Hessian, which describes the local
ellipsoidal deformation of the peak. Defining
\begin{equation}
H_{ij}=-\partial_i\partial_j\zeta_G(0),
\qquad
H_{ii}=x_{\rm B}\sigma_2,
\end{equation}
the traceless Hessian is
\begin{equation}
T_{ij}
=
H_{ij}
-
\frac{x_{\rm B}\sigma_2}{3}\delta_{ij},
\qquad
T_{ii}=0.
\label{eq:Tij_definition_detailed}
\end{equation}
This sector has five independent components and is carried by the quadrupolar
modes \(\ell=2\). The leading quadrupolar spectral kernel is proportional to
\(k^2\), because the Hessian contains two spatial derivatives. Hence
\begin{equation}
q_{2,0}(k)=\frac{k^2}{\sigma_2},
\qquad
\mathcal{R}_{2,0}(r)=
\int\ddlnk\,\Pz(k)q_{2,0}(k)j_2(kr).
\label{eq:q20_BBKS_detailed}
\end{equation}
No subtraction of the height direction is needed, since the \(\ell=2\) angular
sector is automatically orthogonal to the monopole sector. Finally, the typical BBKS profile including height, spherical curvature and local ellipsoidal
deformation can therefore be written as
\begin{equation}
\overline{\zeta}_{G}^{\rm BBKS}(\bm x)
=
\nu \mathcal{R}_\nu(r)
+
n_x \mathcal{R}_x(r)
+
\sum_{m=-2}^{2}
n_{2m}\mathcal{R}_{2,0}(r)Y_{2m}(\hat{\bm x}).
\label{eq:BBKS_profile_detailed}
\end{equation}
The coefficients \(n_{2m}\) are the five action-normalized components of the
traceless Hessian.

It is useful to check that Eq.~\eqref{eq:BBKS_profile_detailed} reproduces the
usual BBKS conditional mean profile (see Eq.~(7.8) in \cite{Bardeen:1985tr}). Let
\begin{equation}
\xi(r)
\equiv
\langle \zeta_G(\bm 0)\zeta_G(\bm r)\rangle
=
\int\ddlnk\,\Pz(k)j_0(kr),
\qquad
\psi(r)=\frac{\xi(r)}{\sigma_0^2},
\end{equation}
and introduce
\begin{equation}
R_{\star}=\sqrt{3}\frac{\sigma_1}{\sigma_2},
\qquad
\varrho=\frac{r}{R_{\star}}.
\label{eq:definitionsRstar}
\end{equation}
Derivatives below are taken with respect to \(\varrho\). The height direction
gives
\begin{equation}
\mathcal{R}_\nu(r)=\sigma_0\psi(r).
\end{equation}
Using Eq.~\eqref{eq:q_x_detailed}, the independent curvature direction gives
\begin{equation}
\mathcal{R}_x(r)
=
-\frac{\sigma_0}{\gamma_{\rm BBKS}\sqrt{1-\gamma_{\rm BBKS}^2}}
\left(
\gamma_{\rm BBKS}^2\psi+\frac{\nabla_*^2\psi}{3}
\right),
\label{eq:Rx_BBKS_form}
\end{equation}
where \(\nabla_*^2\) denotes the Laplacian with respect to \(\varrho\). Therefore
\begin{equation}
\nu \mathcal{R}_\nu(r)+n_x \mathcal{R}_x(r)
=
\sigma_0
\left[
\frac{\nu}{1-\gamma_{\rm BBKS}^2}
\left(
\psi+\frac{\nabla_*^2\psi}{3}
\right)
-
\frac{x_{\rm B}/\gamma_{\rm BBKS}}{1-\gamma_{\rm BBKS}^2}
\left(
\gamma_{\rm BBKS}^2\psi+\frac{\nabla_*^2\psi}{3}
\right)
\right].
\label{eq:spherical_BBKS_recovered}
\end{equation}
This is the spherically averaged BBKS peak profile at fixed \(\nu\) and \(x_{\rm B}\). The quadrupolar radial function can also be written in the standard BBKS form,
\begin{equation}
\mathcal{R}_{2,0}(r)
=
-\frac{\sigma_0}{\gamma_{\rm BBKS}}
\left(
\frac{\psi'}{\varrho}
-
\frac{\nabla_*^2\psi}{3}
\right),
\label{eq:R20_BBKS_form}
\end{equation}
where the prime denotes \(d/d\varrho\). 

The BBKS variables \(e\) and \(p\) parameterize the traceless part of the
local Hessian. Equivalently, if \(\lambda_i\) are the eigenvalues of the curvature matrix,
ordered as \(\lambda_1\geq\lambda_2\geq\lambda_3\),
\[
H_{ij}\equiv -\partial_i\partial_j\zeta_G(0),
\qquad
x_{\rm B}
=
\frac{\lambda_1+\lambda_2+\lambda_3}{\sigma_2},
\]
then, in the principal-axis frame,
\[
e=
\frac{\lambda_1-\lambda_3}{2\sigma_2 x_{\rm B}},
\qquad
p=
\frac{\lambda_1-2\lambda_2+\lambda_3}{2\sigma_2 x_{\rm B}}.
\]
Thus
\[
\lambda_1=
\frac{\sigma_2x_{\rm B}}{3}(1+3e+p),
\qquad
\lambda_2=
\frac{\sigma_2x_{\rm B}}{3}(1-2p),
\qquad
\lambda_3=
\frac{\sigma_2x_{\rm B}}{3}(1-3e+p).
\]
The trace \(x_{\rm B}\) contributes to the spherical profile in Eq.~\eqref{eq:spherical_BBKS_recovered}, while
the traceless combinations proportional to \(e\) and \(p\) are encoded in the
quadrupolar \(\ell=2\) coefficients. In the principal-axis frame, we choose the
five quadrupolar coefficients such that

\begin{equation}
\sum_{m=-2}^{2}n_{2m}Y_{2m}(\hat{\bm x})
=
-\frac{5}{2}x_{\rm B}\mathcal A_{ep}(\hat{\bm x}),
\label{eq:n2m_to_ep}
\end{equation}
where \(\mathcal A_{ep}\) is the usual BBKS angular function written in terms of
the ellipticity \(e\) and prolateness \(p\). With the convention stated above, the \(m\)-sum in
Eq.~\eqref{eq:n2m_to_ep} denotes the five real quadrupolar components of the
traceless Hessian. In the principal-axis frame, and with the convention for
\(\mathcal{A}_{ep}\) used here, only two of them are nonzero. Writing
\[
Y_{20}
=
\frac{\sqrt 5}{2}
\left(3\frac{z^2}{r^2}-1\right),
\qquad
Y^{(c)}_{22}
=
\frac{\sqrt{15}}{2}
\frac{x^2-y^2}{r^2},
\]
one has
\[
\mathcal{A}_{ep}
=
\frac{3e+p}{\sqrt 5}Y_{20}
+
\sqrt{\frac{3}{5}}(e-p)Y^{(c)}_{22}.
\]
Therefore
\[
n_{20}
=
-\frac{\sqrt 5}{2}x_{\rm B}(3e+p),
\qquad
n^{(c)}_{22}
=
-\frac{\sqrt{15}}{2}x_{\rm B}(e-p),
\]
while the \(m=1\) components and the sine-type \(m=2\) component vanish in the
principal-axis frame.

With the normalization used in this
work,
\begin{equation}
\int\frac{d\Omega}{4\pi}\,\mathcal A_{ep}(\hat{\bm x})=0,
\qquad
\int\frac{d\Omega}{4\pi}\,\mathcal A_{ep}^2(\hat{\bm x})
=
\frac{4}{5}(3e^2+p^2).
\label{eq:Aep_normalization_section2}
\end{equation}
Then the quadrupolar contribution becomes
\begin{equation}
\sum_{m=-2}^{2}n_{2m}\mathcal{R}_{2,0}(r)Y_{2m}(\hat{\bm x})
=
\sigma_0\,\frac{5}{2}\frac{x_{\rm B}}{\gamma_{\rm BBKS}}
\left(
\frac{\psi'}{\varrho}
-
\frac{\nabla_*^2\psi}{3}
\right)
\mathcal A_{ep}(\hat{\bm x}).
\label{eq:quadrupole_BBKS_recovered}
\end{equation}
Combining the spherical and quadrupolar pieces gives
\begin{equation}
\frac{\bar{\zeta}_G^{\rm BBKS}(\bm x)}{\sigma_0}
=
\frac{\nu}{1-\gamma_{\rm BBKS}^2}
\left(
\psi+\frac{\nabla_*^2\psi}{3}
\right)
-
\frac{x_{\rm B}/\gamma_{\rm BBKS}}{1-\gamma_{\rm BBKS}^2}
\left(
\gamma_{\rm BBKS}^2\psi+\frac{\nabla_*^2\psi}{3}
\right)
+
\frac{5}{2}\frac{x_{\rm B}}{\gamma_{\rm BBKS}}
\left(
\frac{\psi'}{\varrho}
-
\frac{\nabla_*^2\psi}{3}
\right)
\mathcal A_{ep}(\hat{\bm x}).
\label{eq:BBKS_standard_profile_recovered}
\end{equation}
This is the standard BBKS conditional mean profile written in the present Fourier--Bessel notation. The quadrupolar action is fixed by the norm of the five coefficients. With the angular normalization used here, the full Gaussian action of the BBKS sector is therefore
\begin{equation}
W_{\rm BBKS}
=
\nu^2+n_x^2+\sum_{m=-2}^{2}n_{2m}^2 = \nu^2+n_x^2+5x_{\rm B}^2(3e^2+p^2).
\label{eq:W_BBKS_detailed}
\end{equation}
Thus the BBKS height, independent spherical curvature and local ellipsoidal
variables are recovered as the first action-normalized directions in the
\(\ell=0\) and \(\ell=2\) sectors. All additional finite-action modes orthogonal
to these directions represent genuine residual profile dispersion beyond the
local BBKS peak variables.

\subsection{Finite-action dispersion beyond BBKS and PBH mass function estimation}
\label{subsec:finite_action_dispersion_BBKS}

We now define precisely what is meant by shape dispersion beyond the BBKS peak variables. The BBKS construction fixes the local behavior of the field through second order around the peak: the height, the gradient, the spherical curvature, and the traceless Hessian. However, these quantities do not determine the full curvature profile. Infinitely many coherent configurations can share the same BBKS variables while differing in their radial structure or in higher angular multipoles. In the present formalism, these additional degrees of freedom are simply the orthogonal complement of the BBKS directions in the Gaussian-action metric. They are finite-action deformations that leave the chosen BBKS variables fixed. We therefore write

\begin{equation}
\zeta_G(\bm x)
=
\overline{\zeta}_{G}^{\rm BBKS}(\bm x)
+
\chi^{\perp}(\bm x),
\label{eq:zeta_BBKS_plus_chi_detailed}
\end{equation}
where the superscript $\perp$ means that the residual field does not change the BBKS variables already fixed. In spectral language, this means that the residual kernels are orthogonal to the BBKS kernels in the Gaussian inner product. For example, residual spherical modes must not change the height or the curvature. Therefore they satisfy
\begin{equation}
\int \ddlnk\,\Pz(k)q_{0 \alpha}^{\perp}(k)q_{\nu}(k)
=
0,
\label{eq:orth_height_detailed}
\end{equation}
\begin{equation}
\int \ddlnk\,\Pz(k)q_{0 \alpha}^{\perp}(k)q_x(k)
=
0.
\label{eq:orth_curvature_detailed}
\end{equation}
Similarly, residual quadrupolar modes at fixed Hessian must satisfy
\begin{equation}
\int \ddlnk\,\Pz(k)q_{2 \alpha}^{\perp}(k)q_{2,0}(k)
=
0.
\label{eq:orth_quadrupole_detailed}
\end{equation}

Equivalently, in the derivative basis $q_{\ell \alpha}$, fixing the BBKS variables removes the lowest radial directions in the sectors $\ell=0$, $\ell=1$, and $\ell=2$. For peaks defined directly in $\zeta_G$, the residual index sets are
\begin{equation}
\ell=0:\qquad \alpha\geq 2,
\label{eq:residual_indices_l0}
\end{equation}
\begin{equation}
\ell=1:\qquad \alpha\geq 1,
\label{eq:residual_indices_l1}
\end{equation}
\begin{equation}
\ell=2:\qquad \alpha\geq 1,
\label{eq:residual_indices_l2}
\end{equation}
and
\begin{equation}
\ell\geq 3:\qquad \alpha\geq 0.
\label{eq:residual_indices_lge3}
\end{equation}
The exclusions have a direct interpretation. The mode $q_{0,0}$ fixes the peak height, while $q_{0,1}$ fixes the independent spherical curvature $x_{\rm B}$ at fixed height. The mode $q_{1,0}$ fixes the gradient and is set to zero when the peak is centered at the origin. Finally, $q_{2,0}$ fixes the traceless Hessian, namely the local ellipsoidal deformation. The remaining modes therefore describe genuine residual dispersion beyond the BBKS variables. These conditions avoid double counting. Without them, a mode called ``dispersion'' could simply change the height, curvature, or local ellipsoid already included in the reference BBKS profile.

The residual field is expanded as
\begin{equation}
\chi^{\perp}(\bm x)
=
\sum_{\ell,m,\alpha}
n_{\ell m \alpha}\mathcal{R}_{\ell \alpha}^{\perp}(r){Y}_{\ell m}(\hat{\bm x}),
\label{eq:chi_perp_expansion_detailed}
\end{equation}
where
\begin{equation}
\mathcal{R}_{\ell \alpha}^{\perp}(r)
=
\int \ddlnk\,\Pz(k)q_{\ell \alpha}^{\perp}(k)j_{\ell}(kr),
\label{eq:R_perp_detailed}
\end{equation}
and
\begin{equation}
\int \ddlnk\,\Pz(k)q_{\ell \alpha}^{\perp}(k)q_{\ell \eta}^{\perp}(k)
=
\delta_{\alpha \eta}.
\label{eq:q_perp_normalization_detailed}
\end{equation}
The additional action is therefore
\begin{equation}
\Delta W
=
\sum_{\ell,m,\alpha}n_{\ell m \alpha}^2.
\label{eq:dispersion_cost_detailed}
\end{equation}
The coefficients \(n_{\ell m \alpha}\) in the residual sum are standardized Gaussian
amplitudes of coherent deformations orthogonal to the BBKS sector. They should
not be confused with the pointwise conditional variance of the field around a
peak (see Eq.~(7.9) in \cite{Bardeen:1985tr}).\footnote{
The pointwise conditional variance in \cite{Bardeen:1985tr} is a local quantity: it gives the root mean square (rms)
fluctuation of the field at each radius after imposing the peak constraints.
It does not, however, define a coherent profile deformation. In particular,
because the field becomes decorrelated from the peak constraints at large
radius, the pointwise conditional rms generally approaches the unconditional
rms rather than decaying. Therefore a profile obtained by adding the same sign
of this rms envelope at all radii would not represent a typical finite-action
realization. This is analogous to the distinction emphasized in
Ref.~\cite{Escriva:2025ftp}. In the present construction, the variables \(n_{\ell m \alpha}\)
instead multiply complete action-normalized modes, orthogonal to the fixed
BBKS variables, with finite Gaussian cost
$\Delta W$.}
A single realization is obtained by drawing independent coefficients for these
coherent modes, not by adding the same-sign pointwise rms envelope at every
radius.

A practical way to generate radial functions is to start from the derivative sequence
\begin{equation}
k^{\ell},
\qquad
k^{\ell+2},
\qquad
k^{\ell+4},
\qquad
\ldots
\label{eq:kernel_sequence_detailed}
\end{equation}
and orthogonalize it with respect to $\inner{f}{g}$. The first normalized function is
\begin{equation}
q_{\ell,0}(k)
=
{k^{\ell}\over \sigma_{\ell}},
\label{eq:q_l0_detailed}
\end{equation}
and the next one is
\begin{equation}
q_{\ell,1}(k)
=
{ k^{\ell+2}-(\sigma_{\ell+1}^2/\sigma_{\ell}^2)k^{\ell}
\over
\left(\sigma_{\ell+2}^2-\sigma_{\ell+1}^4/\sigma_{\ell}^2\right)^{1/2} }.
\label{eq:q_l1_detailed}
\end{equation}
For $\ell=0$, the first two directions are the height and curvature sectors, and the following directions describe extra spherical radial-shape dispersion. For $\ell=2$, the first direction is the BBKS Hessian mode, and the following ones describe quadrupolar radial dispersion at fixed Hessian. For $\ell\geq 3$, the modes describe higher multipolar deformations beyond the local ellipsoid.

The reason higher multipoles are automatically beyond the BBKS Hessian is the small-radius expansion around the peak centre $r \rightarrow 0$,
\begin{equation}
j_{\ell}(kr)
=
{(kr)^{\ell}\over (2\ell+1)!!}
+
{\cal O}(r^{\ell+2}).
\label{eq:bessel_small_r_detailed}
\end{equation}
Thus $\ell=3$ begins at cubic order in $r$, $\ell=4$ begins at quartic order, and so on. These modes do not modify the height, the gradient, or the Hessian at the origin.

Combining the BBKS sector and the residual modes, the total action is
\begin{equation}
W
=
\nu^2+n_x^2+
\sum_{m=-2}^{2}n_{2m}^2
+
\overset{\perp}{\sum}_{\ell,m,\alpha} n_{\ell m \alpha}^{2}.
\label{eq:W_total_detailed}
\end{equation}

This equation summarizes the statistical content of the construction. The BBKS variables define the local peak data and the corresponding
conditional mean profile, which serves as the reference configuration,
while the residual coefficients \(n_{\ell m\alpha}\) describe coherent
departures from this profile with a controlled Gaussian cost. These residual modes are the quantities whose effect on the collapse threshold and PBH abundance we wish to assess. In the next section we specialize this general picture to the monochromatic limit. In that limit the radial freedom inside each angular multipole collapses to a single spherical Bessel envelope, so the remaining dispersion is purely angular. This provides a simple setting in which the BBKS ellipsoid can be extended to higher multipoles.

The finite-action decomposition also gives a natural prescription for including
profile dispersion in PBH abundance calculations using \cite{Bardeen:1985tr}.  Let
\(\bm n=\{n_{\tau}\}\) denote a set of action-normalized coherent dispersion
variables around the chosen reference profile.  These variables describe
residual finite-action deformations and should not be confused with the BBKS
local peak variables, such as the height \(\nu\), the spherical curvature
\(x_{\rm B}\), or the ellipsoidal variables \(e\) and \(p\).  After conditioning
on the variables that define the reference profile, the Gaussian probability
density for the remaining coherent shape degrees of freedom is
\begin{equation}
P_{\rm sh}(\bm n)
=
\frac{1}{(2\pi)^{N/2}}
\exp\left[-\frac12\sum_{\tau} n_\tau^2\right] .
\label{eq:Pshape_general}
\end{equation}

At fixed peak height \(\nu\), we define
\begin{equation}
x_*(\nu)\equiv \nu \,  \gamma_{\rm BBKS},
\label{eq:xstar_BBKS}
\end{equation}
which is the centre of the conditional Gaussian distribution of the
stochastic spherical-curvature variable \(x_{\rm B}\). The relevant peak-theory measure is the BBKS differential number density of peaks, whose joint distribution in peak height and spherical curvature
is given by
\begin{equation}
{\cal N}_{\rm pk}^{\rm BBKS}(\nu,x_{\rm B})
=
\frac{1}{(2\pi)^2R_{\star}^3}
\exp\left(-\frac{\nu^2}{2}\right)
\frac{
\exp\left[
-\frac{\left[x_{\rm B}-x_*(\nu)\right]^2}
{2(1-\gamma_{\rm BBKS}^2)}
\right]}
{\sqrt{2\pi(1-\gamma_{\rm BBKS}^2)}}
f_{\rm BBKS}(x_{\rm B}).
\label{eq:BBKS_joint_peak_density}
\end{equation}
Here \(x_{\rm B}\) is the normalized spherical curvature of the peak, \(f_{\rm BBKS}(x)\) is the BBKS curvature weight enforcing the maximum
condition and $\gamma_{\rm BBKS}$ and $R_{\star}$ were introduced in Eqs.~\eqref{eq:height_curvature_overlap} and \eqref{eq:definitionsRstar}, respectively. Explicitly,
\begin{align}
f_{\rm BBKS}(x_{\rm B})
&=
\frac{x_{\rm B}^3-3x_{\rm B}}{2}
\left[
\operatorname{erf}\!\left(\sqrt{\frac{5}{2}}\,x_{\rm B}\right)
+
\operatorname{erf}\!\left(\sqrt{\frac{5}{8}}\,x_{\rm B}\right)
\right]
\nonumber\\
&\quad
+
\sqrt{\frac{2}{5\pi}}
\left[
\left(\frac{31}{4}x_{\rm B}^2+\frac{8}{5}\right)
\exp\!\left(-\frac{5x_{\rm B}^2}{8}\right)
+
\left(\frac{x_{\rm B}^2}{2}-\frac{8}{5}\right)
\exp\!\left(-\frac{5x_{\rm B}^2}{2}\right)
\right] .
\label{eq:fBBKS_definition}
\end{align}
The height-only differential peak density is obtained by integrating over the
curvature variable,
\begin{equation}
{\cal N}_{\rm pk}^{\rm BBKS}(\nu)
=
\int_0^\infty dx_{\rm B}\,
{\cal N}_{\rm pk}^{\rm BBKS}(\nu,x_{\rm B}) .
\label{eq:BBKS_height_peak_density_integral}
\end{equation}
Equivalently,
\begin{equation}
{\cal N}_{\rm pk}^{\rm BBKS}(\nu)
=
\frac{1}{(2\pi)^2R_{\star}^3}
\exp\left(-\frac{\nu^2}{2}\right)
G_{\rm BBKS}
\!\left(\gamma_{\rm BBKS},x_*(\nu)\right).
\label{eq:BBKS_height_peak_density}
\end{equation}
where the curvature-integrated BBKS function is defined by
\begin{equation}
G_{\rm BBKS}(\gamma_{\rm BBKS},x_*)
\equiv
\int_0^\infty {\rm d}x_{\rm B}\,
f_{\rm BBKS}(x_{\rm B})
\frac{
\exp\left[
-\frac{(x_{\rm B}-x_*)^2}{2(1-\gamma_{\rm BBKS}^2)}
\right]}
{\sqrt{2\pi(1-\gamma_{\rm BBKS}^2)}} .
\label{eq:GBBKS_definition}
\end{equation}
The cumulative peak density above a threshold is then
\begin{equation}
N_{\rm pk}^{\rm BBKS}(>\nu_c)
=
\int_{\nu_c}^{\infty}d\nu\,
{\cal N}_{\rm pk}^{\rm BBKS}(\nu).
\label{eq:BBKS_cumulative_density}
\end{equation}

If the local ellipsoidal variables are retained explicitly, the BBKS curvature
weight can be resolved into a conditional distribution of ellipticity and
prolateness,
\begin{equation}
{\cal N}_{\rm pk}^{\rm BBKS}(\nu,x_{\rm B},e,p)
=
{\cal N}_{\rm pk}^{\rm BBKS}(\nu,x_{\rm B})\,
P_{\rm ell}(e,p\mid x_{\rm B}),
\label{eq:BBKS_joint_peak_density_ep}
\end{equation}

where
\begin{equation}
P_{\rm ell}(e,p\mid x_{\rm B})
=
\frac{3^2\,5^{5/2}}{\sqrt{2\pi}}\,
\frac{x_{\rm B}^8}{f_{\rm BBKS}(x_{\rm B})}
\,e\,(e^2-p^2)\,(1-2p)
\left[(1+p)^2-9e^2\right]
\exp\left[
-\frac{5}{2}x_{\rm B}^2(3e^2+p^2)
\right].
\end{equation}
for $(e,p)\in\mathcal D_{ep}$, and zero otherwise, with
\begin{equation}
\int_{\mathcal D_{ep}}de\,dp\,
P_{\rm ell}(e,p\mid x_{\rm B})=1.
\label{eq:Pell_normalization}
\end{equation}
The allowed domain is
\begin{equation}
{\cal D}_{ep}
=
\left\{
0\leq e\leq \frac14,\ -e\leq p\leq e
\right\}
\cup
\left\{
\frac14\leq e\leq \frac12,\ 3e-1\leq p\leq e
\right\}.
\label{eq:ep_domain}
\end{equation}
Thus \(f_{\rm BBKS}(x_{\rm B})\) already contains the integrated contribution
of the anisotropic Hessian sector.  Introducing \(P_{\rm ell}(e,p\mid x_{\rm B})\) only resolves this weight into ellipsoidal configurations; it should not be
counted a second time. In this general notation the collapse threshold $\mu_c$ may depend on the curvature,
ellipticity, prolateness, and residual coherent shape variables,
\begin{equation}
\mu_c
=
\mu_c(x_{\rm B},e,p,\bm n),
\qquad
\nu_c(x_{\rm B},e,p,\bm n)
=
\frac{\mu_c(x_{\rm B},e,p,\bm n)}{\sigma_0}.
\label{eq:shape_threshold_general}
\end{equation}
At the level of a cumulative threshold estimate, the shape-dispersed abundance
is obtained by integrating the BBKS peak abundance over the local non-amplitude
peak variables and over the coherent dispersion variables,
\begin{align}
\beta_{\rm disp}
&\propto
\int
\left(\prod_{\tau=1}^{N}{\rm d}n_\tau\right)
P_{\rm sh}(\bm n)
\int_0^\infty dx_{\rm B}
\int_{\mathcal D_{ep}}de\,dp\,
P_{\rm ell}(e,p\mid x_{\rm B})
\int_{\nu_c(x_{\rm B},e,p,\bm n)}^\infty
d\nu\,
{\cal N}_{\rm pk}^{\rm BBKS}(\nu,x_{\rm B}) .
\label{eq:beta_disp_general}
\end{align}
This expression makes explicit the competition between the Gaussian cost of
realizing a coherent deformation from $P_{\rm sh}(\bm n)$ and the exponential gain associated with a
lower collapse threshold from ${\cal N}_{\rm pk}^{\rm BBKS}(\nu,x_{\rm B})$.  The dominant contribution is therefore not
necessarily the reference profile, nor necessarily the profile with the lowest
threshold, but the configuration that optimizes the combined statistical weight.

For the PBH mass function, the cumulative threshold estimate is replaced by the
differential peak density together with the map from peak variables to PBH mass.
We denote this map by
\begin{equation}
M
=
{\cal M}(\nu,x_{\rm B},e,p,\bm n),
\label{eq:general_mass_map}
\end{equation}
In the usual critical-collapse approximation when $\nu \rightarrow \nu_c$ \cite{Evans:1994pj,Niemeyer:1999ak} one may
write 
\begin{equation}
{\cal M}(\nu,x_{\rm B},e,p,\bm n)
=
\mathcal{K}(x_{\rm B},e,p,\bm n)\,
M_H(x_{\rm B},e,p,\bm n) \sigma_0^{\gamma_{\rm cr}}
\left[
\nu-\nu_c(x_{\rm B},e,p,\bm n)
\right]^{\gamma_{\rm cr}} ,
\label{eq:critical_mass_map_general}
\end{equation}
where $\gamma_{\rm cr} \approx  0.356$ for a radiation-dominated Universe and with the understanding that this expression applies only on the supercritical
side of the threshold.  The dependence of \(M_H\) on the non-amplitude peak
variables reflects the fact that the local curvature and shape of the peak fix,
or contribute to fixing, the characteristic compaction scale of the perturbation. In the numerical implementation below, we also allow $M_H$ to acquire an
explicit dependence on the peak amplitude through the amplitude-dependent
location of the compaction maximum, $r_m$.

The shape-dispersed PBH mass function can then be written in the general
delta-function form
\begin{align}
\frac{df_{\rm PBH}}{d\ln M}
&=
\frac{M}{\rho_{\rm DM}}
\int
\left(\prod_{\tau=1}^{N}{\rm d}n_\tau\right)
P_{\rm sh}(\bm n)
\int_0^\infty dx_{\rm B}
\int_{\mathcal D_{ep}}de\,dp\,
P_{\rm ell}(e,p\mid x_{\rm B})
\int d\nu\,
{\cal N}_{\rm pk}^{\rm BBKS}(\nu,x_{\rm B})
\nonumber\\
&\hspace{1.5cm}\times
\Theta\!\left[\nu-\nu_c(x_{\rm B},e,p,\bm n)\right]\,
\delta_D
\!\left[
\ln M-\ln {\cal M}(\nu,x_{\rm B},e,p,\bm n)
\right] .
\label{eq:mass_function_shape_disp_general_delta}
\end{align}
Here the BBKS factor counts peaks in the Gaussian field, the conditional
distribution \(P_{\rm ell}\) weights the local ellipsoidal shapes, the Gaussian
factor \(P_{\rm sh}\) weights the residual coherent deformations, the step
function imposes the collapse condition, and the Dirac delta implements the map
from peak variables to the PBH mass.

If the mass map can be inverted for the peak height at fixed
\((x_{\rm B},e,p,\bm n)\), we can define $M
= {\cal M}\!\left(\nu_M(x_{\rm B},e,p,\bm n),
x_{\rm B},e,p,\bm n \right)$, and using the Jacobian of the transformation (see for instance \cite{Yoo:2018kvb}) then Eq.~\eqref{eq:mass_function_shape_disp_general_delta} becomes
\begin{align}
\frac{df_{\rm PBH}}{d\ln M}
&=
\frac{M}{\rho_{\rm DM}}
\int
\left(\prod_{\tau=1}^{N}{\rm d}n_\tau\right)
P_{\rm sh}(\bm n)
\int_0^\infty dx_{\rm B}
\int_{\mathcal D_{ep}}de\,dp\,
P_{\rm ell}(e,p\mid x_{\rm B})
\nonumber\\
&\hspace{1.5cm}\times
{\cal N}_{\rm pk}^{\rm BBKS}
\!\left(
\nu_M(x_{\rm B},e,p,\bm n),
x_{\rm B}
\right)
\left|
\frac{\partial\ln {\cal M}}
{\partial\nu}
\right|^{-1}_{x_{\rm B},e,p,\bm n}
\nonumber\\
&\hspace{1.5cm}\times
\Theta
\!\left[
\nu_M(x_{\rm B},e,p,\bm n)
-
\nu_c(x_{\rm B},e,p,\bm n)
\right] .
\label{eq:mass_function_shape_disp_general_inverted}
\end{align}
The integration over \(\nu\) has been
performed through the critical-collapse mass map, leaving an integral over the
BBKS curvature, the BBKS ellipsoidal variables, and the finite-action
shape-dispersion variables. The total fraction of dark matter in the form of PBHs, including shape
dispersion, is then given by $f_{\rm PBH,tot}^{\rm disp}
=
\int d\ln M\,
(d f_{\rm PBH}/d\ln M)$.

\section{Monochromatic case: from the BBKS ellipsoid to higher multipoles}
\label{sec:monochromatic_case_general}

In this section we specialize the constrained finite-action multipolar framework to the monochromatic limit and connect it directly with the ellipsoidal profiles used in previous numerical studies of PBH formation and abundance estimation \cite{Escriva:2024hsv,Escriva:2024rsk}. The monochromatic spectrum is written as
\begin{equation}
\Pz(k)=A_\zeta\,\delta_{\rm D}\!\left(\ln\frac{k}{k_p}\right),
\label{eq:mono_spectrum_section}
\end{equation}
where $A_\zeta=\sigma_0^2$ and $k_p$ is the characteristic scale. The spectral moments are then $\sigma_j^2=A_\zeta k_p^{2j}$, so that $\gamma_{\rm BBKS}=\frac{\sigma_1^2}{\sigma_0\sigma_2}=1$. This implies that the peak height and the spherical curvature are perfectly correlated. Equivalently, every realization satisfies $\nabla^2\zeta_G=-k_p^2\zeta_G$.

Therefore, once the central height has been fixed, the spherical curvature is fixed as well. The independent curvature direction $q_x(k)$ introduced for a finite-width spectrum is singular in the strict monochromatic limit, because its normalization contains $\sqrt{1-\gamma_{\rm BBKS}^2}$. For this reason the monochromatic case must be treated separately, rather than by substituting $\gamma_{\rm BBKS}=1$ in the finite-width BBKS formulas. In the same way, each angular multipole has only one radial envelope, because all functions $k^\ell,k^{\ell+2},k^{\ell+4},\ldots$ are proportional on the support of the spectrum. Thus the monochromatic limit allows angular dispersion, but not independent radial dispersion inside a fixed multipole.

Introducing the dimensionless radius $u = k_p r$, the spherical monochromatic profile is
\begin{equation}
\zeta_{\rm sp}(r)=\mu j_0(u)=\mu\frac{\sin u}{u},
\label{eq:spherical_mono_profile_section}
\end{equation}
with $\mu=\nu\sigma_0$ where $\nu$ denotes the usual peak height in units of $\sigma_0$. The previous ellipsoidal construction is recovered by adding the BBKS quadrupolar sector. In the principal-axis frame, the angular function associated with ellipticity $e$ and prolateness $p$ is \cite{Bardeen:1985tr}
\begin{equation}
\mathcal A_{ep}(\hat{\bm x})=3e\left[1-\sin^2\theta\left(1+\sin^2\phi\right)\right]
+p\left[1-3\sin^2\theta\cos^2\phi\right].
\label{eq:Aep_spherical_section}
\end{equation}
Equivalently, in Cartesian form,
\begin{equation}
\mathcal A_{ep}(\hat{\bm x})=\frac{3e}{r^2}(z^2-y^2)+p\left[1-3\left(\frac{x}{r}\right)^2\right].
\label{eq:Aep_cartesian_section}
\end{equation}
This function is purely quadrupolar and satisfies
\begin{equation}
\int\frac{d\Omega}{4\pi}\mathcal A_{ep}(\hat{\bm x})=0.
\end{equation}
The ellipsoidal contribution to the profile may be written as
\begin{equation}
\zeta_2(\bm x)=\mu\,\mathcal F_2(u)\,\mathcal A_{ep}(\hat{\bm x}),
\label{eq:zeta2_F2_section}
\end{equation}
where
\begin{equation}
\mathcal F_2(u)=\frac{5}{2u^3}\left[3u\cos u+(u^2-3)\sin u\right].
\label{eq:F2_section}
\end{equation}
Using
\begin{equation}
j_2(u)=\frac{(3-u^2)\sin u-3u\cos u}{u^3},
\end{equation}
we obtain
\begin{equation}
\mathcal F_2(u)=-\frac{5}{2}j_2(u),
\end{equation}
and therefore
\begin{equation}
\zeta_G^{(\ell=0,2)}(\bm x)=\zeta_{\rm sp}(r)+\zeta_2(\bm x)=\mu j_0(u)-\frac{5}{2}\mu\,\mathcal A_{ep}(\hat{\bm x})j_2(u).
\label{eq:l02_profile_section}
\end{equation}
This is exactly the monochromatic BBKS ellipsoidal profile: the $\ell=0$ part controls the spherical profile, while the $\ell=2$ part fixes the traceless Hessian and therefore the local ellipsoidal deformation.

The extension beyond the ellipsoidal approximation is immediate. Since each angular multipole has a unique radial envelope in the monochromatic limit, the most general finite-action profile truncated at some $\ell_{\rm max}$ can be written as
\begin{equation}
\zeta_G(\bm x)=\mu\left[j_0(u)-\frac{5}{2}\mathcal A_{ep}(\hat{\bm x})j_2(u)+\sum_{\ell=3}^{\ell_{\rm max}}\mathcal D_\ell(\hat{\bm x})j_\ell(u)\right],
\label{eq:general_mono_profile_section}
\end{equation}
with
\begin{equation}
\mathcal D_\ell(\hat{\bm x})=a_{\ell 0}\,Y_{\ell 0}(\hat{\bm x})+
\sum_{m=1}^{\ell}\left[a^{(c)}_{\ell m}\,Y^{(c)}_{\ell m}(\hat{\bm x})+a^{(s)}_{\ell m}\, Y^{(s)}_{\ell m}(\hat{\bm x})\right].
\label{eq:Dl_realbasis_section}
\end{equation}

Here \(Y_{\ell0}\), \(Y^{(c)}_{\ell m}\) and \(Y^{(s)}_{\ell m}\), with \(m>0\),
denote an orthonormal real spherical-harmonic basis. For \(m>0\) we define
\[
Y^{(c)}_{\ell m}
=
\frac{Y_{\ell m}+(-1)^mY_{\ell,-m}}{\sqrt{2}}
=
\sqrt{2}\,\mathrm{Re}\,Y_{\ell m},
\]
and
\[
Y^{(s)}_{\ell m}
=
\frac{Y_{\ell m}-(-1)^mY_{\ell,-m}}{i\sqrt{2}}
=
\sqrt{2}\,\mathrm{Im}\,Y_{\ell m},
\]
where the last equalities use the convention
\(Y_{\ell,-m}=(-1)^mY_{\ell m}^{\ast}\). Together with \(Y_{\ell0}\), these
functions form a real orthonormal basis with respect to the angular-average
measure \(d\Omega/(4\pi)\).

Relative to a chosen polar axis, the \(m=0\) mode is axisymmetric, since it
has no azimuthal dependence. The real modes with \(m>0\), represented by
\(Y^{(c)}_{\ell m}\) and \(Y^{(s)}_{\ell m}\), contain \(\cos(m\phi)\)- and
\(\sin(m\phi)\)-type angular dependence and therefore break axisymmetry. The quadrupolar sector is special. A general \(\ell=2\) realization is
equivalent to a traceless symmetric tensor and can always be diagonalized by
a spatial rotation. In its principal-axis frame it is described by the BBKS
ellipticity and prolateness parameters, \(e\) and \(p\). Thus the familiar
BBKS ellipsoid is not a restriction to a particular quadrupolar orientation,
but rather the principal-axis representation of a general quadrupole. For \(\ell\geq 3\), the higher multipoles represent angular structure beyond
the local Hessian. After the quadrupole has been used to define the
principal-axis frame of the peak, the orientation of the higher multipoles
relative to this frame becomes an additional physical degree of freedom.
Different \(m\)-components at fixed \(\ell\) then correspond to different
azimuthal patterns with respect to the BBKS ellipsoid, even though they share
the same monochromatic radial envelope \(j_\ell(u)\).

A practical parametrization for numerical work is obtained by keeping the overall amplitude $\mu$ explicit and interpreting the coefficients $a_{\ell m}^{(c,s)}$ as relative shape amplitudes. More precisely, if $n_{\ell m}$ denotes the action-normalized amplitude of a monochromatic angular mode, then in this section we use $a_{\ell m}\equiv n_{\ell m}/ \nu$, and analogously for the real cosine and sine components. In the present relative parametrization the shape is kept fixed while $\mu$ is varied to determine the collapse threshold. Two representative examples are
\begin{equation}
\zeta_G(r,\theta,\phi)=\mu\left[j_0(u)-\frac{5}{2}\mathcal A_{ep}(\theta,\phi)j_2(u)+a_{30}\,Y_{30}(\theta)j_3(u)+a_{40}\,Y_{40}(\theta)j_4(u)\right],
\label{eq:axisymmetric_example_section}
\end{equation}
for an axisymmetric extension, and
\begin{equation}
\zeta_G(r,\theta,\phi)=\mu\left[j_0(u)-\frac{5}{2}\mathcal A_{ep}(\theta,\phi)j_2(u)+a^{(c)}_{31}\,Y^{(c)}_{31}(\theta,\phi)j_3(u)+a^{(c)}_{44}\,Y^{(c)}_{44}(\theta,\phi)j_4(u)\right],
\label{eq:nonaxisymmetric_example_section}
\end{equation}
for a non-axisymmetric extension. The first example goes beyond the BBKS ellipsoid while preserving axisymmetry, whereas the second explicitly illustrates how $m\neq0$ modes introduce azimuthal structure.

The local meaning of the different multipoles follows from the small-$u$ expansion
\begin{equation}
j_\ell(u)=\frac{u^\ell}{(2\ell+1)!!}+\mathcal O(u^{\ell+2}).
\label{eq:small_u_section}
\end{equation}
Therefore,
\begin{equation}
\mu\,\mathcal D_\ell(\hat{\bm x})j_\ell(u)=\mu\,\mathcal D_\ell(\hat{\bm x})\frac{(k_pr)^\ell}{(2\ell+1)!!}+\mathcal O(r^{\ell+2}).
\label{eq:small_u_profile_section}
\end{equation}
The $\ell=2$ contribution starts at order $r^2$ and modifies the Hessian, which is why it defines the local ellipsoid. By contrast, all $\ell\geq3$ contributions start at order $r^\ell$ and leave the height, gradient, and Hessian unchanged at the origin. They are therefore genuine deformations beyond the BBKS ellipsoid.

\begin{figure}[t]
    \centering
    \includegraphics[width=0.72\textwidth]{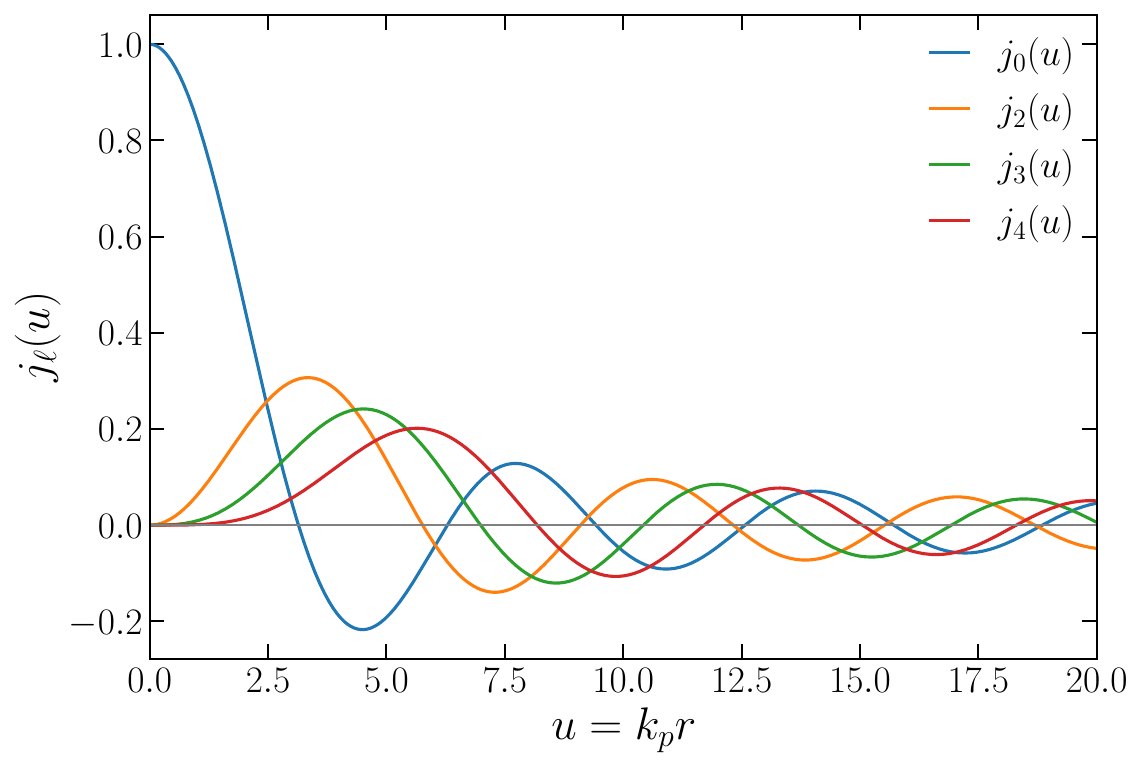}
    \caption{Radial envelopes for the first multipoles in the monochromatic case. Since the power spectrum is supported at a single scale $k_p$, each angular sector has a unique radial shape $j_\ell(u)$.}
    \label{fig:mono_radial_envelopes}
\end{figure}

\begin{figure}[p]
    \centering
    \includegraphics[width=0.96\textwidth]{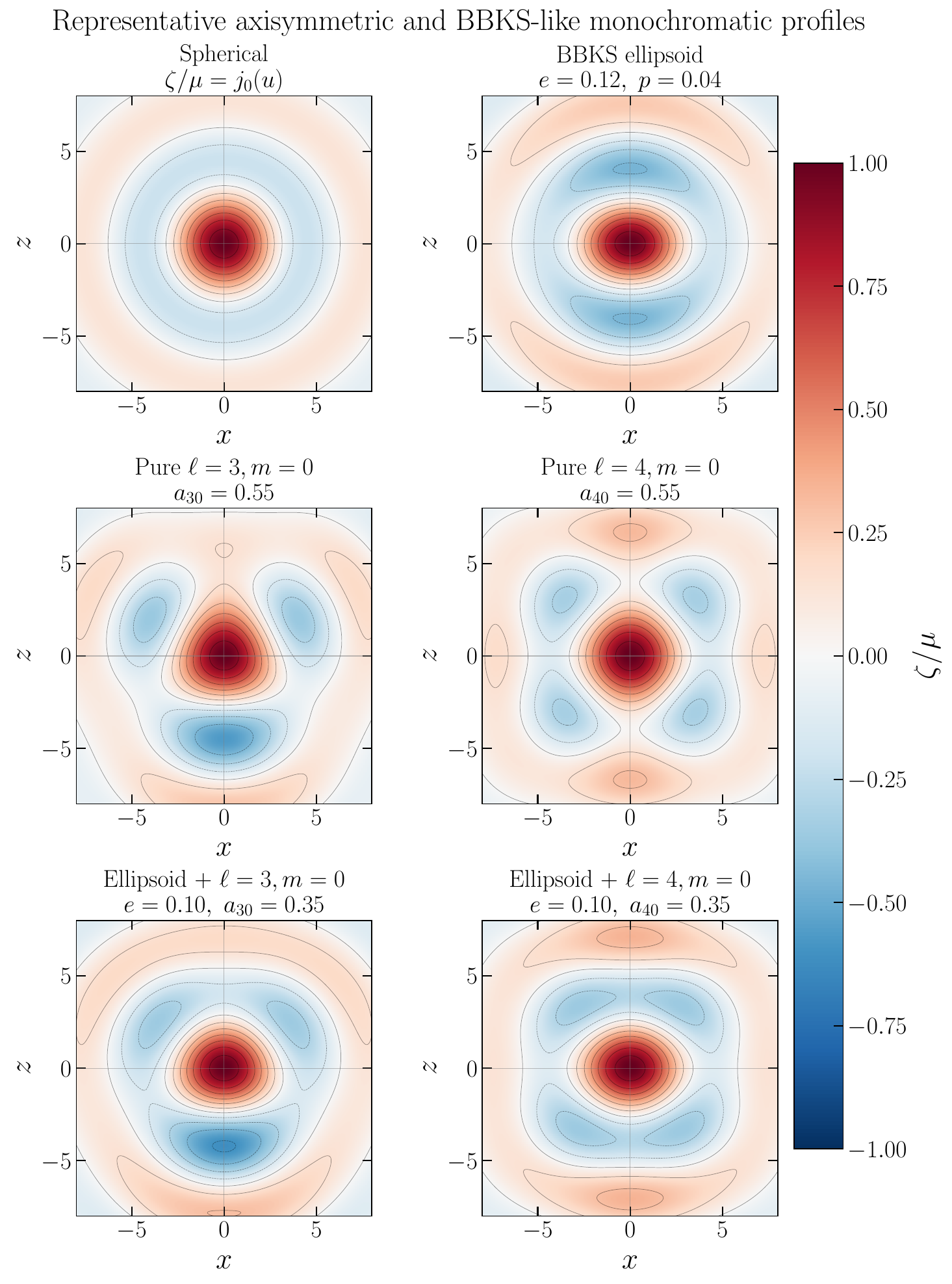}
    \caption{Representative axisymmetric and BBKS-like monochromatic profiles, shown as $x$--$z$ slices of $\zeta/\mu$. The first row shows the spherical profile and the BBKS ellipsoid. The second row shows pure $\ell=3,m=0$ and $\ell=4,m=0$ contributions, which preserve axisymmetry but go beyond the local ellipsoidal shape. The third row shows examples where these higher multipoles are superposed on the BBKS ellipsoid.}
    \label{fig:mono_shapes_xz}
\end{figure}

\begin{figure}[p]
    \centering
    \includegraphics[width=0.96\textwidth]{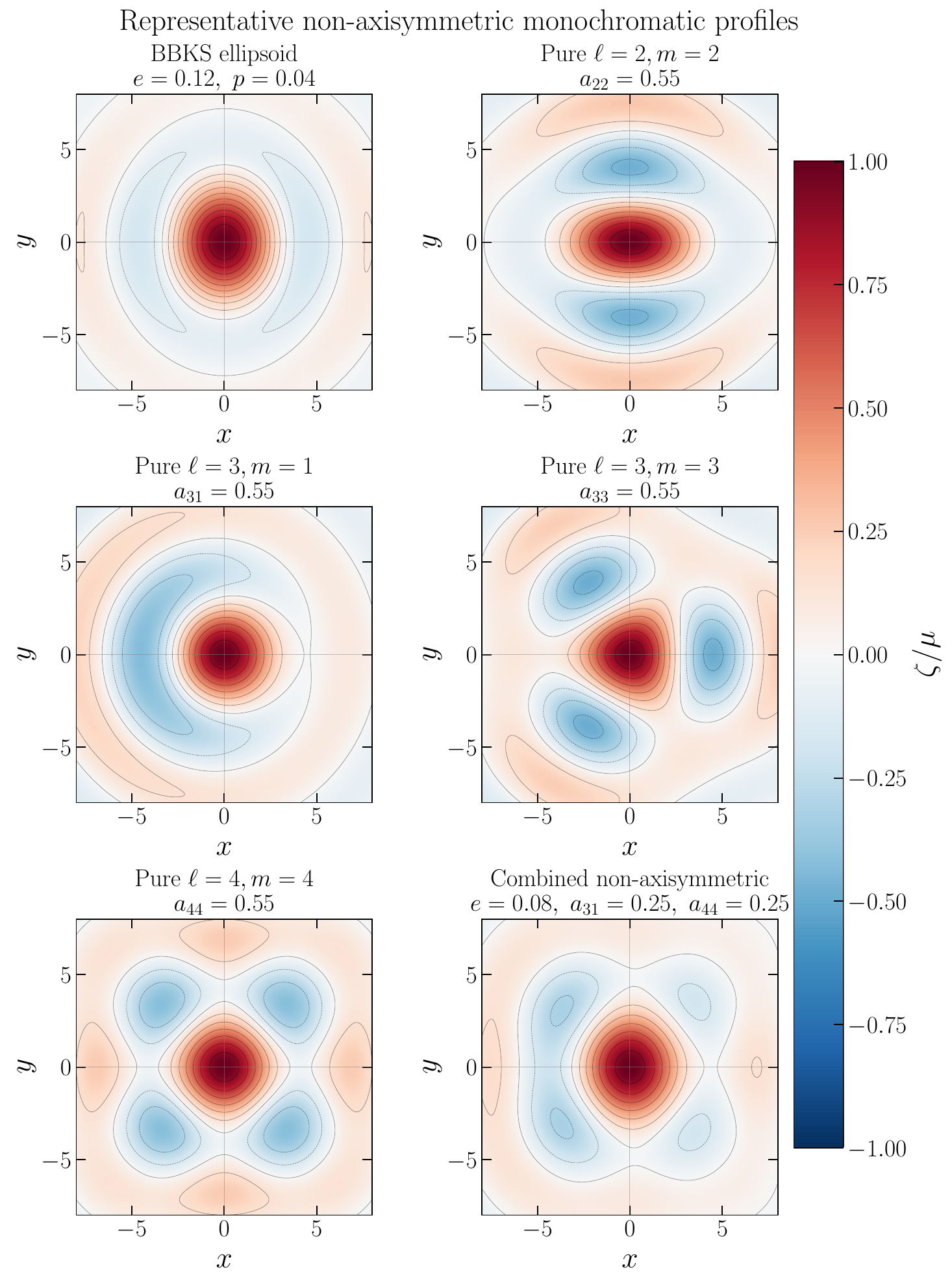}
    \caption{Representative non-axisymmetric monochromatic profiles, shown as $x$--$y$ slices of $\zeta/\mu$. The panels illustrate how modes with $m\neq0$ introduce azimuthal structure. The $\ell=2,m=2$ example is a non-axisymmetric quadrupole, but it still belongs to the BBKS quadrupolar sector and can be rotated into a principal-axis ellipsoid. Genuine post-BBKS angular deformations start at $\ell\geq3$; the octupolar and hexadecapolar examples show such higher-order distortions, and the last panel illustrates a combined realization beyond the BBKS ellipsoid.}
    \label{fig:mono_shapes_xy}
\end{figure}

Figure~\ref{fig:mono_radial_envelopes} shows the radial envelopes $j_\ell(u)$ for the first relevant multipoles. Figure~\ref{fig:mono_shapes_xz} displays representative axisymmetric and BBKS-like configurations in an $x$--$z$ slice. Figure~\ref{fig:mono_shapes_xy} shows representative non-axisymmetric realizations in an $x$--$y$ slice, making explicit the role of modes with $m\neq0$.
The contributions retained in Eq.~\eqref{eq:general_mono_profile_section} are summarized in Table~\ref{tab:mono_contributions_section}. The table is useful because it makes explicit which piece controls the familiar ellipsoidal deformation and which pieces correspond to genuinely new non-spherical structures.
\begin{table}[t]
\centering
\small
\begin{tabular}{p{0.17\textwidth}p{0.34\textwidth}p{0.15\textwidth}p{0.24\textwidth}}
\hline
Contribution & Explicit form for $\zeta/\mu$ & Local order & Physical meaning \\
\hline
Spherical part & $j_0(u)$ & $1+\mathcal O(r^2)$ & Height and, in the monochromatic case, also the spherical curvature. \\
BBKS ellipsoid & $-\frac{5}{2}\mathcal A_{ep}(\hat{\bm x})j_2(u)$ & $\mathcal O(r^2)$ & Traceless Hessian; local ellipsoidal deformation described by $e$ and $p$. \\
Axisymmetric higher mode & $a_{\ell 0}\,Y_{\ell 0}j_\ell(u)$ & $\mathcal O(r^\ell)$ & Higher non-spherical correction preserving axisymmetry. \\
Non-axisymmetric higher mode & $a^{(c,s)}_{\ell m}\,Y^{(c,s)}_{\ell m}j_\ell(u)$, $m>0$ & $\mathcal O(r^\ell)$ & Azimuthally varying deformation beyond the ellipsoid. \\
General higher multipole & $\mathcal D_\ell(\hat{\bm x})j_\ell(u)$ & $\mathcal O(r^\ell)$ & Complete $\ell\geq3$ angular structure at the scale $k_p^{-1}$. \\
\hline
\end{tabular}
\caption{Explicit monochromatic decomposition of the curvature profile.}
\label{tab:mono_contributions_section}
\end{table}
The Gaussian quadratic action of the monochromatic coherent configuration is equally simple. Since the monochromatic spectrum removes radial freedom, the action depends only on the angular amplitudes. In the relative-amplitude parametrization of Eq.~\eqref{eq:general_mono_profile_section}, one finds
\begin{equation}
W=\nu^2\left[1+5(3e^2+p^2)+\sum_{\ell=3}^{\ell_{\rm max}}\left(a_{\ell 0}^2+\sum_{m=1}^{\ell}\left[(a^{(c)}_{\ell m})^2+(a^{(s)}_{\ell m})^2\right]\right)\right].
\label{eq:mono_action_full_section}
\end{equation}
The first term is the spherical cost, the second term is the BBKS ellipsoidal cost, and the remaining terms are the costs of the higher multipoles. Equation~\eqref{eq:mono_action_full_section} should be understood as the Gaussian quadratic action of the specified coherent configuration. 

The collapse threshold is then a function of the chosen shape parameters. Schematically, one may write
\begin{equation}
\mu_c=\mu_c\left(e,p,\{a_{\ell 0}\},\{a^{(c)}_{\ell m},a^{(s)}_{\ell m}\}\right),
\label{eq:mu_c_mono_shape_section}
\end{equation}
and the corresponding critical action is obtained by evaluating Eq.~\eqref{eq:mono_action_full_section} at the collapse threshold. Thus a non-spherical realization enhances PBH formation only if the reduction in the threshold compensates for the Gaussian cost of the deformation. In this way, the monochromatic framework provides a direct generalization of the previously studied ellipsoidal family to a wider set of non-spherical peak shapes, including both axisymmetric and genuinely azimuthally structured realizations.

The monochromatic limit therefore gives a clean angular classification of
non-spherical shapes.  In the next section we keep this angular classification
but replace the fixed monochromatic envelopes by finite-width,
power-spectrum-weighted radial envelopes.  This illustrates the first step
towards the full finite-width problem, where each angular multipole contains a
tower of independent radial shape directions.

\section{Finite-width spectra and non-spherical shape dispersion}
\label{sec:finite_width_nonspherical}

The monochromatic construction of Sec.~\ref{sec:monochromatic_case_general} is useful
because it gives a particularly simple connection between angular multipoles and
real-space shapes. In the spherical sector, the monochromatic profile contains
only one radial shape. When non-spherical perturbations are included, however, each angular multipole
is multiplied by its own fixed monochromatic radial envelope,
\begin{equation}
\zeta_G({\bm x})
=
\mu
\left[
j_0(k_pr)
+
\sum_{\ell\geq 2,m}
a_{\ell m}Y_{\ell m}(\hat{\bm x})j_\ell(k_pr)
\right].
\end{equation}
Thus the higher functions \(j_2,j_3,j_4,\ldots\) should not be interpreted as
additional spherical profiles. They are the radial envelopes associated with
non-spherical angular harmonics.

The monochromatic limit is nevertheless degenerate from the point of view of
radial dispersion. Once the amplitude of a given angular multipole is specified,
its radial dependence is fixed to be \(j_\ell(k_pr)\). For a finite-width
spectrum this is no longer true. The same angular harmonic
\(Y_{\ell m}(\hat{\bm x})\) can be accompanied by a power-spectrum-weighted
radial envelope. Therefore finite spectral width does not simply change the
spherical mean profile; it also modifies the morphology of non-spherical
perturbations.

In this section we illustrate this effect for the finite-width exponential
spectrum
\begin{equation}
{\cal P}_{\zeta_G}(k)
=
A_\zeta
\sqrt{\frac{2}{\pi}}
\frac{k^3}{\tilde{\kappa}^3}
\exp\left(-\frac{k^2}{2\tilde{\kappa}^2}\right),
\qquad
k>0 .
\end{equation}
The spectral moments are
\begin{equation}
\sigma_j^2
=
\int d\ln k\,k^{2j}{\cal P}_{\zeta_G}(k)
=
A_\zeta(2j+1)!!\,\tilde{\kappa}^{2j}.
\label{eq:11}
\end{equation}
In particular,
\begin{equation}
\sigma_0^2=A_\zeta,
\qquad
\sigma_1^2=3A_\zeta\tilde{\kappa}^2,
\qquad
\sigma_2^2=15A_\zeta\tilde{\kappa}^4.
\label{eq:22}
\end{equation}
The BBKS height-curvature correlation is therefore
\begin{equation}
\gamma_{\rm BBKS}
=
\frac{\sigma_1^2}{\sigma_0\sigma_2}
=
\sqrt{\frac{3}{5}},
\label{eq:33}
\end{equation}
which is smaller than unity. This should be contrasted with the monochromatic
limit, where \(\gamma_{\rm BBKS}=1\) and the height and spherical curvature are perfectly
correlated.

At fixed peak height, the BBKS curvature variable is not fixed deterministically
for a finite-width spectrum. Before imposing the full peak constraint, its
Gaussian conditional distribution is
\begin{equation}
P(x_{\rm B}|\nu)
\propto
\exp\left[-\frac{1}{2}
\frac{(x_{\rm B}-\gamma_{\rm BBKS}\nu)^2}
{(1-\gamma_{\rm BBKS}^2)}
\right].
\end{equation}
If the full BBKS peak measure is imposed, this Gaussian conditional density
must be supplemented by the usual BBKS peak-weighting factor and by the
conditions selecting local maxima. In the illustrative profiles below we do
not sample the full peak-weighted distribution. Instead, we use the representative height-conditioned configuration obtained
by setting $n_x=0$ in the orthogonalized curvature direction. We use $
x_{\rm B}
=
x_*(\nu)
=
\gamma_{\rm BBKS}\nu$ for the finite-width BBKS quadrupolar
sector in the examples shown below. For illustration, we take the leading action-normalized radial direction in
each multipole sector to be
\begin{equation}
q_{\ell0}(k)
=
\frac{k^\ell}{\sigma_\ell}.
\end{equation}
This gives
\begin{equation}
\int d\ln k\,
{\cal P}_{\zeta_G}(k)
q_{\ell0}^2(k)
=
1 .
\end{equation}
The corresponding real-space radial envelope is
\begin{equation}
\mathcal{R}_{\ell0}(r)
=
\int d\ln k\,
{\cal P}_{\zeta_G}(k)
q_{\ell0}(k)j_\ell(kr).
\end{equation}
Dividing by \(\sigma_0\), we define the dimensionless envelope
\begin{equation}
{\cal J}_\ell(r)
\equiv
\frac{\mathcal{R}_{\ell0}(r)}{\sigma_0}.
\end{equation}
Introducing
\begin{equation}
\rho
=
\frac{\tilde{\kappa} r}{\sqrt{2}},
\qquad
t=\frac{k}{\tilde{\kappa}},
\end{equation}
one obtains
\begin{equation}
{\cal J}_\ell(\rho)
=
\frac{1}{\sqrt{(2\ell+1)!!}}
\int_0^\infty dt\,
\sqrt{\frac{2}{\pi}}\,
t^{\ell+2}
e^{-t^2/2}
j_\ell(\sqrt{2}\,t\rho).
\end{equation}
For the exponential spectrum considered here, this integral can be evaluated
explicitly:
\begin{equation}
{\cal J}_\ell(\rho)
=
\frac{(\sqrt{2}\rho)^\ell}{\sqrt{(2\ell+1)!!}}
e^{-\rho^2}.
\end{equation}

For this leading action-normalized radial direction, the finite-width
exponential spectrum replaces the monochromatic envelope \(j_\ell(k_pr)\) by
the power-spectrum-weighted envelope \({\cal J}_\ell(\rho)\).  

The finite-width analogue of the monochromatic non-spherical profile can then
be written schematically as
\begin{equation}
\frac{\zeta_G({\bm x})}{\mu}
=
{\cal J}_0(\rho)
-
\frac{5}{2}\gamma_{\rm BBKS}\,
{\cal A}_{ep}(\hat{\bm x}){\cal J}_2(\rho)
+
\sum_{\ell=3}^{\ell_{\rm max}}
\mathcal{D}_\ell(\hat{\bm x}){\cal J}_\ell(\rho),
\end{equation}
where
\begin{equation}
\mathcal{D}_\ell(\hat{\bm x})
=
a_{\ell0}Y_{\ell0}(\hat{\bm x})
+
\sum_{m=1}^{\ell}
\left[
a_{\ell m}^{(c)}Y_{\ell m}^{(c)}(\hat{\bm x})
+
a_{\ell m}^{(s)}Y_{\ell m}^{(s)}(\hat{\bm x})
\right].
\end{equation}
The finite-width spectrum does not create new angular harmonics by itself.  The
angular functions are the same as in the monochromatic construction.  What
changes is the radial structure multiplying each harmonic.  For the leading
action-normalized direction \(q_{\ell0}\), this radial structure is described
by \({\cal J}_\ell(\rho)\).  More generally, finite spectral width opens a
radial shape space within each \(\ell\) sector, so that non-spherical
perturbations are specified not only by their angular multipole but also by
their radial deformation direction.

The coefficients shown in the figures below should therefore be understood as
illustrative coherent shape amplitudes, rather than as samples from the full
BBKS peak-weighted distribution. A complete peak calculation would require the
joint distribution of \(\nu\), \(x_{\rm B}\), \(e\), \(p\), and the
higher multipolar coefficients. In the present section we use these profiles
only to visualize how a finite-width spectrum modifies the real-space
morphology of non-spherical perturbations.

\begin{figure}[!htbp]
\centering
\includegraphics[width=0.78\textwidth]{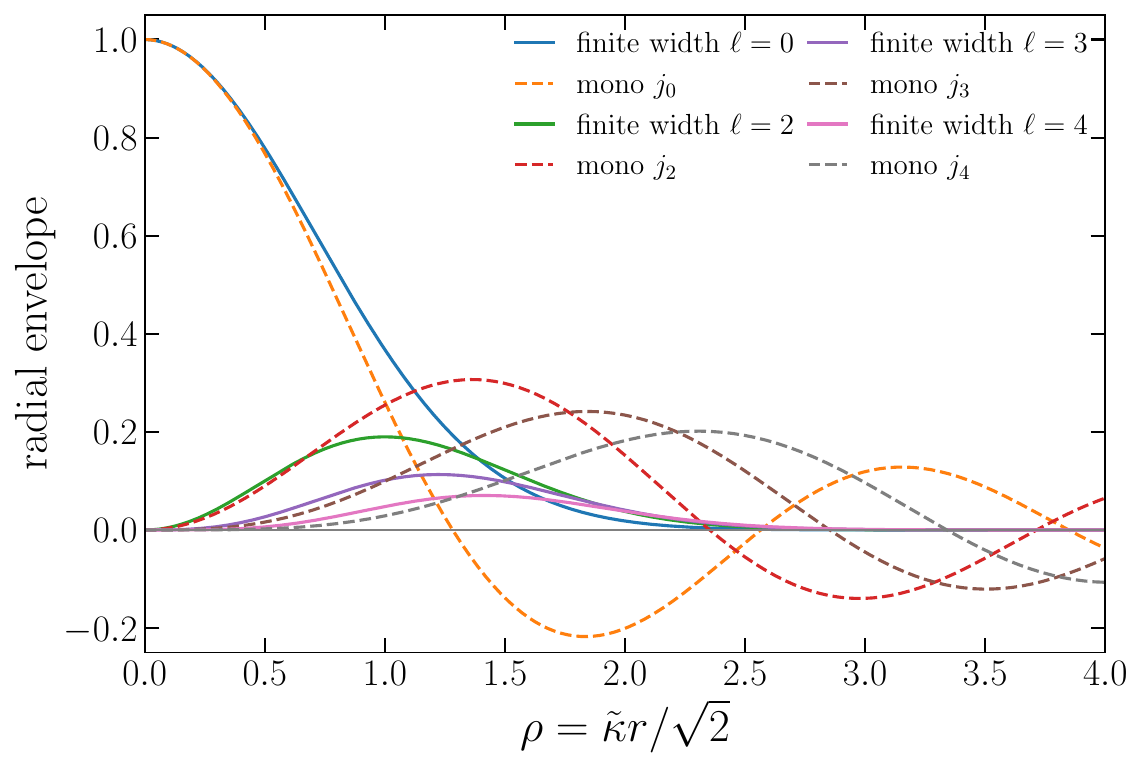}
\caption{
Radial envelopes multiplying different angular multipoles. The dashed curves
show the monochromatic envelopes \(j_\ell(k_pr)\). Only the \(\ell=0\) curve is
the spherical monochromatic profile; the higher-\(\ell\) curves multiply
non-spherical angular harmonics. The solid curves show the corresponding
finite-width envelopes \({\cal J}_\ell(\rho)\) for the exponential spectrum.
}
\label{fig:finite_width_radial_envelopes}
\end{figure}

Figure~\ref{fig:finite_width_radial_envelopes} compares the first finite-width
radial envelopes with the corresponding monochromatic envelopes. The dashed
curves are not additional spherical profiles. Only the \(\ell=0\) dashed curve
is the monochromatic spherical profile. The curves with \(\ell\geq2\) are the
fixed radial envelopes multiplying the angular harmonics \(Y_{\ell m}\) in the
monochromatic non-spherical construction. For visual comparison, the
monochromatic wavenumber is chosen so that the central curvature of
\(j_0(k_pr)\) matches that of \(e^{-\rho^2}\).

\begin{figure}[!htbp]
\centering
\includegraphics[width=\textwidth]{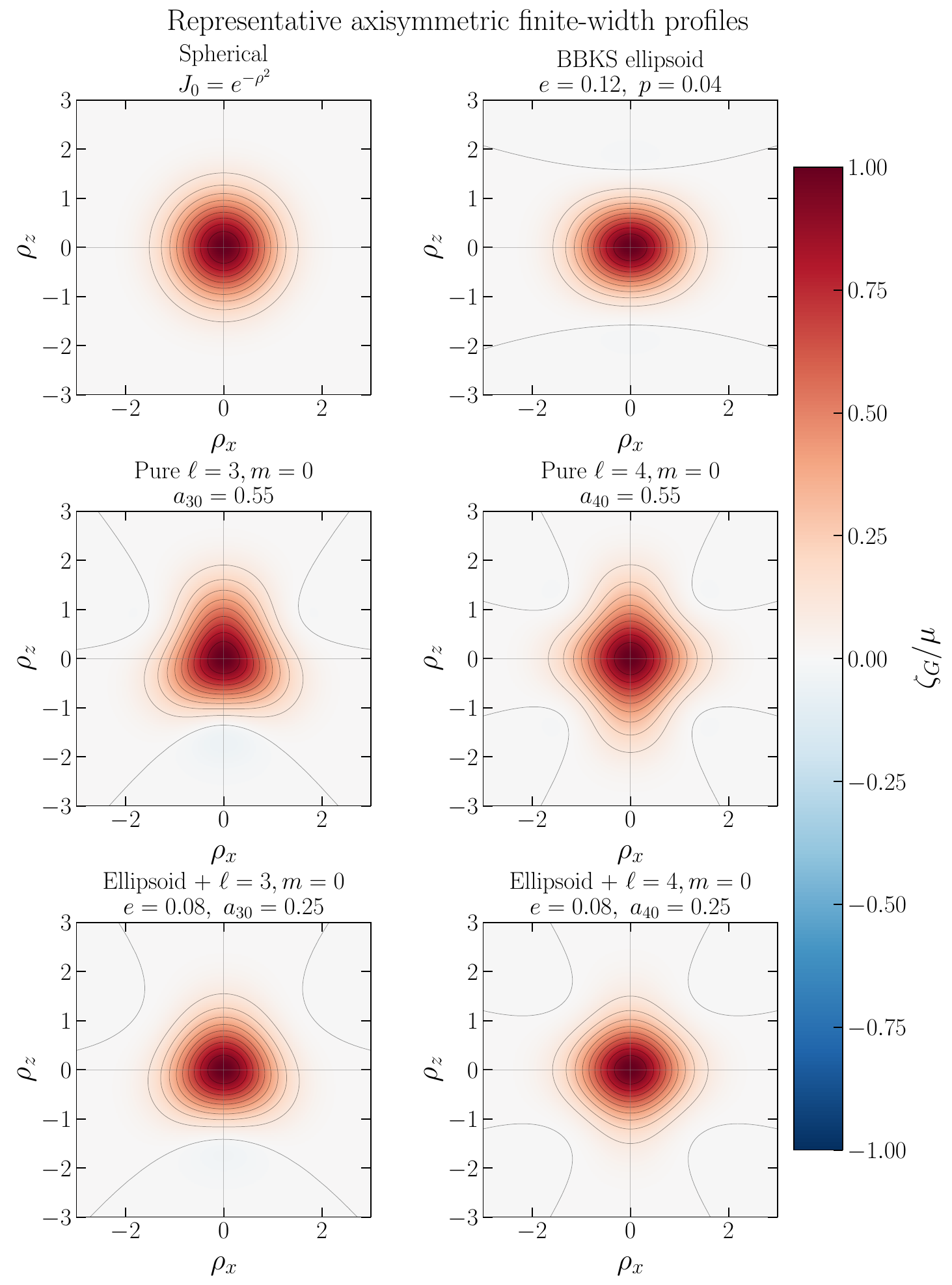}
\caption{
Representative axisymmetric and BBKS-like profiles for the finite-width
exponential spectrum, shown as slices of \(\zeta_G/\mu\). The coordinates are
\(\rho_x=\tilde{\kappa} x/\sqrt{2}\) and \(\rho_z=\tilde{\kappa} z/\sqrt{2}\). The \(\ell=3,m=0\) and
\(\ell=4,m=0\) examples show higher multipolar deformations beyond the local
ellipsoid, with finite-width radial envelopes \({\cal J}_3\) and
\({\cal J}_4\).
}
\label{fig:finite_width_axisymmetric_shapes}
\end{figure}

\begin{figure}[!htbp]
\centering
\includegraphics[width=\textwidth]{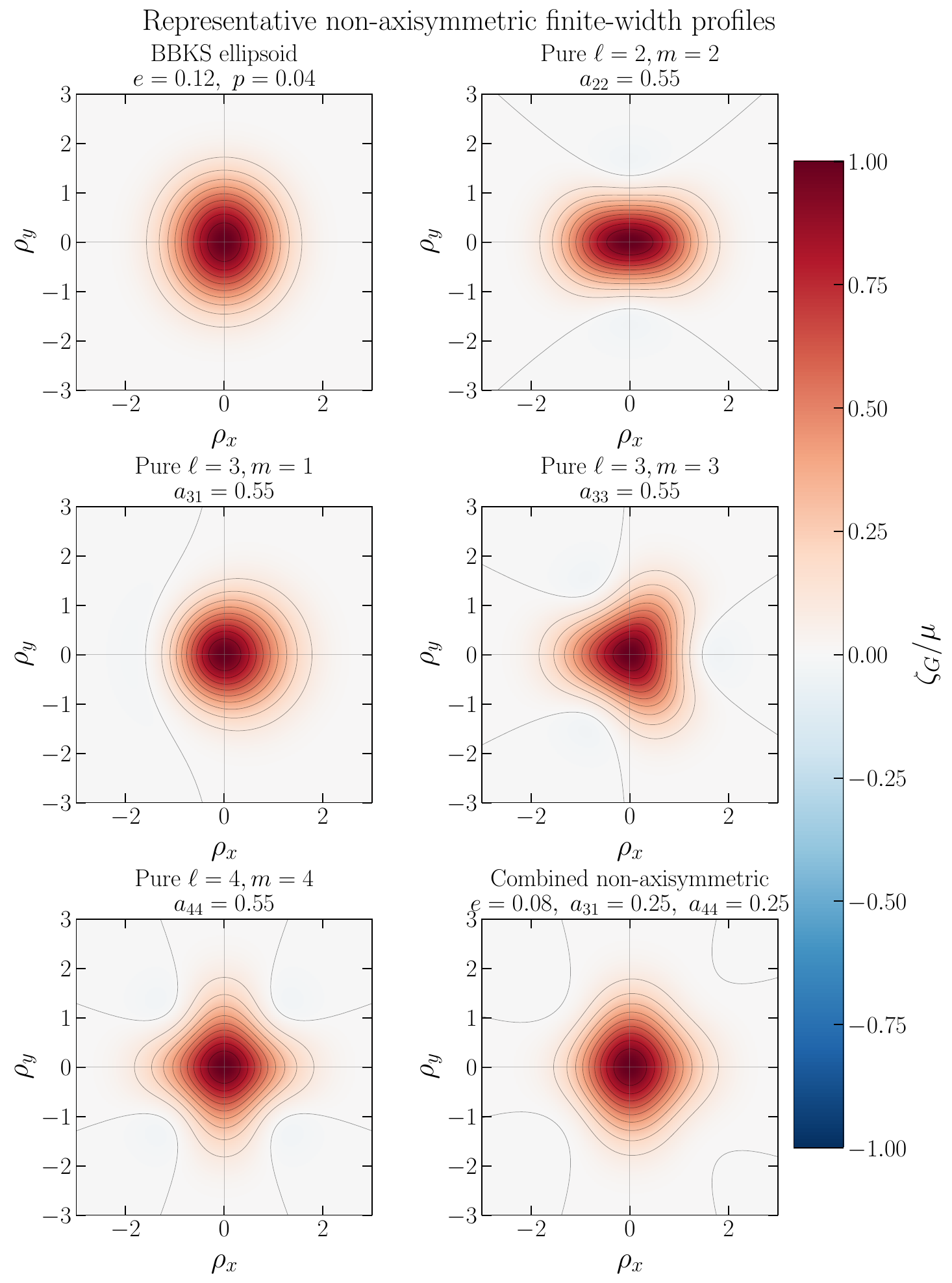}
\caption{
Representative non-axisymmetric profiles for the finite-width exponential
spectrum, shown as \(\rho_x\)--\(\rho_y\) slices of \(\zeta_G/\mu\), with
\(\rho_i=\tilde{\kappa} X_i/\sqrt{2}\). The BBKS quadrupolar sector uses $x_{\rm B}=x_*(\nu)=\gamma_{\rm BBKS}\nu$. The pure \(\ell=2,m=2\) example is included as
a quadrupolar BBKS-sector reference; it is not a genuine post-BBKS mode, since a
general quadrupole can be rotated into the principal-axis ellipsoid. Genuine
post-BBKS angular deformations start at \(\ell\geq3\). Compared with the
monochromatic case, the angular patterns are the same, but their radial
envelopes are modified by the finite width of the spectrum.
}
\label{fig:finite_width_nonaxisymmetric_shapes}
\end{figure}

Figures~\ref{fig:finite_width_axisymmetric_shapes} and
\ref{fig:finite_width_nonaxisymmetric_shapes} show representative finite-width
non-spherical profiles. They are the direct analogue of the monochromatic shape
illustrations of Sec.~\ref{sec:monochromatic_case_general}, but with \(j_\ell\) replaced
by \({\cal J}_\ell\). For the BBKS quadrupolar sector we use the height-conditioned representative
curvature $x_{\rm B}=x_*(\nu)=\gamma_{\rm BBKS}\nu$. The axes are the dimensionless coordinates
\[
\rho_i=\frac{\tilde{\kappa} X_i}{\sqrt{2}},
\]
and the plots are restricted to \(|\rho_i|\leq3\), where the finite-width
exponential envelopes have most of their support.
The profiles in this section are therefore used only to illustrate how finite
spectral width modifies the radial envelopes of non-spherical multipoles; the
one-dimensional spherical split mode used in the numerical simulations is
introduced separately in Sec.~\ref{sec:numerical}.

\section{Numerical examples}
\label{sec:numerical}
In this section, we evaluate the effect of shape dispersion on gravitational collapse.  Due to the complexity of relativistic simulations in \(3+1\) dimensions, we restrict the numerical analysis to spherical symmetry and leave the non-spherical case for future work. For this purpose, we consider two examples: a sharply peaked power spectrum with finite width (A), and a finite-width scale-invariant spectrum with a tunable bandwidth (B).  To follow the full nonlinear gravitational collapse of the fluctuations, we use the public
SPriBHoS codes~\cite{Escriva:2019sim,Escriva:2025eqc}\footnote{The codes are publicly available in the GitHub repository~\cite{escriva_github}.}.  We refer the reader to Appendix~\ref{appendix:numerics} for a brief description of the numerical methodology. Throughout this section, $r$ denotes the conformally flat radial coordinate
defined by the long-wavelength metric in Eq.~\eqref{eq:LW_metric}.

\subsection{Case A: sharply peaked finite-width spectrum with logarithmic non-Gaussianity}
\label{sec:caseA}

We retain only the monopole sector, $\ell=0$, and choose the constraint set
\begin{equation}
\cC=\left\{\zeta_G(0)=\mu\right\}.
\label{eq:sph_gauss_constraint_set}
\end{equation}
Thus the central amplitude is fixed, while the radial shape is allowed to vary through a coherent action-normalized deformation. This is the simplest nontrivial realization of the general expansion of Eq.~\eqref{eq:intro-master}. 

We consider the dimensionless curvature power spectrum
\footnote{The spectrum is normalized such that
\[
\sigma_0^2=\int {\cal P}_{\zeta_G}(k)\,d\ln k=A_{\zeta}.
\]
For comparison, a lognormal spectrum with the same variance may be written as
\[
{\cal P}_{\rm LN}(k)
=
\frac{A_{\zeta}}{\sqrt{2\pi}\Delta}
\exp\left[-\frac{\ln^2(k/k_\ast)}{2\Delta^2}\right].
\]
Matching also the peak height of the spectrum used here, whose maximum is at
\(k_{\rm peak}=\sqrt{3}\tilde{\kappa}\), gives the peak-height-matched effective
lognormal width
\[
\Delta_{\rm eff}
=
\frac{e^{3/2}}{6\sqrt{3}}
\simeq 0.43 .
\]
}

\begin{equation}
\Pz(k)=
A_{\zeta}\sqrt{\frac{2}{\pi}}\,\frac{k^3}{\tilde{\kappa}^3}
\exp\left(-\frac{k^2}{2\tilde{\kappa}^2}\right),
\qquad k>0,
\label{eq:sph_gauss_power_spectrum}
\end{equation}
where $A_{\zeta}$ fixes the variance of the Gaussian field and $\tilde{\kappa}$ fixes the characteristic scale. The maximum of the spectrum is located at $k_{\rm peak}=\sqrt{3}\,\tilde{\kappa}$. Some statistical parameters were already defined in Eqs.~\eqref{eq:11},\eqref{eq:22},\eqref{eq:33}.

Although the numerical family constructed below fixes only the height, this value is useful when comparing with the full BBKS height-curvature sector. The spherical two-point function\footnote{In some PBH abundance calculation frameworks, an explicit transfer
function is applied to the power spectrum when defining the statistical correlators (see for instance \cite{DeLuca:2023tun}). We do
not introduce such an additional transfer kernel. Our curvature profiles are specified as
superhorizon initial data, where the curvature perturbation is conserved at leading
order in the gradient expansion, and the subsequent evolution through horizon re-entry and
nonlinear collapse is determined by relativistic numerical simulations.} is
\begin{equation}
\xi(r)\equiv\left\langle \zeta_G(0)\zeta_G(r)\right\rangle
=
\int_0^\infty \ddlnk\,\Pz(k)j_0(kr).
\label{eq:sph_gauss_xi_def}
\end{equation}
For Eq.~\eqref{eq:sph_gauss_power_spectrum} the integral gives
\begin{equation}
\xi(r)=A_\zeta\exp\left(-\frac{\tilde{\kappa}^2r^2}{2}\right).
\label{eq:sph_gauss_xi}
\end{equation}
It is convenient to introduce $x\equiv \tilde{\kappa} r/{\sqrt{2}}$ and $r_{\rm width} \equiv \sqrt{2}/\tilde{\kappa}$. Then the normalized correlator is
\begin{equation}
\Psi(r)\equiv \frac{\xi(r)}{\sigma_0^2}=e^{-x^2}.
\label{eq:sph_gauss_Psi}
\end{equation}

We first rewrite the fixed-height mean profile in the notation of the general formalism. The action-normalized height direction is
\begin{equation}
q_\nu(k)=\frac{1}{\sigma_0}=\frac{1}{\sqrt{A_{\zeta}}}.
\label{eq:sph_gauss_qnu}
\end{equation}
The associated radial function is
\begin{equation}
\mathcal{R}_\nu(r)=\int_0^\infty \ddlnk\,\Pz(k)q_\nu(k)j_0(kr)
=\sigma_0\,\Psi(r)=\sigma_0\,e^{-x^2}.
\label{eq:sph_gauss_Rnu}
\end{equation}
Then the conditional mean profile can be written as
\begin{equation}
\overline\zeta_G(r)=\nu \mathcal{R}_\nu(r)=\mu e^{-x^2}
=\mu\exp\left(-\frac{\tilde{\kappa}^2r^2}{2}\right).
\label{eq:sph_gauss_mean_profile}
\end{equation}
Equivalently, in spectral space,
\begin{equation}
\overline B_{00}(k)=\mu\frac{\Pz(k)}{\sigma_0^2}.
\label{eq:sph_gauss_Bbar}
\end{equation}

We now introduce a coherent spherical deformation around this mean profile. In the notation of the general formalism, a deformation along a chosen
action-normalized monopole direction \(q_{0\alpha}\) is written as
\begin{equation}
\Delta B_{00}(k)
=
n\,\Pz(k)q_{0\alpha}(k),
\label{eq:sph_gauss_DeltaB_general}
\end{equation}
where \(n\) is the amplitude associated with the selected radial direction and
\begin{equation}
\int_0^\infty \ddlnk\,\Pz(k)q_{0 \alpha}(k)q_{0 \eta}(k)=\delta_{\alpha \eta}.
\label{eq:sph_gauss_q_norm_general}
\end{equation}
The corresponding real-space radial mode is
\begin{equation}
\mathcal{R}_{0 \alpha}(r)=\int_0^\infty \ddlnk\,\Pz(k)q_{0 \alpha}(k)j_0(kr).
\label{eq:sph_gauss_R0a_general}
\end{equation}

Following Ref.~\cite{Escriva:2025ftp}, we use a split-spectrum ansatz to construct a coherent
finite-action deformation at fixed central amplitude. The split divides the spectrum into long- and short-wavelength parts with equal variance weight. This is not a cost-minimizing choice, (every action-normalized direction has the same quadratic cost $n^2$), but rather a representative direction that maximizes the real-space profile dispersion and avoids the destructive interference produced by rapidly alternating signs in $k$-space. Here we use it as a representative direction in shape space.

\begin{equation}
q_{0,{\rm split}}(k)=\frac{1}{\sigma_0}\operatorname{sign}(\overline k-k).
\label{eq:sph_gauss_qsplit}
\end{equation}
The splitting scale $\overline k$ is chosen so that the two domains contain equal variance,
\begin{equation}
\int_0^{\overline k}\ddlnk\,\Pz(k)
=
\int_{\overline k}^{\infty}\ddlnk\,\Pz(k)
=\frac{\sigma_0^2}{2}.
\label{eq:sph_gauss_equal_variance}
\end{equation}
This immediately implies
\begin{equation}
\int_0^\infty \ddlnk\,\Pz(k)q_{0,{\rm split}}(k)q_\nu(k)=0,
\label{eq:sph_gauss_qsplit_orth_height}
\end{equation}
so the split deformation preserves the central amplitude. It also satisfies
\begin{equation}
\int_0^\infty \ddlnk\,\Pz(k)q_{0,{\rm split}}^2(k)=1.
\label{eq:sph_gauss_qsplit_norm}
\end{equation}
Therefore the coefficient $n$ is already an action-normalized Gaussian coordinate, with
\begin{equation}
\Delta W=n^2,
\qquad
P_{\rm sh}(n)=\frac{1}{\sqrt{2\pi}}e^{-n^2/2}.
\label{eq:sph_gauss_branch_weight}
\end{equation}

The radial function associated with the split mode is
\begin{equation}
\mathcal{R}_{0,{\rm split}}(r)=\int_0^\infty \ddlnk\,\Pz(k)q_{0,{\rm split}}(k)j_0(kr).
\label{eq:sph_gauss_Rsplit_def}
\end{equation}
Using Eq.~\eqref{eq:sph_gauss_qsplit}, this is
\begin{equation}
\mathcal{R}_{0,{\rm split}}(r)=
\frac{1}{\sigma_0}\left[
\int_0^{\overline k}\ddlnk\,\Pz(k)j_0(kr)
-
\int_{\overline k}^{\infty}\ddlnk\,\Pz(k)j_0(kr)
\right].
\label{eq:sph_gauss_Rsplit_integral}
\end{equation}
Equivalently, define the normalized long- and short-wavelength correlators
\begin{equation}
\Psi_+(r)=\frac{2}{\sigma_0^2}\int_0^{\overline k}\ddlnk\,\Pz(k)j_0(kr),
\label{eq:sph_gauss_Psi_plus_def}
\end{equation}
\begin{equation}
\Psi_-(r)=\frac{2}{\sigma_0^2}\int_{\overline k}^{\infty}\ddlnk\,\Pz(k)j_0(kr).
\label{eq:sph_gauss_Psi_minus_def}
\end{equation}
They obey
\begin{equation}
\Psi_+(0)=\Psi_-(0)=1,
\qquad
\Psi(r)=\frac{1}{2}\left[\Psi_+(r)+\Psi_-(r)\right].
\label{eq:sph_gauss_half_correlator_properties}
\end{equation}
Then
\begin{equation}
\mathcal{R}_{0,{\rm split}}(r)=\frac{\sigma_0}{2}\left[\Psi_+(r)-\Psi_-(r)\right]
=\sigma_0\left[\Psi_+(r)-\Psi(r)\right].
\label{eq:sph_gauss_Rsplit_psipm}
\end{equation}

For the spectrum in Eq.~\eqref{eq:sph_gauss_power_spectrum}, we introduce the spectral variable
\begin{equation}
y=\frac{k}{\sqrt{2}\tilde{\kappa}},
\label{eq:sph_gauss_y_spec_def}
\end{equation}
so that
\begin{equation}
\Pz(k)\ddlnk=A_{\zeta}\frac{4}{\sqrt\pi}\,y^2e^{-y^2}\,dy.
\label{eq:sph_gauss_weight_y}
\end{equation}
The equal-variance condition then becomes
\begin{equation}
\operatorname{erf}(\overline y)-\frac{2}{\sqrt\pi}\overline y e^{-\overline y^2}=\frac{1}{2},
\label{eq:sph_gauss_ybar_equation}
\end{equation}
which can be solved numerically to obtain
\begin{equation}
\overline y\simeq1.087652,
\qquad
\overline k=\sqrt2\tilde{\kappa}\overline y\simeq1.538172\,\tilde{\kappa}.
\label{eq:sph_gauss_ybar_value}
\end{equation}
The variable $y$ is the spectral integration variable, while $x=\tilde{\kappa} r/\sqrt2$ is the dimensionless radius.

The long-wavelength correlator has the closed form
\begin{equation}
\Psi_+(r)=
2e^{-x^2}\operatorname{Re}\left[\operatorname{erf}(\overline y+ix)\right]
-
\frac{2}{\sqrt\pi}e^{-\overline y^2}\frac{\sin(2\overline y x)}{x}.
\label{eq:sph_gauss_Psi_plus_closed}
\end{equation}
The limit $x\to0$ is regular and should be understood through
\begin{equation}
\frac{\sin(2\overline y x)}{x}\longrightarrow 2\overline y.
\end{equation}
Combining Eqs.~\eqref{eq:sph_gauss_Rsplit_psipm} and \eqref{eq:sph_gauss_Psi_plus_closed}, we obtain
\begin{equation}
\mathcal{R}_{0,{\rm split}}(r)=\sigma_0\,\mathcal{G}(x),
\label{eq:sph_gauss_Rsplit_G}
\end{equation}
where
\begin{equation}
\mathcal{G}(x)=
e^{-x^2}\left\{2\operatorname{Re}\left[\operatorname{erf}(\overline y+ix)\right]-1\right\}
-
\frac{2}{\sqrt\pi}e^{-\overline y^2}\frac{\sin(2\overline y x)}{x}.
\label{eq:sph_gauss_G_def}
\end{equation}
On the other hand, by construction,
\begin{equation}
\mathcal{G}(0)=0,
\qquad
\mathcal{G}(x)=\mathcal O(x^2)\quad (x\to0).
\label{eq:sph_gauss_G_regular}
\end{equation}
Thus the split mode leaves the central height fixed and defines a smooth spherical profile.

The final one-parameter family of profiles can now be written exactly as in the general finite-action expansion,
\begin{equation}
\zeta_G(r,n)=\nu \mathcal{R}_\nu(r)+n\mathcal{R}_{0,{\rm split}}(r).
\label{eq:sph_gauss_profile_general_notation}
\end{equation}
Using Eqs.~\eqref{eq:sph_gauss_Rnu} and \eqref{eq:sph_gauss_Rsplit_G}, this becomes
\begin{equation}
\zeta_G(r,n)=\mu e^{-x^2}+n\sigma_0\,\mathcal{G}(x),
\qquad
x=\frac{\tilde{\kappa} r}{\sqrt2}.
\label{eq:sph_gauss_profile_family}
\end{equation}
The parameter $n$ is the finite-action shape coordinate. The reference
branch $n=0$ corresponds to the conditional mean configuration under
the chosen fixed-height conditioning, for which
$x_{\rm B}=x_*(\nu)=\gamma_{\rm BBKS}\nu$. For $n\neq0$, however, the split mode is allowed to change the spherical curvature of the profile. This is intentional: the goal of this first spherical study is to test how coherent radial-shape deformations at fixed central amplitude affect the collapse threshold. If one instead wanted a residual spherical mode at fixed height and fixed BBKS curvature, one should project the split direction orthogonally to $q_x$,
\begin{equation}
q_{0,{\rm split}}^{\perp}(k)=
\frac{q_{0,{\rm split}}(k)-\inner{q_{0,{\rm split}}}{q_x}\,q_x(k)}{\left[1-\inner{q_{0,{\rm split}}}{q_x}^{2}\right]^{1/2}},
\label{eq:sph_gauss_projected_split}
\end{equation}
as is done for the finite top-hat spectrum in Sec.~\ref{sec:subcaseB22}, where this projected construction is shown to yield the same qualitative behaviour as the unprojected one.

The corresponding radial function \(\mathcal{R}_{0,{\rm split}}^\perp(r)\) would describe
a genuine residual radial-shape deformation at fixed BBKS height and curvature.
For simplicity, we do not use this projected version in the present numerical
example.

For the collapse calculation it is useful to write the compaction functions (introduced first in \cite{Shibata:1999zs}) associated with Eq.~\eqref{eq:sph_gauss_profile_family}. The compaction function $\mathcal{C}$ in the comoving gauge is defined as twice the mass excess over the areal radius (see appendix~\ref{appendix:numerics}), and has been shown to be a useful quantity to characterize the threshold for black hole formation. At leading order in gradient expansion and for radiation domination we use
\begin{equation}
\mathcal{C}_{\ell}(r,n)=-\frac{4}{3}r\zeta_G'(r,n),
\label{eq:sph_gauss_Cl_def}
\end{equation}
\begin{equation}
\mathcal{C}(r,n)=\mathcal{C}_{\ell}(r,n)-\frac{3}{8}\mathcal{C}_{\ell}^2(r,n)
\label{eq:sph_gauss_C_def}
\end{equation}
where we denote $\mathcal{C}_{\ell}$ as the linear component of the compaction function. Since $r\,d/dr=x\,d/dx$, Eq.~\eqref{eq:sph_gauss_profile_family} gives
\begin{equation}
r\zeta_G'(r,n)=-2\mu x^2e^{-x^2}+n \,\sigma_0\,x \mathcal{G}'(x).
\label{eq:sph_gauss_rzeta_prime}
\end{equation}
Therefore
\begin{equation}
\mathcal{C}_\ell(r,n)=
\frac{8}{3}\mu x^2e^{-x^2}
-
\frac{4}{3}n\, \sigma_0\,x \mathcal{G}'(x),
\label{eq:sph_gauss_Cl_profile}
\end{equation}
and
\begin{equation}
\mathcal{C}(r,n)=
-\frac{2}{3}\left[-2\mu x^2e^{-x^2}+n \,\sigma_0\,x \mathcal{G}'(x)\right]
\left[2-2\mu x^2e^{-x^2}+n \, \sigma_0\,x\mathcal{G}'(x)\right].
\label{eq:sph_gauss_C_profile}
\end{equation}

In addition, we will use the long-wavelength density perturbation, denoted by \(\tilde{\rho}\), as a diagnostic of the
initial overdensity and compensation structure, its definition is given in Eq.~\eqref{eq:perturbations}.

On the other hand, a useful closed expression for $\mathcal{G}'(x)$ is obtained by defining
\begin{equation}
Q(x)=2\operatorname{Re}\left[\operatorname{erf}(\overline y+ix)\right]-1,
\qquad
\alpha=\frac{2}{\sqrt\pi}e^{-\overline y^2}.
\label{eq:sph_gauss_Q_alpha_def}
\end{equation}
Then
\begin{equation}
\mathcal{G}(x)=e^{-x^2}Q(x)-\alpha\frac{\sin(2\overline yx)}{x},
\label{eq:sph_gauss_G_Qalpha}
\end{equation}
and
\begin{equation}
\mathcal{G}'(x)=
-2xe^{-x^2}Q(x)
+
\alpha\frac{(2x^2+1)\sin(2\overline yx)-2\overline yx\cos(2\overline yx)}{x^2}.
\label{eq:sph_gauss_Gprime}
\end{equation}
For the Gaussian reference branch, $n=0$, one recovers
\begin{equation}
\mathcal{C}_{\ell}(r,0)=\frac{8}{3}\mu x^2e^{-x^2},
\label{eq:sph_gauss_Cl_mean}
\end{equation}
whose maximum occurs at $x=1$, i.e., $r_m(n=0)=\sqrt2/\tilde{\kappa}$ with $\mathcal{C}_{\ell}(r_m,0)=8 \mu e^{-1}/3$. For $n\neq0$, the maximum of $\mathcal{C}_\ell(r,n)$, and therefore the characteristic scale used in the numerical initial data, should be determined numerically. The collapse simulations will provide a branch-dependent threshold $\mu_c=\mu_c(n)$. At leading exponential order, the relative statistical importance of the branch $n$ is controlled by the critical action
\begin{equation}
W_c(n)=\frac{\mu_c^2(n)}{\sigma_0^2}+n^2
=\nu_{c}^2(n)+n^2.
\label{eq:sph_gauss_Phi}
\end{equation}
Therefore a deformed branch is statistically favoured over the reference branch only if the decrease in the collapse threshold compensates for the additional coherent-shape cost $n^2$. We define \(s\equiv  n \sigma_0\), thus $s$ is the physical amplitude of the coherent Gaussian curvature deformation, while $n$ is its action-normalized Gaussian coordinate.

In addition, we evaluate the effect of non-Gaussianities on the Gaussian fluctuation, which have been shown to have a significant impact on PBH formation scenarios; see
Ref.~\cite{Pi:2024lsu} for a review.  We use a logarithmic local template for non-Gaussianity~\cite{Atal:2019cdz,Pi:2022ysn}\footnote{It is important to remark that Eq.~\eqref{eq:template_log} is used here as a phenomenological local
non-Gaussian template, rather than as an exact relation derived from a specific inflationary
model. In Ref.~\cite{Escriva:2025ftp}, the explicit USR plateau calculation showed that the usual logarithmic map is modified by the evolution of the field and momentum perturbations. Comparing the
non-generalized logarithmic template with the numerical \(\delta N\) result, Ref.~\cite{Escriva:2025ftp}
found deviations of about \(8\%\) in \(\mu_c\) for the Gaussian $\zeta_G$, while a generalized template reduced these differences to about \(3\%\). The threshold and abundance estimates presented below should therefore be understood as conditional on this assumed local map, with \(1-\beta_{\rm NG}\zeta_G(r,s)>0\) imposed throughout the adiabatic branch.}. Specifically, the physical curvature
profile is written as
\begin{equation}
    \zeta[\zeta_G]= -\frac{1}{\beta_{\rm NG}}\log\left[1-\beta_{\rm NG} \, \zeta_G(r,s)\right],
    \label{eq:template_log}
\end{equation}
for $\beta_{\rm NG}=0$, Eq.~\eqref{eq:template_log} is understood in the continuous limit, $\zeta=\zeta_G$. For $\beta_{\rm NG}\neq0$, the logarithmic map modifies the linear compaction
function according to
\begin{equation}
\mathcal{C}_{\ell}(r,s;\beta_{\rm NG})
=
\frac{
\displaystyle
\frac{8}{3}\mu x^2e^{-x^2}
-\frac{4}{3}s\,x\,\mathcal{G}'(x)
}{
\displaystyle
1-\beta_{\rm NG}
\left[
\mu e^{-x^2}+s\,\mathcal{G}(x)
\right]
},
\label{eq:Cl_NG_caseA}
\end{equation}
In the limit $\beta_{\rm NG}\to0$, these expressions reduce to the
Gaussian results given above.

The cases with \(\beta_{\rm NG}>0\) were first studied numerically in
Refs.~\cite{Atal:2019erb,Escriva:2023uko,Escriva:2025ftp}, where it was also shown that the vacuum-bubble scenario, namely PBH production from vacuum bubbles, can arise and become dominant for \(\beta_{\rm NG} \gtrsim 3.1\). The cases with negative \(\beta_{\rm NG}\) were studied in Refs.~\cite{Shimada:2024eec,Inui:2024fgk}, using a monochromatic power spectrum. 

For each shape, we then perform a numerical simulation to determine the collapse
threshold \(\mu_c\) as a function of \(\beta_{\rm NG}\). For
$\beta_{\rm NG}\neq0$, the logarithmic map is real only if
$1-\beta_{\rm NG}\zeta_G(r,s)>0$ throughout the profile. The boundary of
the adiabatic domain is therefore determined by
$\max_r[\beta_{\rm NG}\zeta_G(r,s)]=1$. In general, this defines a
shape-dependent limiting amplitude
$\mu_{\rm sing}=\mu_{\rm sing}(s)$. For $s=0$, the singular boundary
satisfies $\mu_{\rm sing}=1/\beta_{\rm NG}$ whenever the corresponding
amplitude branch reaches the singularity. For the positive-amplitude
branch considered here, this boundary is relevant for
$\beta_{\rm NG}>0$. Beyond it, the logarithmic map becomes singular and
the vacuum-bubble channel can be triggered.

As shown in Refs.~\cite{Escriva:2023uko,Escriva:2025ftp}, numerical simulations
are necessary in this vacuum-bubble production regime in order to determine accurately the
corresponding critical conditions and PBH mass functions. In the present work we restrict ourselves to the adiabatic channel, corresponding to the collapse of super-horizon curvature fluctuations.

\begin{figure}[t]
\centering
\includegraphics[width=1.\textwidth]{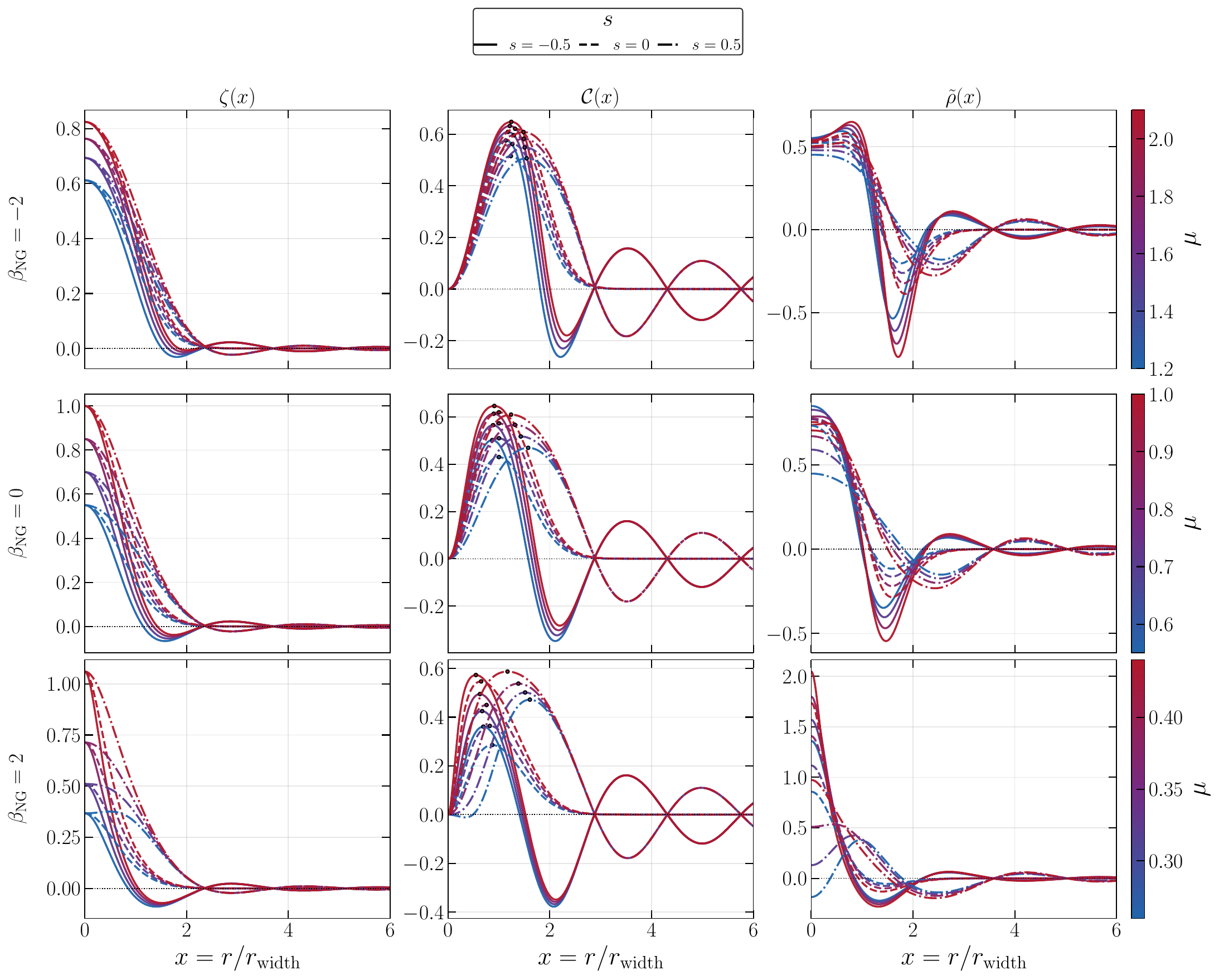}
\caption{
The rows correspond to different values of the
logarithmic non-Gaussianity parameter, \(\beta_{\rm NG}=-2,0,2\), while the
columns show, from left to right, the curvature profile \(\zeta(x)\), the compaction function \({\cal C}(x)\), and the density contrast
\(\tilde{\rho}(x)\) perturbation.  The radial coordinate is normalized as
\(x=r/r_{\rm width}\).  Different line styles indicate the split amplitude
\(s=-0.5,0,0.5\), and the colour scale gives the corresponding amplitude \(\mu\).
}
\label{fig:initial_shape_profiles}
\end{figure}

Figure~\ref{fig:initial_shape_profiles} illustrates how the initial curvature
profile and the associated compaction function are modified by the dispersion
degree of freedom. For fixed $\beta_{\rm NG}$, the parameter $\mu$ controls the overall amplitude of the perturbation, while the parameter $s$ changes
the shape around the reference profile. The branch $s=0$ corresponds
to the conditional mean configuration under the chosen conditioning,
which we use as the reference branch, whereas $s=\pm0.5$ introduces
a deformation of the radial profile. The main effect of the dispersion mode is not simply a rescaling of the profile amplitude.  Instead, it changes the radial structure of the perturbation.  In particular, the peak of the compaction function is displaced, and the outer compensation region is modified. 

The \(\tilde{\rho}(x)\) panels provide the corresponding long-wavelength
density diagnostic: they make explicit how the same shape deformation redistributes the central overdensity and the surrounding compensated underdensity. The shapes are mainly characterized by a positive overdensity surrounded by an underdense region, although for the case ($\beta_{\rm NG}=2$) it is possible to obtain configurations with a central underdensity.

The comparison between the different rows shows the role of the logarithmic
non-Gaussian map.  For negative $\beta_{\rm NG}$ the profiles are effectively
enhanced at larger values of $\mu$, while for positive $\beta_{\rm NG}$ the same
range of curvature amplitudes is reached with smaller values of $\mu$.  The
non-Gaussian transformation therefore changes the relation between the Gaussian
seed amplitude and the physical curvature perturbation.  Nevertheless, for each
fixed value of $\beta_{\rm NG}$, the dispersion parameter $s$ continues to
control the profile shape and the degree of compensation. The compaction profiles also show that sufficiently large shape deformations can
generate oscillatory or overcompensated tails. These features are not present in
the reference profile and arise from the dispersion basis itself.

\begin{figure}[t]
\centering
\includegraphics[width=1.\textwidth]{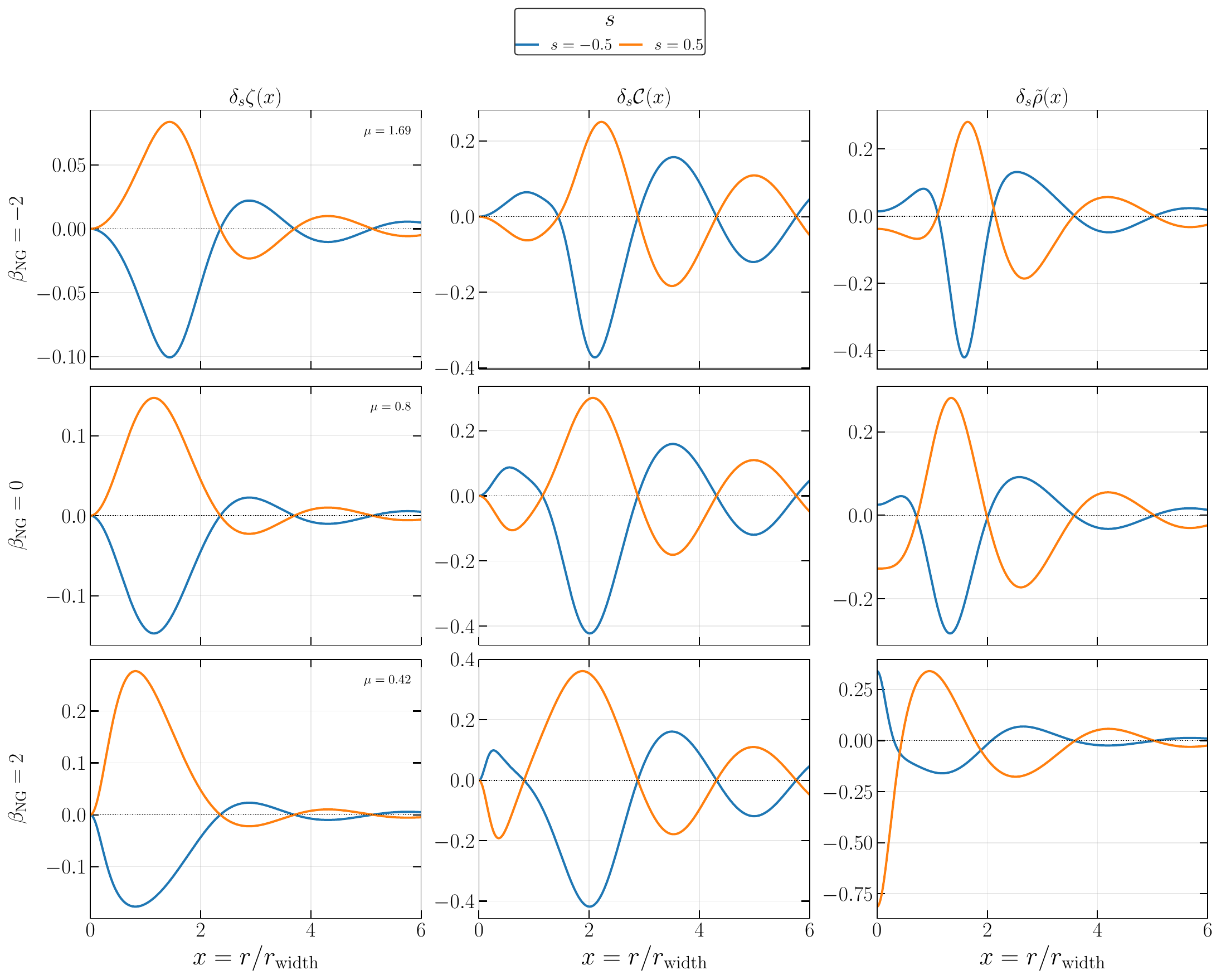}
\caption{
Difference between the critical profiles with positive and negative split
deformation and the corresponding reference profile at \(s=0\), for the
finite-width exponential spectrum.  The rows correspond to
\(\beta_{\rm NG}=-2,0,2\), while the columns show, from left to right, the
differences in the curvature profile, \(\delta_s\zeta(x)\), the compaction function, \(\delta_s{\mathcal{C}}(x)\), and the density contrast perturbation,
\(\delta_s\tilde{\rho}(x)\).  The radial coordinate is normalized as
\(x=r/r_{\rm width}\).  The blue and orange curves correspond to
\(s=-0.5\) and \(s=0.5\), respectively.  The value of \(\mu\) indicated in each
row gives the threshold amplitude of the reference branch.
}
\label{fig:dispersion_delta_profiles}
\end{figure}

Figure~\ref{fig:dispersion_delta_profiles} isolates the contribution of the
shape-dispersion mode by subtracting the reference $s=0$ profile at fixed
$\mu$ and fixed $\beta_{\rm NG}$.  In this way, the plotted quantities,
\begin{equation}
\delta_s\zeta(x)=\zeta_s(x)-\zeta_{s=0}(x),\qquad
\delta_s \mathcal{C}(x)=\mathcal{C}_s(x)-\mathcal{C}_{s=0}(x),\qquad
\delta_s\tilde{\rho}(x)=\tilde{\rho}_s(x)-\tilde{\rho}_{s=0}(x).
\end{equation}
show only the deformation induced by the parameter $s$. Increasing the power-spectrum amplitude at fixed $n$ increases the amount of shape
dispersion $s$.

The curvature perturbation is modified in a radially dependent way.  The
positive and negative values of $s$ produce opposite deformations around the reference profile, with the largest deviations occurring near the transition between
the central overdense region and the outer compensation region.  This confirms
that the dispersion mode does not act as a simple amplitude renormalization of
$\zeta(x)$, but changes the shape of the profile itself.

The corresponding variation of the compaction function is more pronounced,
because $\mathcal{C}(x)$ depends on radial derivatives of the curvature
perturbation.  As a result, even a moderate deformation in $\zeta(x)$ can produce a sizeable shift in the compaction profile, including changes in the peak
position, peak height, and compensated tail.  This explains why the collapse threshold $\mu_c(s)$ is sensitive to the dispersion parameter as we will see later: the relevant quantity for PBH formation is controlled by the full radial
structure of \(\mathcal{C}(x)\), especially around its dominant maximum, rather
than only by the central value of the curvature perturbation.

The comparison among different values of $\beta_{\rm NG}$ shows that the
logarithmic non-Gaussian map changes the response of the curvature profiles to
the same Gaussian-seed deformation. Therefore, the impact of the dispersion
mode and the effect of local non-Gaussianity are not independent: the
non-Gaussian map modifies how a fixed shape fluctuation in the seed profile is
translated into the physical curvature and compaction profiles. 

\begin{figure}[t]
    \centering
    \includegraphics[width=1.0\textwidth]{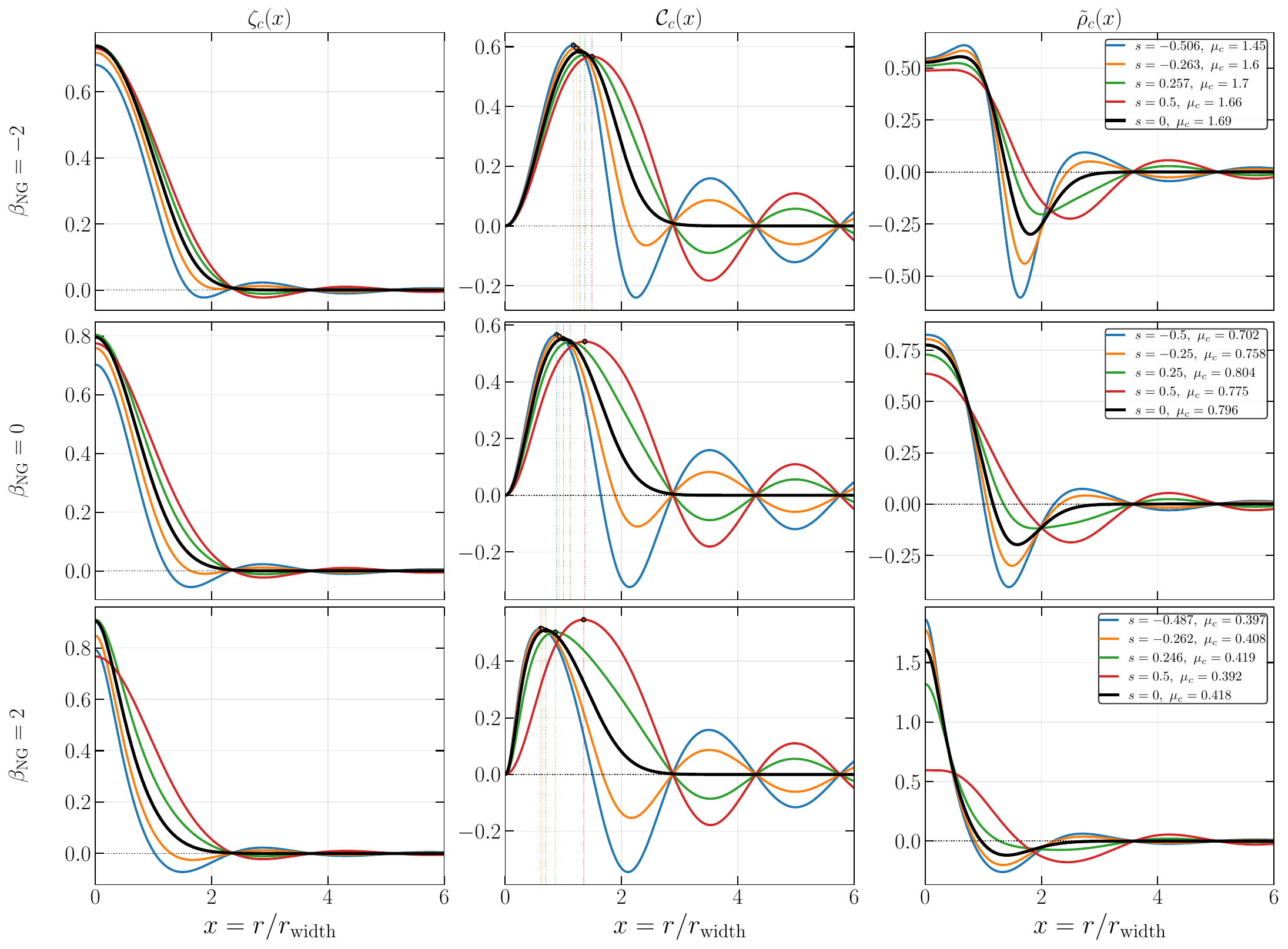}
\caption{Critical profiles. The rows correspond to
\(\beta_{\rm NG}=-2,0,2\), while the columns show, from left to right, the
critical curvature profile \(\zeta_c(x)\), the compaction function
\({\cal C}_c(x)\), and the density contrast perturbation \(\tilde{\rho}_c(x)\).  The radial
coordinate is normalized as \(x=r/r_{\rm width}\).  Different colours
correspond to different values of the split amplitude \(s\), with the
associated collapse threshold \(\mu_c\) indicated in the legends; the black
curve shows the reference profile \(s=0\).  In the middle panels, the vertical
dotted lines mark the position of the maximum of the compaction function for
each branch.}
    \label{fig:threshold_results_initial_shape}
\end{figure}

\begin{figure}[t]
    \centering
    \includegraphics[width=0.48\textwidth]{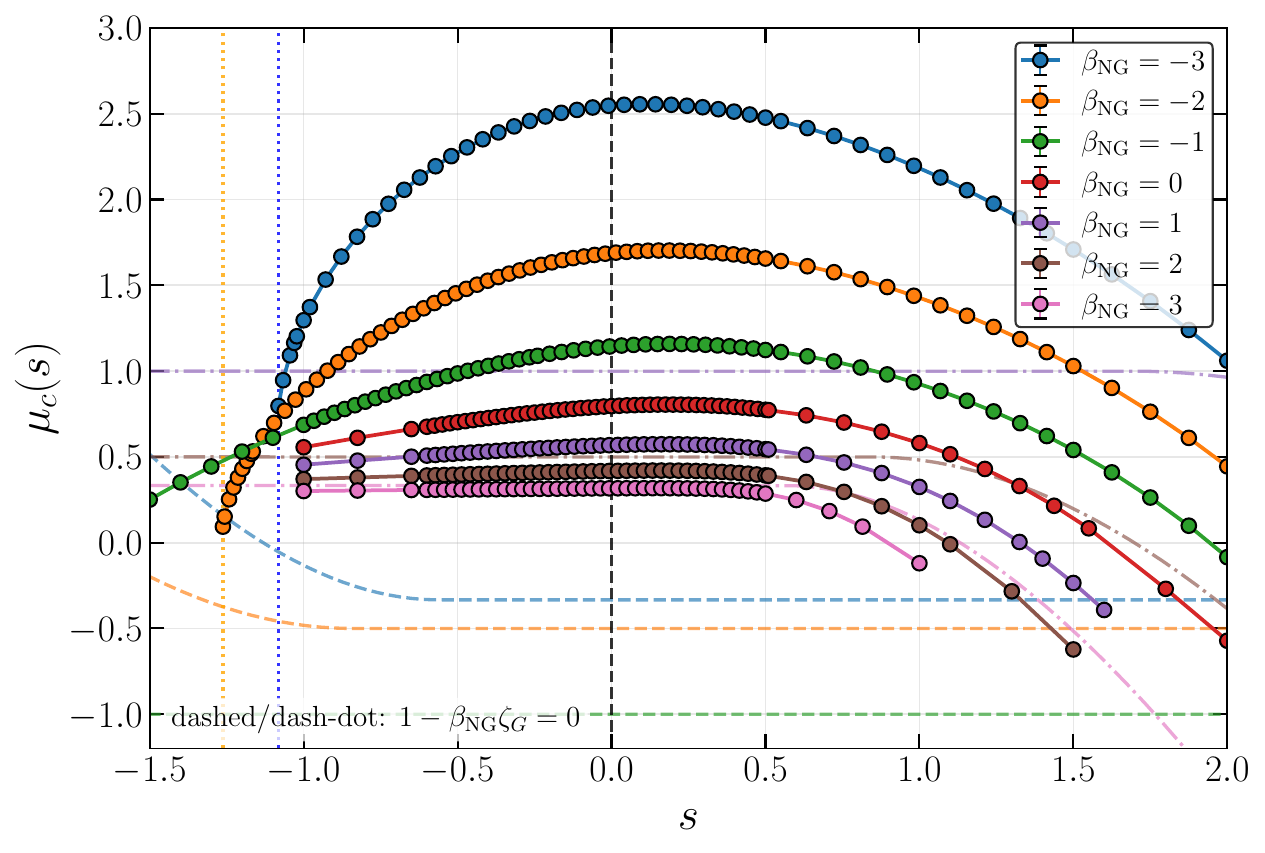}
        \includegraphics[width=0.48\textwidth]{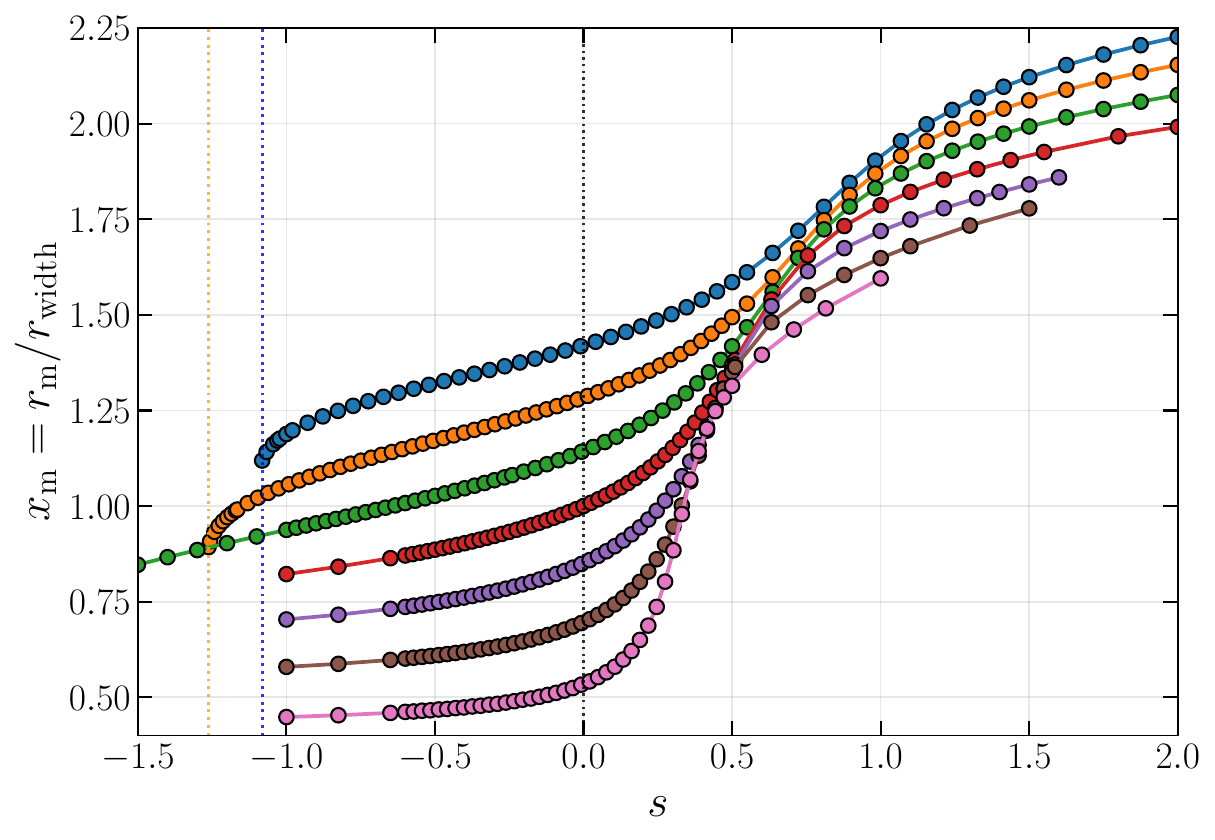}
    \includegraphics[width=0.48\textwidth]{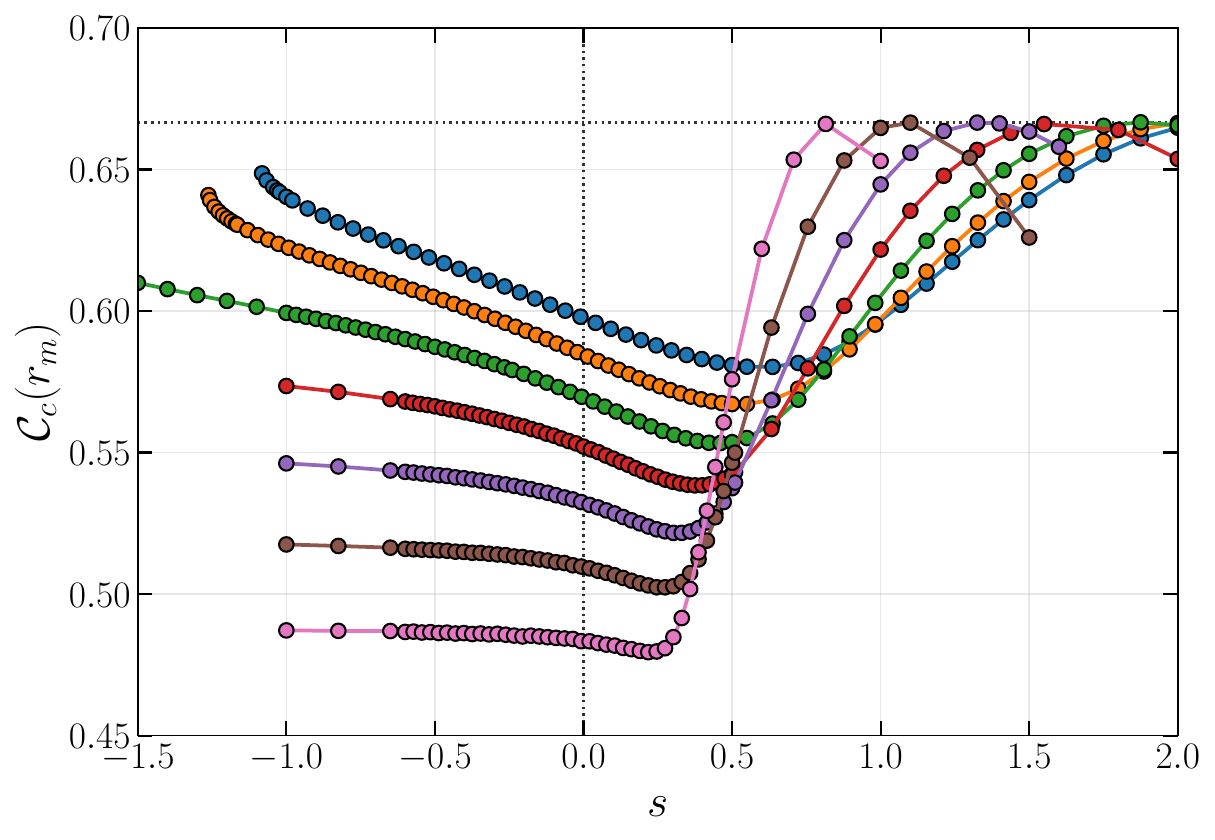}
        \includegraphics[width=0.48\textwidth]{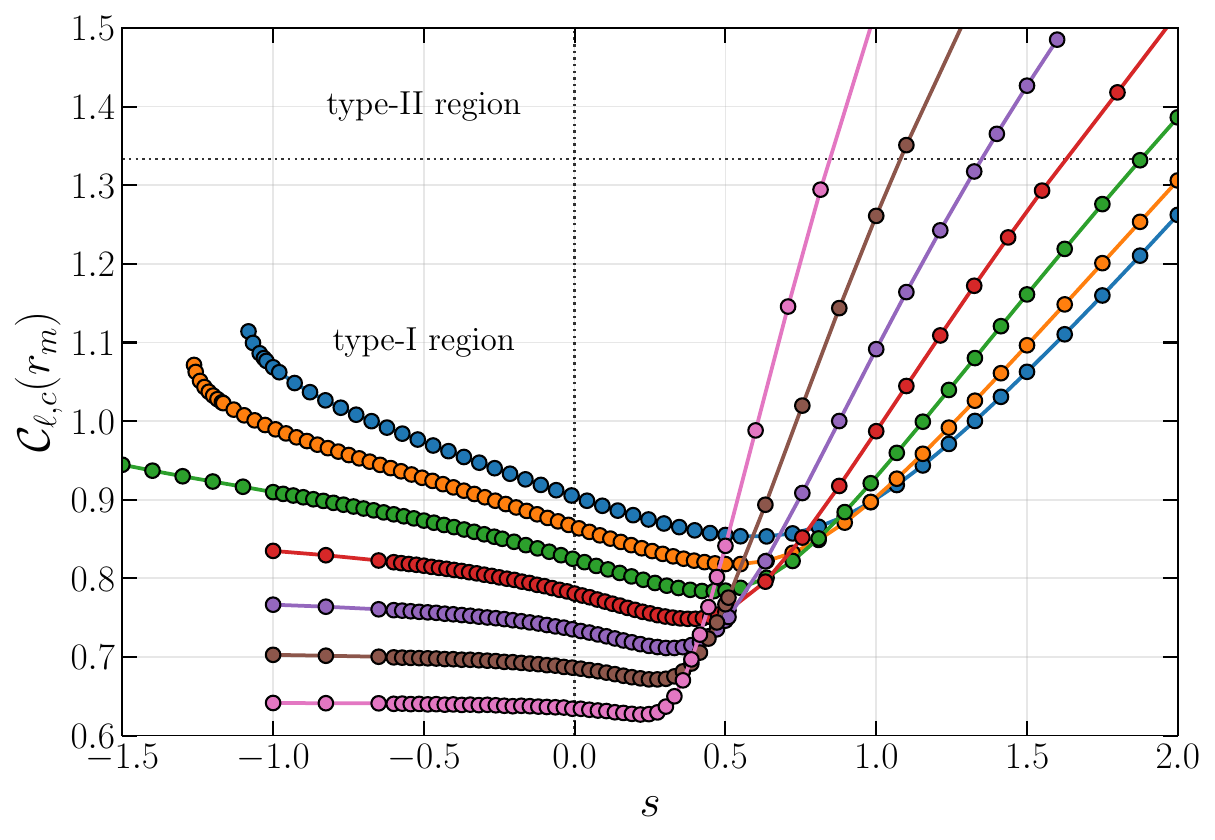}
\caption{
Threshold quantities as a function of the shape-dispersion parameter 
$s$ for several values of the non-Gaussian parameter 
$\beta_{\rm NG}$. 
The upper-left panel shows the threshold amplitude $\mu_c(s)$, while the 
upper-right panel shows the position of the maximum of the compaction function, 
$x_m=r_m/r_{\rm width}$. 
The lower-left and lower-right panels show, respectively, the nonlinear and 
linear compaction peaks at threshold,
$\mathcal{C}_{c}(r_m,s)$ and
$\mathcal{C}_{\ell,c}(r_m,s)$.
Different colours correspond to different values of $\beta_{\rm NG}$.
The vertical dotted lines mark reference values of
\(s\): the black line denotes the zero-dispersion branch, \(s=0\),
while the coloured lines indicate \(s=s_{\rm div}\) for
\(\beta_{\rm NG}=-3\) and \(\beta_{\rm NG}=-2\). The shaded/annotated region marks the domain where 
the critical initial data correspond to type-I fluctuations. 
The dashed and dash-dotted curves in the $\mu_c(s)$ panel indicate the 
regularity boundary of the logarithmic non-Gaussian map,
$1-\beta_{\rm NG}\zeta_G=0$.
}
\label{fig:threshold_results}
\end{figure}

\begin{figure}[t]
    \centering
    \includegraphics[width=1.0\textwidth]{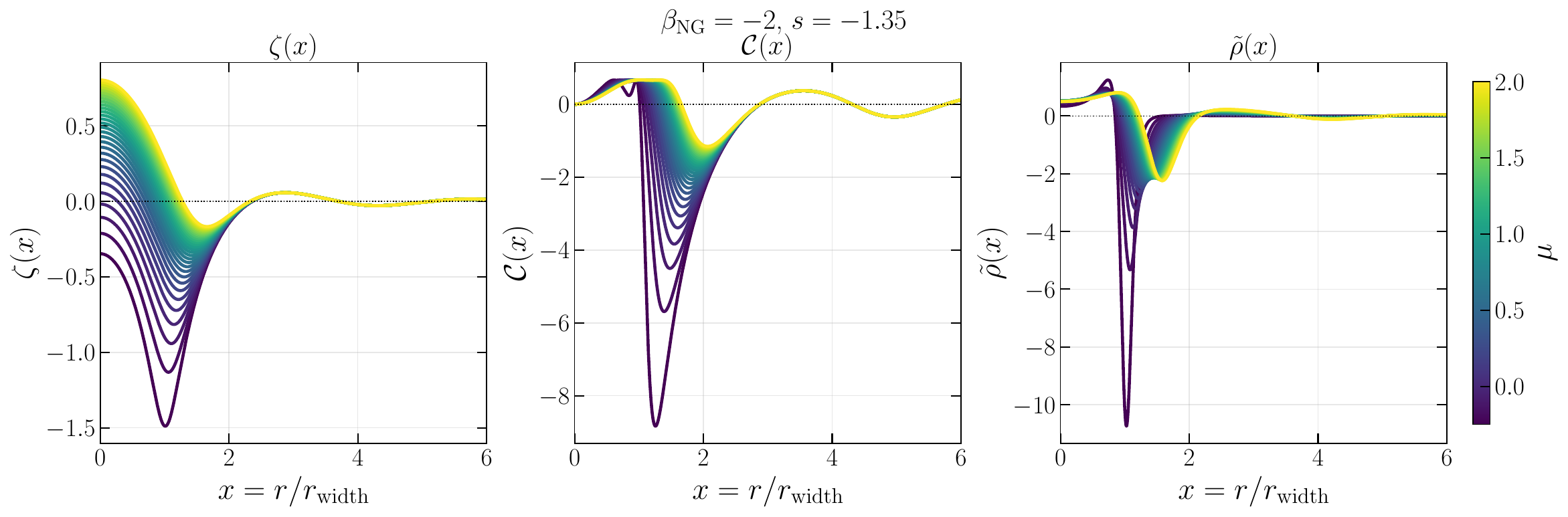}
    \includegraphics[width=0.5\textwidth]{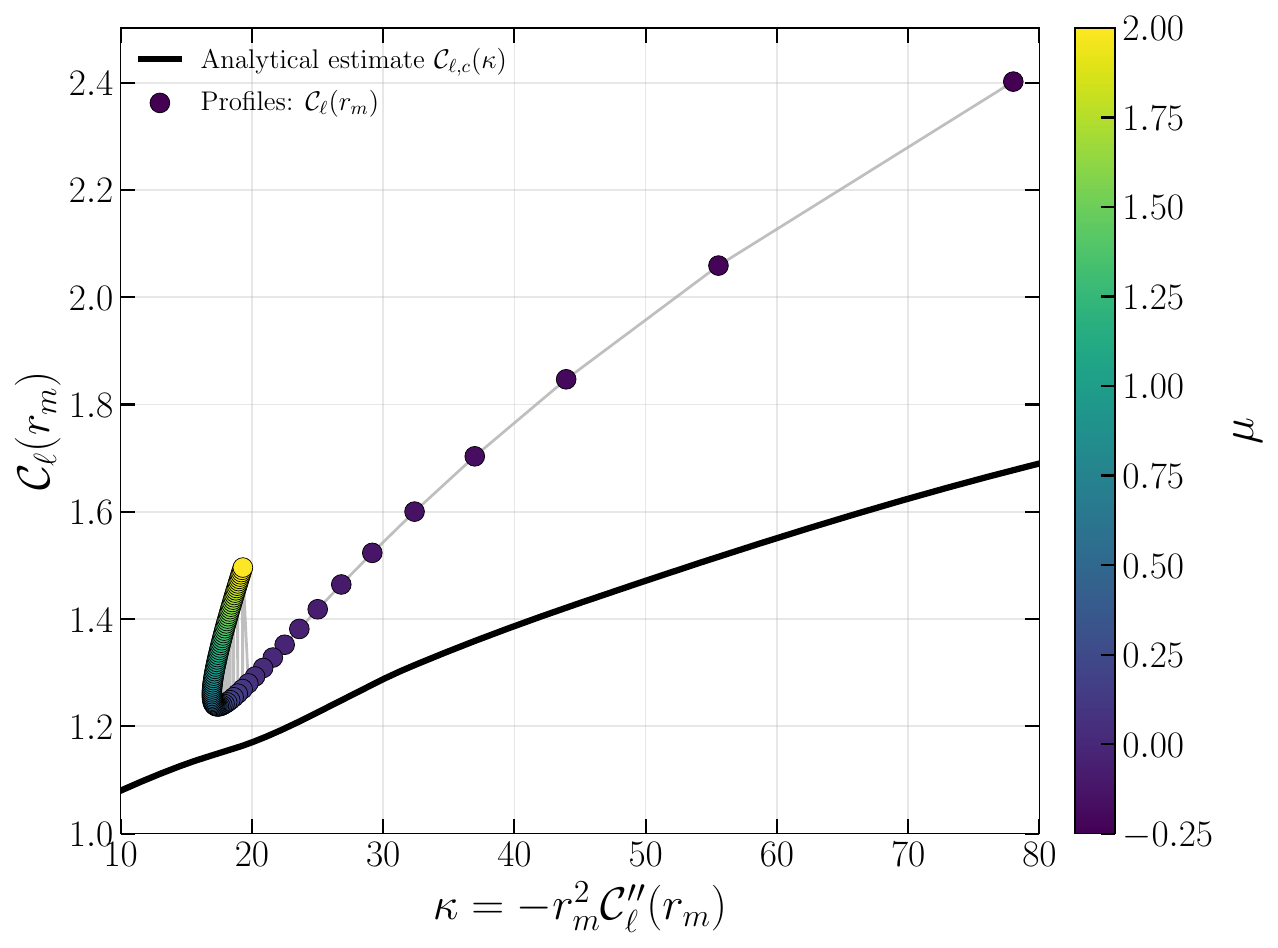}
\caption{Representative all-collapsing branch for the logarithmic non-Gaussian case \(\beta_{\rm NG}=-2\) and fixed shape deformation \(s=-1.35\).
The upper panels show the curvature profile \(\zeta(x)\), the compaction function \(\mathcal{C}(x)\), and the long-wavelength density perturbation \(\tilde{\rho}(x)\), with \(x=r/r_{\rm width}\), for Gaussian amplitudes in the range \(-0.25\leq\mu\leq2\). The colour scale denotes \(\mu\).
The lower panel shows the corresponding peak linear compaction
\(\mathcal{C}_{\ell}(r_m)\) as a function of
\(\kappa=-r_m^2\mathcal{C}_{\ell}^{\prime\prime}(r_m)\), compared with the analytical estimate from Ref.~\cite{Escriva:2025rja}.}
\label{fig:beta_m2_sminus135_all_collapse}
\end{figure}

Through relativistic numerical simulations, we determine the threshold \(\mu_c(s)\). The associated critical profiles for different parameter values are shown in Fig.~\ref{fig:threshold_results_initial_shape}. Fig.~\ref{fig:threshold_results} shows the collapse threshold and
the associated compaction quantities as the coherent shape parameter \(s\) is
varied. The threshold amplitude \(\mu_c(s)\) depends nontrivially on the
shape of the Gaussian seed profile.  For negative \(\beta_{\rm NG}\), larger
values of \(\mu\) are required because the logarithmic map suppresses the
physical curvature perturbation.  Nevertheless, the threshold is still
sensitive to the dispersion parameter, and the lowest values of \(\mu_c\) are
typically reached away from the reference branch \(s=0\).  For positive
\(\beta_{\rm NG}\), the logarithmic map enhances the curvature fluctuation and
the threshold in the Gaussian amplitude is correspondingly smaller. An
important aspect is that we find that, for the cases \(\beta_{\rm NG}=-3\) and
\(\beta_{\rm NG}=-2\), below a given value \(s<s_{\rm div}\) (which corresponds to the vertical coloured dotted line shown in the top-left panel of Fig.~\ref{fig:threshold_results}), all initial
conditions we have numerically tested collapse to form PBHs, independently of the peak value \(\mu\). A
representative example of this behaviour is shown in
Fig.~\ref{fig:beta_m2_sminus135_all_collapse}. We fix
\(\beta_{\rm NG}=-2\) and choose the strongly deformed branch
\(s=-1.35\), for which the numerical scans indicate collapse throughout the
range of amplitudes considered.  For this branch, the configurations are of type II fluctuations (the areal radius at leading order in gradient expansion $\sim r e^{\zeta}$ is not a monotonically increasing function of the radial coordinate \cite{Kopp:2010sh}) for sufficiently negative
amplitudes, \(\mu \lesssim -0.003\), and again for large positive amplitudes,
\(\mu \gtrsim 1.063\).  In the intermediate range,
\(-0.003 \lesssim \mu \lesssim 1.063\), the profiles correspond to type-I fluctuations (the areal radius is a monotonically increasing function of the radial coordinate). Although the central Gaussian amplitude
\(\mu\) is varied from positive values down to small and negative values, the
coherent deformation produces a large peak value of \(\mathcal{C}_{\ell}(r_m)\), which is above the analytical threshold estimate
curve of Ref.~\cite{Escriva:2025rja}. This provides a simple consistency check of the numerical result. This also shows that the peak value \(\mu\) clearly does not determine the critical threshold, and that a criterion based on \(\mathcal{C}_{\ell}\)~\cite{Escriva:2025rja} becomes useful.

The upper-right panel of Figure~\ref{fig:threshold_results} shows that changing \(s\) also changes the characteristic
scale of the perturbation, as measured by the location of the compaction
maximum \(x_m=r_m/r_{\rm width}\), where increasing $s$ shifts the compaction maximum toward larger radii, independently of $\beta_{\rm NG}$.

The lower panels show the thresholds expressed in terms of \(\mathcal{C}_c(r_m)\) and
\(\mathcal{C}_{\ell,c}(r_m)\). As a general trend, the threshold values
decrease as \(s\) increases from the reference branch \(s=0\), while they
become larger for \(s<0\). Only for sufficiently large positive \(s\) do the
thresholds increase significantly again. This behaviour is qualitatively
different from that of the Gaussian-amplitude threshold \(\mu_c\). The reference-branch critical configurations considered here lie in the type-I
region of collapse, while sufficiently large shape deformations can enter
the type-II region, as discussed below.

\begin{figure}[t]
    \centering
    \includegraphics[width=1.0\textwidth]{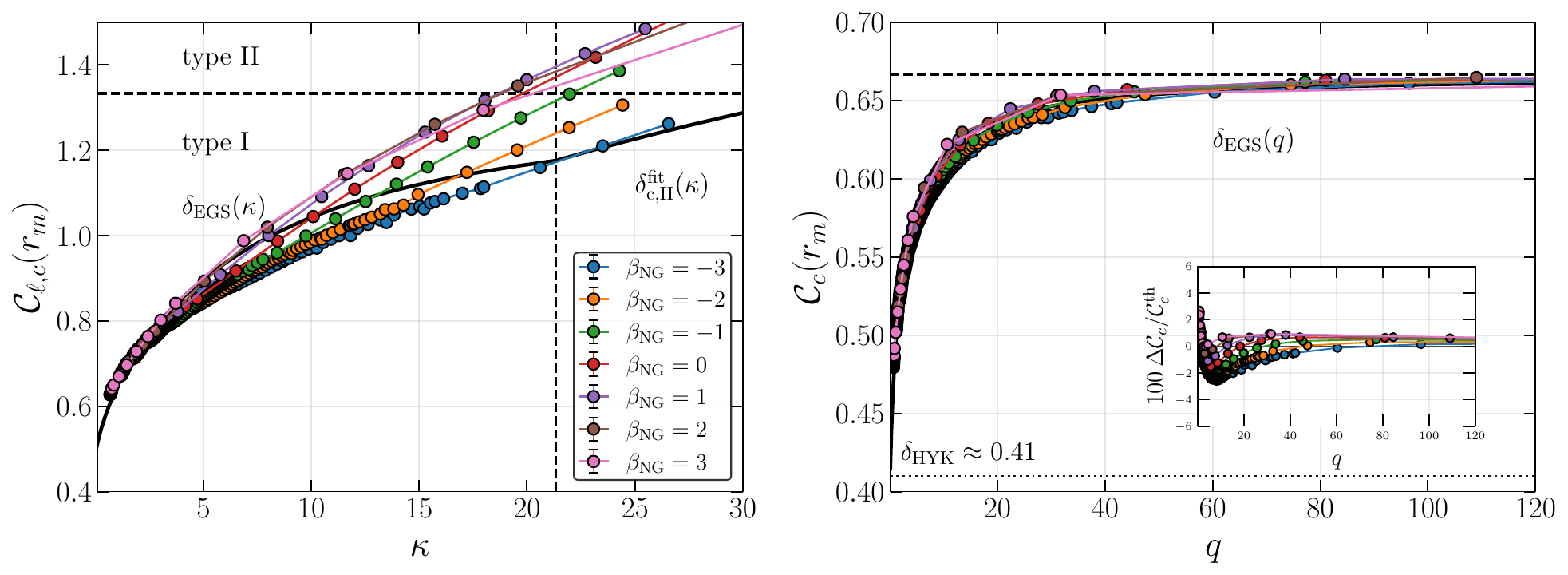}
\caption{
Comparison of the threshold compaction obtained from the shape-dispersion analysis with standard threshold prescriptions based on the profile shape. 
Left panel: linear compaction function threshold $\mathcal{C}_{\ell,c}(r_m)$ as a function of the curvature parameter $\kappa$, compared with the fitting formula $\delta_{\rm EGS}(\kappa)$ for $\kappa \lesssim 21$ and $\delta^{\rm fit}_{\rm c,II}(\kappa)$ for $\kappa \gtrsim 21$ following \cite{Escriva:2025rja} with $\delta^{\rm fit}_{\rm c,II}(\kappa) \equiv \mathcal{C}_{\ell,c}^{\rm fit,II}(\kappa) = a\kappa^b, a \approx 0.51962, b\approx 0.26687$. Right panel: compaction function threshold $\mathcal{C}_{c}(r_m)$ as a function of the shape parameter $q$, compared with the EGS estimate $\delta_{\rm EGS}(q)$. 
The horizontal dotted line indicates the reference value $\delta_{\rm HYK}\simeq 0.41$. 
The coloured points correspond to the thresholds obtained for different values of the logarithmic non-Gaussian parameter $\beta_{\rm NG}$ and different values of the shape-dispersion parameter $s$. 
The inset shows the relative deviation between the numerical threshold and the corresponding analytical or semi-analytical estimate.
}
\label{fig:threshold_results_comparison}
\end{figure}

We also compare our numerical threshold results with the analytical threshold
prescriptions commonly used in the literature: the prescription of
Ref.~\cite{Escriva:2019nsa}, which we denote by
\(\delta_{\rm EGS}(q)\)\footnote{Specifically,
\[
\delta_{\rm EGS}(q)
=
\frac{4}{15}
e^{-1/q}
\frac{
q^{\,1-\frac{5}{2q}}
}{
\Gamma\!\left(\frac{5}{2q}\right)
-
\Gamma\!\left(\frac{5}{2q},\frac{1}{q}\right)
},
\], where $\Gamma(a,z)$ denotes the upper incomplete gamma function.}, and the estimate
\(\delta_{\rm HYK}\) of Ref.~\cite{Harada:2013epa}. The comparison is shown in
Fig.~\ref{fig:threshold_results_comparison}. In the right panel, we show the compaction functions threshold
\(\mathcal{C}_{c}(r_m)\) as a function of the shape parameter $q \equiv -\frac{\tilde{r}_m^2\mathcal{C}_c''(\tilde{r}_m)}
{4\mathcal{C}_c(\tilde{r}_m)}$, introduced in Ref.~\cite{Escriva:2019nsa} for type-I fluctuations. Here $\tilde r$ is the areal radial coordinate, related to the conformally flat radial coordinate $r$ used for the curvature profile through
Eq.~\eqref{eq:rtilde_relation}. In the left panel, we show the
linear compaction function threshold \(\mathcal{C}_{\ell,c}(r_m)\) as a function of $\kappa \equiv -r_m^2\mathcal{C}''_{\ell,c}(r_m)$. This parameter was used in Ref.~\cite{Germani:2023ojx} in the study of the
nonlinear statistics of the compaction function for type-I fluctuations,
and in Ref.~\cite{Escriva:2025rja} to construct an extension of the
threshold prescription into the type-II region in terms of
$\kappa$ and $\mathcal{C}_{\ell,c}$ for certain classes of profiles.

When comparing the numerical results with the analytical estimate of
Ref.~\cite{Escriva:2019nsa}, we find very good agreement, with deviations within
about \(4\%\). For the linear compaction function threshold
$\mathcal{C}_{\ell,c}$, the deviations from the type-I EGS prediction
appear larger, partly because of the nonlinear relation between
$\mathcal{C}_{c}$ and $\mathcal{C}_{\ell,c}$. Within the type-I region,
the EGS estimate for $\mathcal{C}_{c}$ remains accurate at the few-percent level. However, once the critical configurations enter the type-II region, the assumptions underlying the EGS prescription, derived for type-I configurations, are no longer satisfied. Consequently, the threshold $\mathcal{C}_{c}$ need not asymptote to $2/3$ as $\kappa \rightarrow \infty$, as discussed in Ref.~\cite{Escriva:2025rja}. This behaviour is also consistent with the results of Ref.~\cite{Escriva:2025rja}, which showed that the critical threshold lies in the type-I collapse regime for low values of $\kappa$, whereas it enters the type-II collapse regime for sufficiently large $\kappa$. On the other hand, when comparing with the estimate of
Ref.~\cite{Harada:2013epa}, we find a significantly larger deviation. This is
consistent with the comparison and discussion presented in
Ref.~\cite{Escriva:2020tak} for different equations of state. The collapse
threshold is sensitive to the detailed curvature profile and to the nonlinear
dynamics of gravitational collapse, and therefore it cannot be regarded as a
constant value independent of the perturbation shape. Numerical simulations are
therefore essential for an accurate quantitative determination of the threshold.
In particular, simplified estimates or treatments that do not explicitly account
for the profile dependence do not provide a valid general prescription for the
collapse threshold.

\begin{figure}[!htbp]
    \centering
    \includegraphics[width=1.0\textwidth]{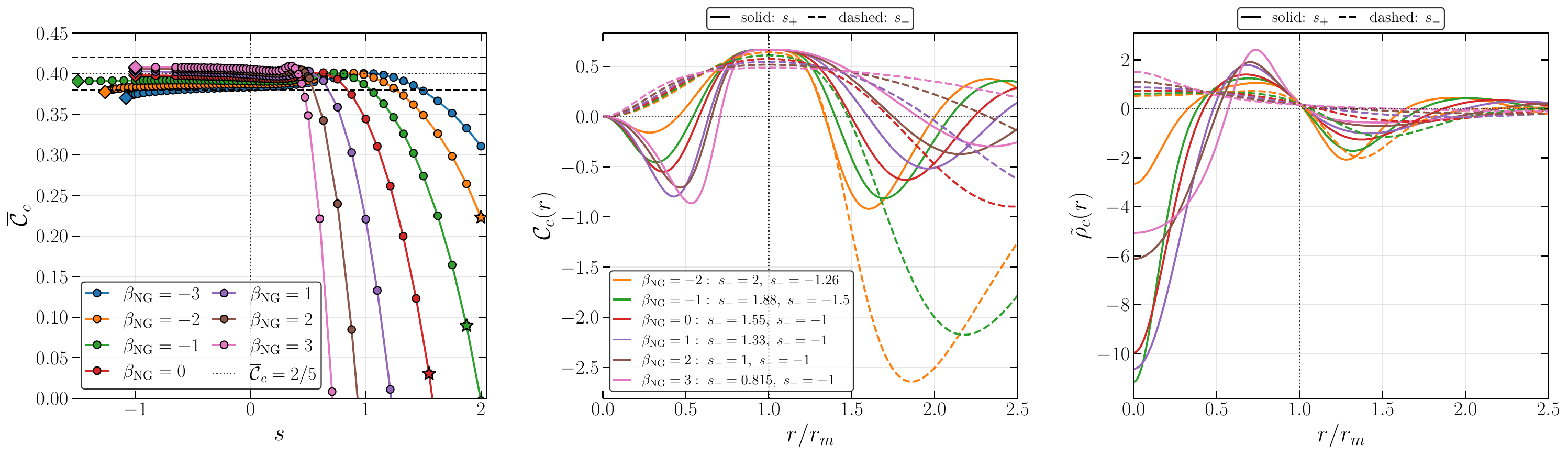}
\caption{Left panel shows the volume-averaged critical compaction
\(\overline{\mathcal C}_c\) as a function of the shape parameter \(s\) for
different values of \(\beta_{\rm NG}\). The horizontal dotted line marks
\(\overline{\mathcal C}_c=2/5\). Middle and right panels show the corresponding critical compaction
function \(\mathcal C_c(r)\) and density perturbation
\(\tilde{\rho}_c(r)\) for representative positive- and negative-\(s\)
configurations. Solid lines correspond to \(s_+\) and dashed lines to \(s_-\).
The radial coordinate is normalized by the corresponding compaction scale
\(r_m\), indicated by the vertical dotted line.}
\label{fig:threshold_averaged_panel}
\end{figure}

As an additional characterization of the critical configurations, we consider the volume-averaged compaction function introduced in \cite{Escriva:2019nsa}. For a profile with areal radius
\(R(r)=a re^{\zeta(r)}\) (where $a$ is the scale factor), it is defined as
\begin{equation}
\overline{\mathcal C}
=
\frac{3}{R_m^3}
\int_0^{R_m}
\mathcal C(R)R^2\,dR ,
\qquad
R_m\equiv R(r_m),
\end{equation}
or, equivalently,
\begin{equation}
\overline{\mathcal C}
=
\frac{3}{r_m^3e^{3\zeta(r_m)}}
\int_0^{r_m}
\mathcal C(r)
\left[1+r\zeta'(r)\right]
e^{3\zeta(r)}r^2\,dr .
\label{eq:Cbar}
\end{equation}
At the collapse threshold we denote this quantity by
\(\overline{\mathcal C}_c\).
For radiation domination, the value
\(\overline{\mathcal C}_c\simeq 2/5\) was found to provide an approximately
universal threshold for a broad class of PBH-forming profiles
\cite{Escriva:2019nsa}.

Figure~\ref{fig:threshold_averaged_panel} shows
\(\overline{\mathcal C}_c\) along the critical branches considered here.
For negative values of \(s\), the averaged critical compaction remains close to \(2/5\), with deviations within \(\sim 5\%\) (denoted by the horizontal dashed lines in the left panel), even for relatively large profile deformations. In contrast, for sufficiently large positive \(s\), sizeable deviations from \(2/5\)
appear. To illustrate the origin of these deviations, the middle and right panels
show the critical compaction and density profiles for representative
positive- and negative-\(s\) configurations. The negative-\(s\) profiles
retain a conventional central overdensity and give
\(\overline{\mathcal C}_c\simeq2/5\). For large positive \(s\), instead,
the profiles can develop a central underdense region surrounded by a positive
mass excess. The inner negative contribution then reduces the averaged
compaction, although the outer overdense region can still collapse and form a
PBH. Similar departures from the ($2/5$) criterion have previously been found for sufficiently large negative non-Gaussianity within a quadratic expansion model, where profiles with a negative mass excess develop near the central region \cite{Escriva:2022pnz}. However, as we will see later, configurations with large positive values of $s$ are highly statistically suppressed.

\subsubsection{Branch-weight diagnostic}

Before computing the full PBH mass function, it is useful to estimate which branches of the threshold curve are statistically favoured. This gives a simple diagnostic of the competition between two effects. On the one hand, large values of $|s|$ are suppressed by the Gaussian cost of the coherent shape deformation. On the other hand, a branch with a smaller collapse threshold $\mu_c(s)$ can be exponentially enhanced because PBHs form from less rare peaks.

The numerical simulations provide the threshold amplitude
$\mu_c=\mu_c(s;\beta_{\rm NG})$ as a function of the physical shape
amplitude $s$. From Eq.~\eqref{eq:beta_disp_general}, we define the
branch-weight diagnostic, up to an overall normalization independent of the
sampled branch,
\begin{equation}
\log \Upsilon_{\rm br}(s;A_{\zeta})
=
-{s^2\over 2A_{\zeta}}
+
\log N_{\rm pk}^{\rm BBKS}
\left(
>
{\mu_c(s;\beta_{\rm NG})\over \sqrt{A_{\zeta}}}
\right)
+
{\rm const.}
\label{eq:logW_branch_weight_insert}
\end{equation}
The first term is the Gaussian penalty for exciting the coherent shape mode,
written in terms of the numerical deformation amplitude \(s\).  Equivalently,
it is \(-n^2/2\), with \(n=s/\sqrt{A_{\zeta}}\).  The second term is the cumulative BBKS peak-abundance factor evaluated at the threshold height required for that
branch to collapse. We also define a normalized relative contribution $w_{\rm rel}$ as 

\begin{equation}
w_{\rm rel}(s)
=
\frac{
\Upsilon_{\rm br}(s;A_\zeta)
}{
\displaystyle
\int_{s_{\min}}^{s_{\max}} ds'\,
\Upsilon_{\rm br}(s';A_\zeta)
},
\qquad
\int_{s_{\min}}^{s_{\max}} ds\,
w_{\rm rel}(s)
=
1.
\label{eq:relative_branch_weight22}
\end{equation}

\begin{figure}[t]
    \centering
    \includegraphics[width=0.48\textwidth]{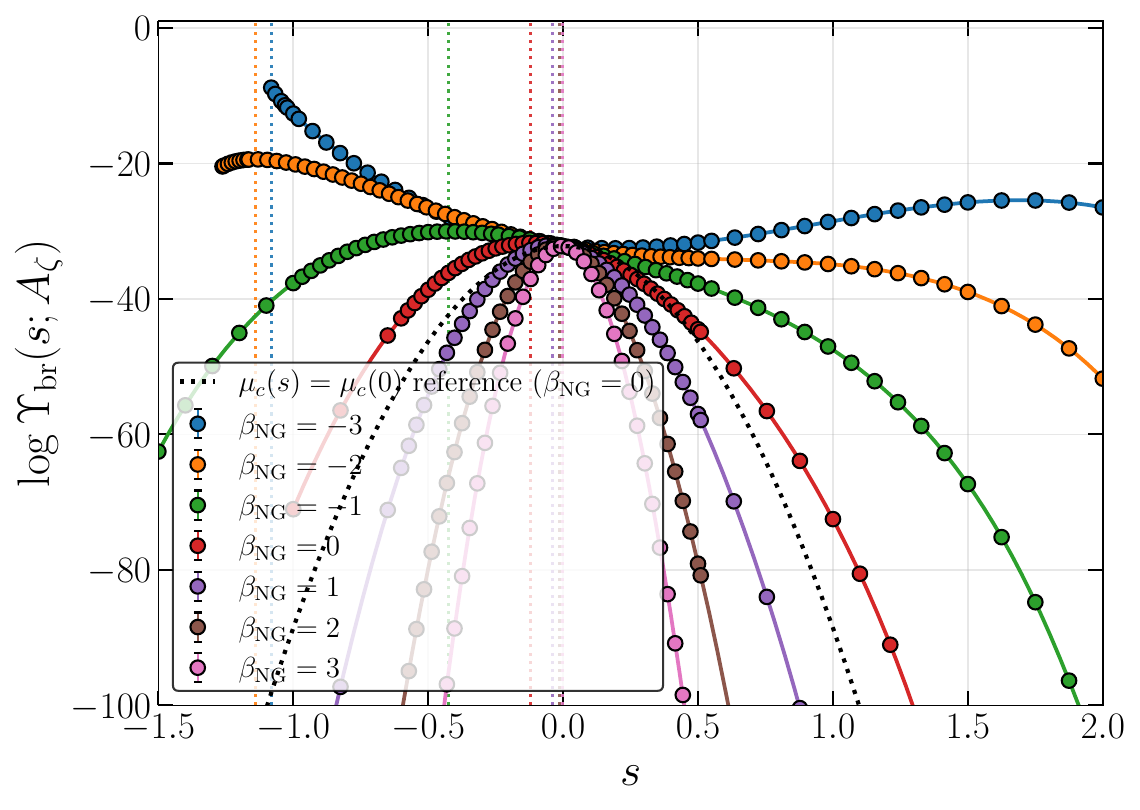}
    \includegraphics[width=0.48\textwidth]{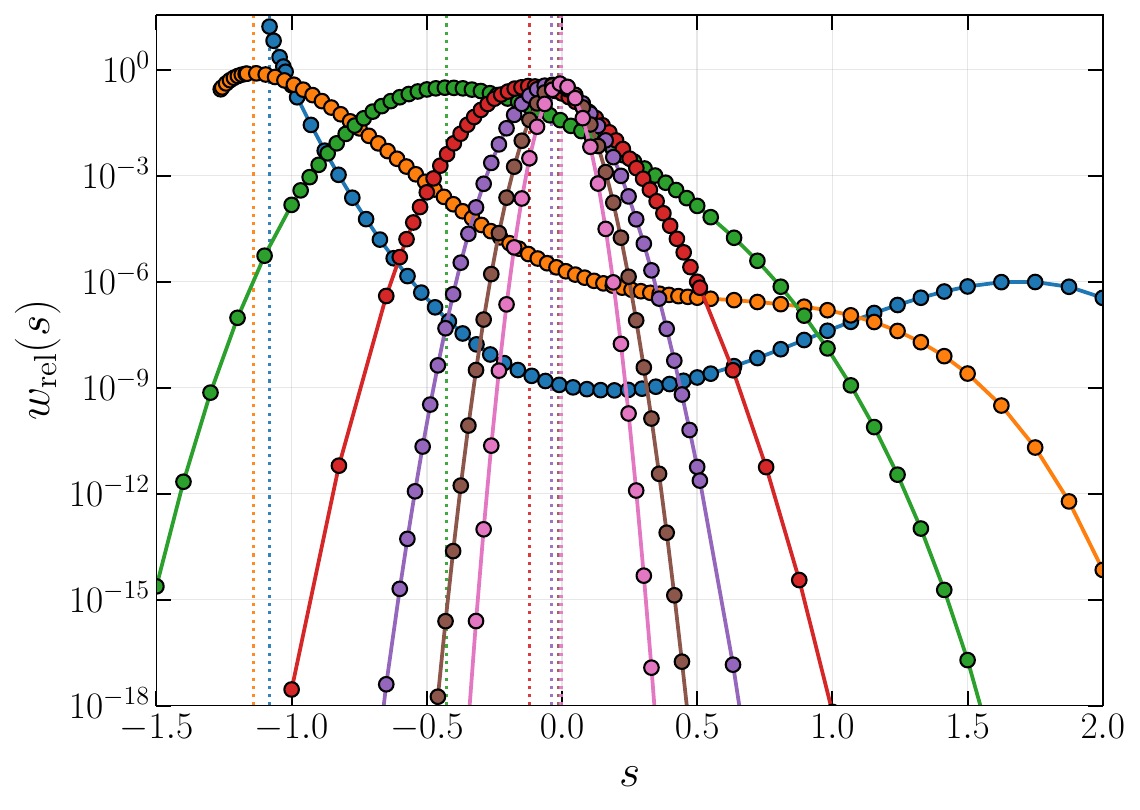}
    \caption{
    Branch-weight diagnostic for the coherent spherical shape-dispersion branches as a function of $s$, for several values of $\beta_{\rm NG}$.
    Left panel: logarithm of the branch weight $\log \Upsilon_{\rm br}(s;A_{\zeta})$, including the Gaussian cost of the shape deformation and the cumulative BBKS peak-abundance factor evaluated at the threshold $\mu_c(s)$.
    The black dotted curve shows the artificial reference case in which the collapse threshold is kept fixed, $\mu_c(s)=\mu_c(0)$, so that the variation with $s$ is due only to the Gaussian shape-probability suppression.
    Right panel: normalized relative-weight density $w_{\rm rel}$ of each sampled branch. The vertical dotted lines indicate, for each \(\beta_{\rm NG}\), the sampled value \(s=s_\star\) that gives the largest normalized relative-weight density \(w_{\rm rel}\).}
    \label{fig:shape_weight}
\end{figure}

Figure~\ref{fig:shape_weight} shows the resulting branch-weight diagnostic. The left panel displays $\log \Upsilon_{\rm br}(s;A_{\zeta})$ for the different values of $\beta_{\rm NG}$. The black dotted curve shows the fixed-threshold reference $\mu_c(s)=\mu_c(0)$, for which the variation with $s$ is entirely due to the Gaussian shape-probability suppression. The deviation of the coloured curves from this reference measures the effect of the threshold variation induced by the coherent shape deformation. Branches for which $\mu_c(s)$ decreases sufficiently can compensate, or even overcome, the Gaussian cost of moving away from $s=0$.

The right panel shows the normalized relative-weight density $w_{\rm rel}$.
The dominant branch is not necessarily the most probable shape $s=0$.
Instead, it is selected by the balance between the rarity of the shape
deformation and the efficiency of collapse. In the examples shown here,
negative values of $\beta_{\rm NG}$ tend to shift the dominant contribution
towards negative $s$, while for positive $\beta_{\rm NG}$ the dominant
region remains closer to the reference branch $s=0$. In these cases, the
abundance is mainly determined by the reference branch, while deviations
from it are not statistically significant enough to dominate. The precise
location depends on the detailed behaviour of the numerical threshold curve
$\mu_c(s;\beta_{\rm NG})$. In this sense and for the cases tested, positive
non-Gaussianity tends to reduce the impact of shape dispersion, since the
dominant contribution remains close to the reference branch. By contrast,
negative non-Gaussianity enhances the effect of shape dispersion: branches
away from the reference profile can become statistically dominant because
the reduction in the collapse threshold compensates the Gaussian cost of
the deformation. We also note that, for the case \(\beta_{\rm NG}=-3\), the branch-weight
diagnostic \(\log \Upsilon_{\rm br}(s;A_{\zeta})\) does not show a clear internal maximum within the
simulated range.  Instead, the weight keeps increasing towards the edge of the
available threshold curve.  This suggests that, for sufficiently strong negative
non-Gaussianity, the dominant contribution may lie close to the boundary of the
allowed or simulated shape domain.  A more detailed scan of this region would be
needed to determine whether the weight is eventually cut off by the regularity
condition, by the breakdown of the adiabatic channel, or by the physical limits
of the profile family.

Finally, it is also interesting to note that, whereas the critical-profile data correspond to type-I fluctuations for the reference profile $s=0$ cases considered, type-II critical profiles can be realized for configurations with large dispersion, particularly for positive values of $s$. However, such configurations are highly statistically suppressed in the cases considered.

\subsubsection{Shape-dispersed PBH abundance and mass function}
\label{subsec:shape_dispersed_mass_function_insert}

We now turn to the full mass-function calculation using
Eq.~\eqref{eq:mass_function_shape_disp_general_inverted}.  For each branch \(n\), the PBH mass is related to the Gaussian amplitude through the critical
collapse relation in Eq.~\eqref{eq:critical_mass_map_general}. A precise determination of the mass spectrum would require dedicated
near-critical simulations for each profile to calibrate the critical-scaling
prefactor and the profile-dependent mass map. However, given the large number of cases
considered in this work, this would be computationally expensive. Moreover, \(\mathcal{K} M_H\) enters multiplicatively and linearly in the
critical-scaling relation, whereas the collapse threshold controls the
exponentially suppressed rare-peak abundance. Consequently, the
sensitivity of the predicted abundance to the collapse threshold
parametrically dominates over that associated with moderate
profile-dependent variations in $\mathcal{K}$. We therefore adopt the effective
value \(\mathcal{K}_{\rm eff}\simeq 6\) to describe PBH masses in the critical
regime, motivated by Ref.~\cite{Escriva:2021pmf}, where values in the
range \(K\sim\mathcal{O}(1\text{--}10)\) were obtained for a
representative set of profiles,

\begin{equation}
M(\mu,s)
=
\mathcal{K}_{\rm eff}M_H(r_m(\mu,s))
\left[\mu-\mu_c(s)\right]^{\gamma_{\rm cr}},
\label{eq:critical_scaling_fixedMH_insert}
\end{equation}
$M_H(r_m)$ is the reference horizon mass associated with
the characteristic scale of the maximum of the compaction function $r_m(\mu,s)$. In the implementation, we may also include a profile-dependent horizon
mass. This accounts for the fact that the position of the compaction maximum
can vary with both the branch \(s\) and the amplitude \(\mu\).  We model this
effect as \(M_H(\mu,s)=M_k(k_{\rm peak})(k_{\rm peak}r_m(\mu,s))^2\), where
\(M_k(k)\simeq 1.22\times10^{13}M_\odot
(g_*/106.75)^{-1/6}(k/{\rm Mpc}^{-1})^{-2}\) is the cosmological horizon mass
when the comoving mode \(k\), given in units of \({\rm Mpc}^{-1}\), re-enters
the horizon~\cite{Tada:2019amh}.  In this work, we place the peak of the power
spectrum in the asteroid-mass range by choosing
\(k_{\rm peak}\simeq1.1\times10^{13}\,{\rm Mpc}^{-1}\), corresponding to
\(M_k(k_{\rm peak})\simeq10^{-13}M_\odot\). The corresponding
Jacobian is
\begin{equation}
\left|
\frac{d\ln M}{d\nu}
\right|^{-1}_{s}
=
\left|
\sigma_0
\left[
\frac{\gamma_{\rm cr}}{\mu-\mu_c(s)}
+
\frac{\partial\ln M_H(\mu,s)}{\partial\mu}
\right]
\right|^{-1}_s.
\label{eq:jacobian_rm_dependent_insert}
\end{equation}

In the present effective calculation, the general mass-function expression is specialized to the case of a single coherent variable describing spherical shape deformations. The numerical
threshold family is labelled by the fixed-height deformation
$s$. For each coherent-shape branch, we do not explicitly sample the BBKS
curvature, ellipticity, or prolateness variables; instead, we use the height-only BBKS peak density
\({\cal N}_{\rm pk}^{\rm BBKS}(\nu)\), in which the curvature variable has
already been integrated out through the standard BBKS function $G_{\rm BBKS}(\gamma_{\rm BBKS},x_*(\nu))$. The mass-function
integrand over the coherent shape variable is therefore
\begin{equation}
{\cal I}(M,n)
=
P_{\rm sh}(n)\,
{\cal N}_{\rm pk}^{\rm BBKS}\!\left(\nu\right)
\left|
\frac{d\ln M}{d\nu}
\right|^{-1}_{\nu=\nu_{M}(M,n)}.
\label{eq:mass_function_integrand_insert}
\end{equation}
The shape-dispersed PBH mass function is obtained by integrating over the
simulated range of the coherent deformation,
\begin{equation}
\frac{df_{\rm PBH}}{d\ln M}
=
\frac{M}{\rho_{\rm DM}}
\int_{n_{\rm min}}^{n_{\rm max}}dn\,
{\cal I}(M,n),
\label{eq:shape_dispersed_mass_function_insert}
\end{equation}
The finite integration interval reflects the fact that the threshold curve has only been determined over a finite simulated range in $n$, chosen to cover the region relevant for the abundance calculation. As a reference, we also compute the no-dispersion mass function by setting
$n=0$, or equivalently $s=0$, and removing the integral over the shape
distribution:
\begin{equation}
\left.\frac{df_{\rm PBH}}{d\ln M}\right|_{n=0}
=
\frac{M}{\rho_{\rm DM}}\,
{\cal N}_{\rm pk}^{\rm BBKS}\!\left(\nu\right)
\left|
\frac{d\ln M}{d\nu}
\right|^{-1}_{n=0}.
\label{eq:no_dispersion_mass_function_insert}
\end{equation}
Comparing Eq.~\eqref{eq:shape_dispersed_mass_function_insert} with
Eq.~\eqref{eq:no_dispersion_mass_function_insert} isolates the effect of the
coherent spherical shape dispersion on the PBH mass function. Additionally, we can define the ratio between the abundance of peaks obtained
with and without including shape dispersion as
\begin{equation}
\frac{\beta_{\rm disp}}{\beta_0}
=
\int_{n_{\min}}^{n_{\max}} dn\,
P_{\rm sh}(n)
\frac{
N_{\rm pk}^{\rm BBKS}(>\nu_c(n))
}{
N_{\rm pk}^{\rm BBKS}(>\nu_{\rm ref})
},
\label{eq:abundance_peak_ratio}
\end{equation}

Since \(s=n\sqrt{A_{\zeta}}\), the dominant branch can be expressed in units of the standard deviation as \(n_\star=s_\star/\sqrt{A_{\zeta}}\).  For the calibration used
here, \(\sqrt{A_{\zeta}}=\mu_c(0)/\nu_{\rm ref}\), with
\(\nu_{\rm ref}\simeq 8.45\), which gives $f_{\rm PBH,tot}^{(0)} \approx 1$.  Hence
\[
n_\star=\frac{s_\star\nu_{\rm ref}}{\mu_c(0)} .
\]
In the full mass-function calculation, the relevant branch is not necessarily
the one where the threshold is minimal. Instead, we define
\begin{equation}
n_\star^{\rm MF}
=
\underset{n}{\rm arg\,max}
\left[
\frac{d f_{\rm PBH,tot}^{\rm disp}}{dn}
\right],
\label{eq:nstar_mass_function_definition}
\end{equation}
with
\begin{equation}
\frac{d f_{\rm PBH,tot}^{\rm disp}}{dn}
=
\int d\ln M\,
\frac{M}{\rho_{\rm DM}}
P_{\rm sh}(n)
{\cal N}_{\rm pk}^{\rm BBKS}\!\left(\nu_M(n)\right)
\left|\frac{d\ln M}{d\nu}\right|^{-1}.
\label{eq:dfPBHtot_dn_definition}
\end{equation}

In addition to the dominant branch, it is useful to quantify how the required
power-spectrum amplitude changes once the shape dispersion is included. We define
$A_{\zeta,0}^{(f_{\rm PBH}=1)}$ as the amplitude that gives unit abundance in the no-dispersion
calculation,
\begin{equation}
f_{\rm PBH,tot}^{(0)}\!\left(A_{\zeta,0}^{(f_{\rm PBH}=1)}\right)=1,
\label{eq:A0_unit_abundance}
\end{equation}
and $A_{\zeta,\rm disp}^{(f_{\rm PBH}=1)}$ as the amplitude that gives unit abundance after
integrating over the coherent shape variable,
\begin{equation}
f_{\rm PBH,tot}^{\rm disp}
\!\left(
A_{\zeta,\rm disp}^{(f_{\rm PBH}=1)}
\right)
=1.
\label{eq:Adisp_unit_abundance}
\end{equation}
We then introduce the amplitude-retuning ratio
\begin{equation}
{\mathcal{Q}}_A
\equiv
\frac{A_{\zeta,\rm disp}^{(f_{\rm PBH}=1)}}{A_{\zeta,0}^{(f_{\rm PBH}=1)}}.
\label{eq:amplitude_retuning_ratio}
\end{equation}
If ${\mathcal{Q}}_A<1$, the inclusion of shape dispersion lowers the amplitude of the
primordial power spectrum required to obtain the same PBH abundance. If
${\mathcal{Q}}_A\simeq1$, the shape dispersion has little impact on the amplitude
normalization.

\begin{table*}[t]
\centering
\scriptsize
\setlength{\tabcolsep}{4pt}
\resizebox{\textwidth}{!}{%
\begin{tabular}{c c c c c c c c c}
\toprule
$\beta_{\rm NG}$
& $\mu_c(0)$
& $s_\star^{\rm MF}$
& $n_\star^{\rm MF}$
& $\nu_c(s_\star^{\rm MF})$
& $\beta_{\rm disp}/\beta_0$
& $f_{\rm PBH,tot}^{\rm disp}/f_{\rm PBH,tot}^{(0)}$
& $A_{\zeta,0}^{(f_{\rm PBH}=1)}$
& ${\mathcal{Q}}_A$
\\
\midrule
$0$  & $0.7957$  & $-0.122$ & $-1.29$ & $8.29$ & $1.90$ & $1.91$ & $8.88\times10^{-3}$ & $0.982$ \\
$1$  & $0.5685$ & $-0.039$ & $-0.58$ & $8.41$ & $1.21$ & $1.21$ & $4.55\times10^{-3}$ & $0.995$ \\
$2$  & $0.4181$ & $-0.014$ & $-0.29$ & $8.44$ & $1.05$ & $1.06$ & $2.47\times10^{-3}$ & $0.998$ \\
$3$  & $0.3167$ & $-0.003$ & $-0.09$ & $8.45$ & $1.01$ & $1.04$ & $1.42\times10^{-3}$ & $1.000$ \\
$-1$ & $1.144$ & $-0.428$ & $-3.16$ & $7.52$ & $11.5$ & $11.9$ & $1.83\times10^{-2}$ & $0.929$ \\
$-2$ & $1.688$ & $-1.172$ & $-5.87$ & $2.56$ & $1.82\times10^{5}$ & $2.78\times10^{5}$ & $3.97\times10^{-2}$ & $0.616$ \\
$-3$ & $2.548$ & $-1.082$ & $-3.59$ & $2.64$ & $3.48\times10^{8}$ & $3.41\times10^{8}$ & $9.10\times10^{-2}$ & $0.339$ \\
\bottomrule
\end{tabular}%
}
\caption{
Summary of the effective one-dimensional shape-dispersed PBH abundance and
mass-function calculation.  The columns \(s_\star^{\rm MF}\) and
\(n_\star^{\rm MF}=s_\star^{\rm MF}/\sqrt{A_{\zeta}}\) identify the coherent shape
branch that dominates the mass-integrated PBH abundance in the baseline
calibration.  The ratio \(\beta_{\rm disp}/\beta_0\) is the integrated BBKS
peak-abundance diagnostic, while
\(f_{\rm PBH,tot}^{\rm disp}/f_{\rm PBH,tot}^{(0)}\) is the corresponding ratio
obtained from the full PBH mass function.  The amplitude $A_{\zeta,0}^{(f_{\rm PBH}=1)}$ is
defined by the no-dispersion condition \(f_{\rm PBH,tot}^{(0)}=1\), and
\({\mathcal{Q}}_A\) measures the reduction of the
power-spectrum amplitude required to obtain the same PBH abundance after
including shape dispersion. }
\label{tab:shape_dispersion_mass_function_summary}
\end{table*}

Table~\ref{tab:shape_dispersion_mass_function_summary} shows that the
branch that dominates the abundance is not, in general, identical to the branch
where \(\mu_c(s)\) is minimal. This distinction is important because the abundance
is controlled by a competition between threshold reduction and Gaussian rarity in
shape space. For positive \(\beta_{\rm NG}\), the threshold variation is modest and
the dominant branch remains close to the reference profile, with
\(|n_\star^{\rm MF}|\lesssim 1\). The resulting enhancement over the no-dispersion
reference is therefore mild.

For negative \(\beta_{\rm NG}\), the threshold curve develops branches with much
smaller values of \(\mu_c(s)\), and the abundance can be dominated by coherent
shape deformations several standard deviations away from the mean. In particular, for \(\beta_{\rm NG}=-2\), the dominant mass-function
contribution comes from the branch with action-normalized shape coordinate
\(n_\star^{\rm MF}\simeq -5.9\). Although this represents a rare coherent profile deformation, the associated reduction in the collapse threshold lowers the required peak height to $\nu_c(s_\star) \simeq 2.56$. This reduction more than compensates for the Gaussian suppression of the shape probability, making this branch the statistically dominant contribution and producing a large enhancement in the integrated PBH abundance. The same qualitative mechanism operates for
\(\beta_{\rm NG}=-3\), where the dominant branch is around
\(n_\star^{\rm MF}\simeq -3.6\). 

The last columns of Table~\ref{tab:shape_dispersion_mass_function_summary}
show that the enhancement produced by shape dispersion can be reinterpreted
as a reduction of the power-spectrum amplitude required to obtain the same PBH
abundance. For positive \(\beta_{\rm NG}\), the shape-dispersed and no-dispersion
amplitude normalizations are very similar, \({\mathcal{Q}}_A\simeq1\), because the
dominant branch remains close to the reference profile. By contrast, for negative
\(\beta_{\rm NG}\), especially \(\beta_{\rm NG}=-2\) and \(\beta_{\rm NG}=-3\), the
threshold reduction along rare shape branches strongly enhances the abundance.
Consequently, the amplitude required to obtain \(f_{\rm PBH}=1\) after including
shape dispersion can be substantially smaller than the amplitude required in the
no-dispersion calculation. For \(\beta_{\rm NG}=-2\) the required amplitude is
reduced to \({\mathcal{Q}}_A\simeq0.62\) of the no-dispersion value, while for
\(\beta_{\rm NG}=-3\) it is reduced to \({\mathcal{Q}}_A\simeq0.34\). We stress that these ratios quantify the enhancement within a one-parameter family of profiles passing through the reference configuration: only the single coherent split direction is integrated, while all orthogonal residual modes are held at their mean. They should therefore be read as the effect of the dominant radial direction, not as a full marginalization over the residual shape space.

These results should be interpreted as a sensitivity test of the effective
shape-dispersed prescription. The retuning ratio \({\mathcal{Q}}_A\) quantifies how much
the inferred primordial power-spectrum amplitude changes once the one-dimensional
shape integration is included.

\begin{figure}[t]
    \centering
    \includegraphics[width=1.0\textwidth]{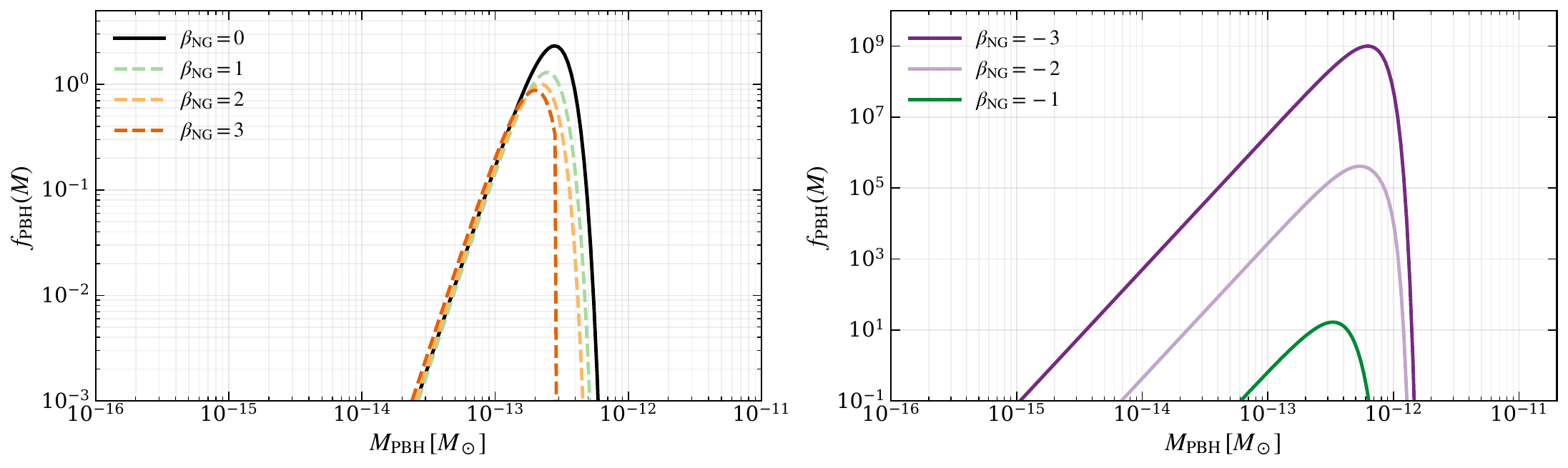}
    \caption{
    Shape-dispersed PBH mass functions for the different non-Gaussian branches, computed using no-dispersion calibration of the power-spectrum amplitude \(A_{\zeta}\) for $s=0$. Left panel: positive and vanishing non-Gaussianity, \(\beta_{\rm NG}=0,1,2,3\).
    Right panel: negative non-Gaussianity, \(\beta_{\rm NG}=-1,-2,-3\).}
    \label{fig:mass_functions_shape_dispersion}
\end{figure}

Figure~\ref{fig:mass_functions_shape_dispersion} shows the shape-dispersed PBH
mass functions obtained for the different non-Gaussian branches considered in
this work. The amplitude \(A_{\zeta}\) is fixed, for each value of \(\beta_{\rm NG}\), by the
no-dispersion calibration \(s=0\), so that the comparison isolates the effect of
integrating over the coherent shape degree of freedom. The left panel shows the
cases \(\beta_{\rm NG}\geq0\), for which the resulting mass functions remain close
to one another and the enhancement relative to the reference branch is modest. This reflects the fact that the threshold variation with \(s\) is not large enough
to overcome the Gaussian suppression of rare shape fluctuations. In this sense, these results confirm that neglecting residual profile dispersion
is a good approximation for the reference branch studied in Ref.~\cite{Escriva:2023uko}, in the
regime where the power spectrum is not very broad and shape dispersion remains
subdominant. They are also
consistent with Ref.~\cite{Escriva:2025ftp}, which found that, for a specific
relatively narrow power spectrum with an Ultra-Slow-Roll (USR) plateau ($\beta_{\rm NG}=3$), the PBH abundance in the adiabatic channel is
dominated by the mean-profile contribution.

The right panel shows the cases \(\beta_{\rm NG}<0\). For these models, the collapse threshold along rare coherent-shape branches can be substantially lower than for the reference branch. Despite the
Gaussian suppression of these rare deformations, the lower threshold
strongly enhances the BBKS peak abundance. As a result, the mass
functions can be enhanced by many orders of magnitude, especially for
\(\beta_{\rm NG}=-2\) and \(\beta_{\rm NG}=-3\). This behaviour illustrates the exponential sensitivity of
PBH production to the collapse threshold once shape dispersion is included. It
also indicates that, for models with negative non-Gaussianity, the effect of
profile dispersion may need to be included in order to obtain reliable abundance
estimates and identify the most representative curvature profiles. In addition,
for such cases the dominant spherical branches correspond to relatively small
effective peak heights, \(\nu_c(s_\star)\simeq 2.56\) and
\(\nu_c(s_\star)\simeq 2.64\), respectively. This suggests that these regimes may
also be sensitive to angular shape degrees of freedom, since the usual high-peak
near-spherical approximation becomes less restrictive at low peak height and
departures from spherical symmetry can become important. The present results
should therefore be viewed as identifying a region where radial shape dispersion
already has a large effect, and where extending the calculation to include BBKS
ellipticity, prolateness and higher multipoles may be relevant.

\subsection{Case B: finite-band scale-invariant spectrum with tunable bandwidth}
\label{sec:caseB}

In the previous example, we showed that, within the one-dimensional profile family considered, shape dispersion associated with a finite-width power spectrum can dominate the abundance in the presence of negative non-Gaussianity, while its impact remains mild in the Gaussian case and
is further suppressed for positive non-Gaussianity. We now consider a second analytically tractable example, where the dimensionless Gaussian curvature power spectrum is constant over a finite interval in wavenumber. This allows us to vary the spectral bandwidth, and hence the available radial-profile dispersion, at fixed total variance.
\begin{equation}
{\cal P}_{\zeta_G}(k)
=
{\cal P}_0\,\Theta(k-k_{-})\Theta(k_{+}-k).
\label{eq:flat_top_hat_spectrum}
\end{equation}
with $k_-<k_+$ and introduce the logarithmic width and the scale
\begin{equation}
L\equiv \ln\left({k_+\over k_-}\right),
\qquad
k_0\equiv \sqrt{k_-k_+},
\qquad
\Delta\equiv {L\over2}.
\label{eq:L_k0_Delta_flat}
\end{equation}
Thus
\begin{equation}
k_-=k_0e^{-\Delta},
\qquad
k_+=k_0e^{\Delta}.
\label{eq:kpm_Delta_flat}
\end{equation}
The total variance is
\begin{equation}
\sigma_0^2
=
\int d\ln k\,{\cal P}_{\zeta_G}(k)
={\cal P}_0L.
\label{eq:sigma0_flat}
\end{equation}
As in the previous example, we denote this variance by
\begin{equation}
A_{\zeta}\equiv \sigma_0^2,
\qquad
{\cal P}_0={A_{\zeta}\over L}={A_{\zeta}\over 2\Delta}.
\label{eq:A_P0_flat}
\end{equation}

For the flat finite-width spectrum, the spectral moments for \(j\geq 1\) are given by
\begin{equation}
\sigma_j^2
=
{\cal P}_0{k_+^{2j}-k_-^{2j}\over 2j}
=
A_{\zeta} k_0^{2j}{\sinh(2j\Delta)\over 2j\Delta}.
\label{eq:moments_flat_result}
\end{equation}
In particular,
\begin{equation}
\sigma_1^2
=
A_{\zeta} k_0^2{\sinh(2\Delta)\over 2\Delta},
\qquad
\sigma_2^2
=
A_{\zeta} k_0^4{\sinh(4\Delta)\over 4\Delta}.
\label{eq:sigma12_flat}
\end{equation}
The BBKS height-curvature correlation is therefore
\begin{equation}
\gamma_{\rm BBKS}
=
{\sigma_1^2\over \sigma_0\sigma_2}
=
{\sinh(2\Delta)/(2\Delta)
\over
\left[\sinh(4\Delta)/(4\Delta)\right]^{1/2}}.
\label{eq:gamma_flat}
\end{equation}
For any finite width, $\gamma_{\rm BBKS}<1$, while in the narrow-width limit $\Delta\to0$ one has $\gamma_{\rm BBKS}\to1$. The BBKS length scale is
\begin{equation}
R_\star
=
\sqrt3{\sigma_1\over\sigma_2}
=
{\sqrt3\over k_0}
\left[
{\sinh(2\Delta)/(2\Delta)
\over
\sinh(4\Delta)/(4\Delta)}
\right]^{1/2}
=
{\sqrt6\over \sqrt{k_-^2+k_+^2}}.
\label{eq:Rstar_flat}
\end{equation}

We retain only the monopole sector and fix the central amplitude,
\begin{equation}
\mathfrak{C}_{\mu}
\equiv
\left\{\zeta_G(0)=\mu\right\}.
\label{eq:height_constraint_flat}
\end{equation}
Equivalently, one may view the no-dispersion reference branch as the BBKS
spherical conditional profile with the independent curvature component set to
zero,

\begin{equation}
x_{\rm B}=x_*(\nu)=\gamma_{\rm BBKS}\nu,
\qquad
n_x=
\frac{x_{\rm B}-\gamma_{\rm BBKS}\nu}
{\sqrt{1-\gamma_{\rm BBKS}^2}}
=0.
\end{equation}

With this choice, the BBKS spherical profile reduces to the height-conditioned mean profile. This is only a representative choice; for a finite-width spectrum the curvature is in general an independent stochastic variable.

The spherical two-point function is
\begin{equation}
\xi(r)
=
\int_{k_-}^{k_+}d\ln k\,{\cal P}_0j_0(kr),
\qquad
j_0(z)={\sin z\over z}.
\label{eq:xi_flat_def}
\end{equation}
Define
\begin{equation}
F(z)\equiv {\rm Ci}(z)-{\sin z\over z},
\qquad
{dF\over dz}={\sin z\over z^2},
\label{eq:F_Ci_flat}
\end{equation}
where ${\rm Ci}(z)$ is the cosine integral function ${\rm Ci}(z)=-\int_{z}^{\infty}( \cos(t)/t)dt$. Then
\begin{equation}
\xi(r)
=
{\cal P}_0\left[F(k_+r)-F(k_-r)\right],
\label{eq:xi_flat_result}
\end{equation}
and the normalized correlator is
\begin{equation}
\Psi(r)
\equiv
{\xi(r)\over \sigma_0^2}
=
{F(k_+r)-F(k_-r)\over L}.
\label{eq:Psi_flat_r}
\end{equation}
Introducing
\begin{equation}
u\equiv k_0r,
\label{eq:u_flat_def}
\end{equation}
we can write
\begin{equation}
\Psi(u)
=
{F(ue^\Delta)-F(ue^{-\Delta})\over 2\Delta}.
\label{eq:Psi_flat_u}
\end{equation}
The fixed-height mean profile is
\begin{equation}
\bar\zeta_G(r)=\mu\Psi(r).
\label{eq:mean_profile_flat}
\end{equation}
In the narrow-width limit,
\begin{equation}
\Psi(u)\longrightarrow uF'(u)={\sin u\over u}=j_0(u),
\label{eq:Psi_flat_narrow}
\end{equation}
so the monochromatic spherical profile is recovered. In the action-normalized notation, the height direction is
$q_\nu(k)=1/\sigma_0$,
as in the example of section \ref{sec:caseA} and the corresponding radial function is

\begin{equation}
\mathcal{R}_\nu(r)
=
\int d\ln k\,{\cal P}_{\zeta_G}(k)q_\nu(k)j_0(kr)
=
\sigma_0 \,\Psi(r).
\label{eq:Rnu_flat}
\end{equation}
Thus
\begin{equation}
\bar\zeta_G(r)=\nu \mathcal{R}_\nu(r),
\qquad
\nu={\mu\over\sqrt{A_{\zeta}}}.
\label{eq:mean_Rnu_flat}
\end{equation}

We now introduce one coherent radial deformation at fixed central amplitude, using the same logic as in section \ref{sec:caseA}. The equal-variance split scale is determined by
\begin{equation}
\int_{k_-}^{\bar k}d\ln k\,{\cal P}_0
=
\int_{\bar k}^{k_+}d\ln k\,{\cal P}_0
={\sigma^2_0 \over2}.
\label{eq:equal_variance_flat}
\end{equation}
For the flat spectrum this gives
\begin{equation}
\bar k=\sqrt{k_-k_+}=k_0.
\label{eq:kbar_flat}
\end{equation}
We choose the sign convention
\begin{equation}
q_{0,{\rm split}}(k)
=
{1\over\sigma_{0}}\,{\rm sign}(k_0-k),
\label{eq:qsplit_flat}
\end{equation}
where reversing the sign is equivalent to replacing $n\to -n$. This mode satisfies Eqs.\eqref{eq:sph_gauss_qsplit_orth_height},\eqref{eq:sph_gauss_qsplit_norm}.

Therefore the coefficient $n$ multiplying this mode is a standard Gaussian variable, with action cost
\begin{equation}
\Delta W=n^2.
\label{eq:action_cost_flat}
\end{equation}
The split radial profile is
\begin{equation}
\mathcal{R}_{0,{\rm split}}(r)
=
{1\over \sigma_0}
\left[
\int_{k_-}^{k_0}d\ln k\,{\cal P}_0j_0(kr)
-
\int_{k_0}^{k_+}d\ln k\,{\cal P}_0j_0(kr)
\right].
\label{eq:Rsplit_flat_def}
\end{equation}
Using Eq.~\eqref{eq:F_Ci_flat}, this becomes
\begin{equation}
\mathcal{R}_{0,{\rm split}}(r)=\sigma_0\,\mathcal{G}_{\rm F}(r),
\label{eq:Rsplit_G_flat}
\end{equation}
where
\begin{equation}
\mathcal{G}_{\rm F}(r)
=
{2F(k_0r)-F(k_-r)-F(k_+r)\over L}.
\label{eq:Gth_flat_r}
\end{equation}
Equivalently,
\begin{equation}
\mathcal{G}_{\rm F}(u)
=
{2F(u)-F(ue^{-\Delta})-F(ue^\Delta)\over 2\Delta}.
\label{eq:Gth_flat_u}
\end{equation}
By construction,
\begin{equation}
\mathcal{G}_{\rm F}(0)=0,
\label{eq:Gth_zero_flat}
\end{equation}
so the split deformation preserves the central height. Near the origin,
\begin{equation}
\Psi(u)=1-{\sinh(2\Delta)\over 12\Delta}u^2
+{\sinh(4\Delta)\over 480\Delta}u^4+\cdots,
\label{eq:Psi_exp_flat}
\end{equation}
while
\begin{equation}
\mathcal{G}_{\rm F}(u)
={\cosh(2\Delta)-1\over 12\Delta}u^2
+{1-\cosh(4\Delta)\over 480\Delta}u^4+
\cdots.
\label{eq:Gth_exp_flat}
\end{equation}
Thus the split keeps the height fixed but changes the spherical curvature. This is intentional: the family is the finite-width analogue of the fixed-height radial-shape deformation used in the previous example, and it is not projected against the BBKS curvature direction. The final one-parameter family of Gaussian profiles is

\begin{equation}
\zeta_G(r,s)
=
\mu\Psi(r)+s\,\mathcal{G}_{\rm F}(r),
\label{eq:zeta_n_flat_explicit}
\end{equation}

with $s\equiv n\sigma_0$. The logarithmic radial derivatives needed for the compaction function are compact. Since
\begin{equation}
u{d\over du}F(au)=j_0(au),
\label{eq:log_der_F_flat}
\end{equation}
one has
\begin{equation}
D\Psi(u)
\equiv
r{d\Psi\over dr}
=
u{d\Psi\over du}
=
{j_0(ue^\Delta)-j_0(ue^{-\Delta})
\over 2\Delta},
\label{eq:DPsi_flat}
\end{equation}
and
\begin{equation}
D \mathcal{G}_{\rm F}(u)
\equiv
r{d\mathcal{G}_{\rm F}\over dr}
=
u{d\mathcal{G}_{\rm F}\over du}
=
{2j_0(u)-j_0(ue^{-\Delta})-j_0(ue^\Delta)
\over 2\Delta}.
\label{eq:DG_flat}
\end{equation}
Therefore,
\begin{equation}
r\zeta_G'(r,s)
=
\mu D\Psi(u)+sD \mathcal{G}_{\rm F}(u).
\label{eq:rzeta_prime_flat}
\end{equation}
For the Gaussian curvature profile, the leading-order linear compaction and the compaction function in the long-wavelength approximation are
\begin{equation}
\mathcal{C}_{\ell}(r,s)
=
-{4\over3}\left[\mu D\Psi(u)+s D \mathcal{G}_{\rm F}(u)\right],
\label{eq:Cl_flat}
\end{equation}
\begin{equation}
\mathcal{C}(r,s)
=
-{2\over3}\left[\mu D \Psi(u)+s D \mathcal{G}_{\rm F}(u)\right]
\left[2+\mu D \Psi(u)+sD \mathcal{G}_{\rm F}(u)\right].
\label{eq:C_flat}
\end{equation}
The compaction-peak position $r_m$ is determined numerically from these expressions. Unlike the Gaussian template of Case A in Sec.~\ref{sec:caseA}, the finite top-hat spectrum generally produces a more oscillatory compensated tail, and the peak location depends on the width $\Delta$.

This finite-width top-hat spectrum interpolates between the monochromatic case and a genuinely finite-band stochastic profile. In the limit $\Delta\to0$, the normalized correlator tends to $j_0(k_0r)$, the split profile tends to zero, and $\gamma_{\rm BBKS}\to1$. Thus the independent spherical radial dispersion disappears, as expected for a monochromatic spectrum. For finite $\Delta$, the field contains a continuum of wavelengths between $k_-$ and $k_+$; the mean profile is the average of $j_0(kr)$ over this band, while $\mathcal{G}_{\rm F}$ measures the coherent difference between the long-wavelength half and the short-wavelength half of the spectrum. Increasing $\Delta$ therefore increases the possibility of radial-shape dispersion at fixed central amplitude. Physically, the increasing profile freedom for broader spectra is associated
with the larger range of wavelengths contributing to the perturbation.
Long-wavelength modes mainly modify the extended environment and compensation
region, whereas shorter-wavelength modes affect the structure closer to the
central peak. The resulting dispersion therefore reflects the relative
contribution of fluctuations across different scales. Since the mode is not projected against the BBKS curvature direction,
this should be interpreted, as in Case A of Sec.~\ref{sec:caseA}, as an
effective fixed-height one-dimensional shape family rather than as a fully
conditional fixed-\((\nu,x_{\rm B})\) construction.

\begin{figure}[!htbp]
    \centering
    \includegraphics[width=1.0\textwidth]{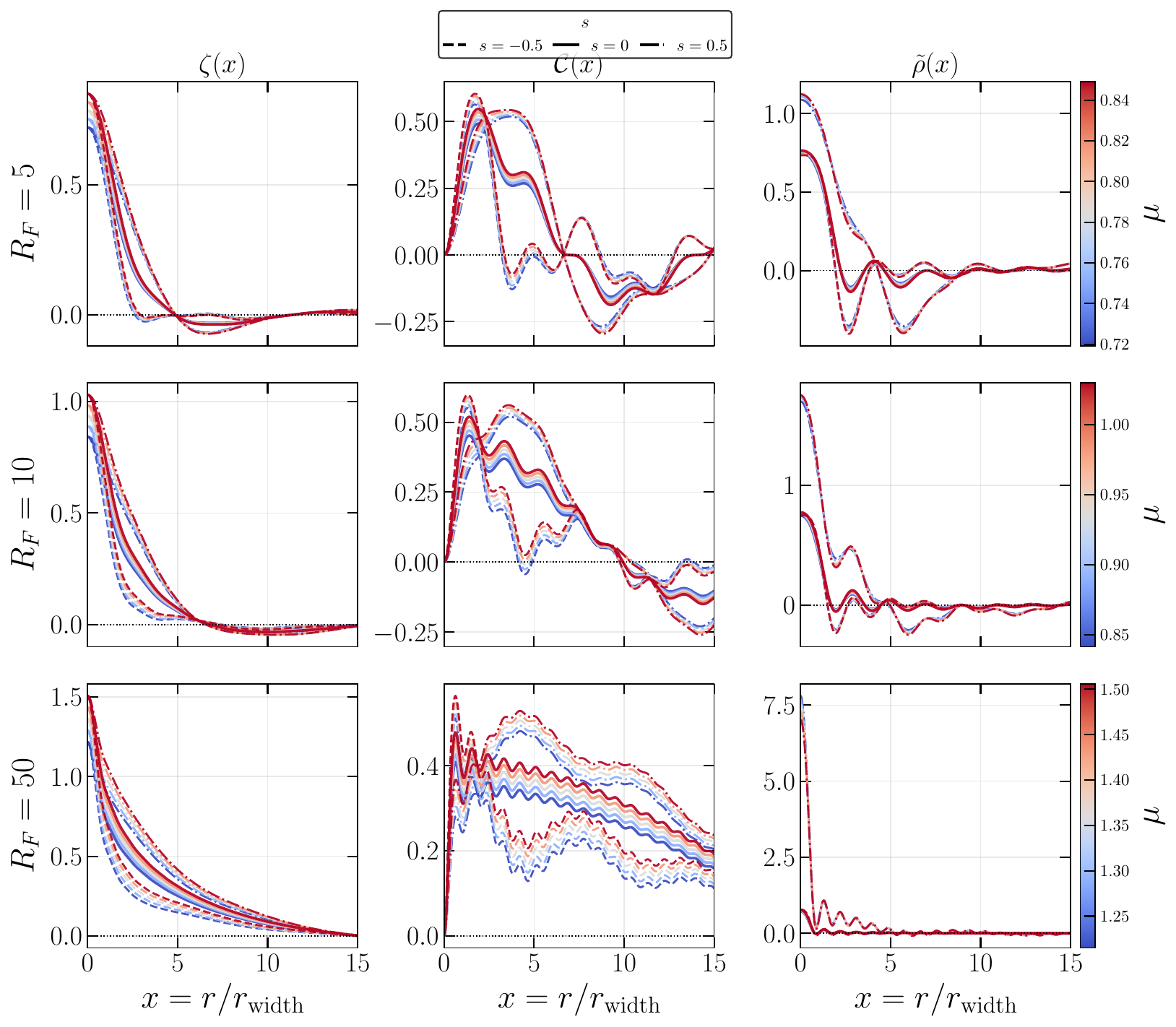}
\caption{Representative initial curvature, compaction, and density profiles for the finite top-hat spectrum using the original, unprojected split deformation.  The rows correspond to the bandwidth ratios \(R_F=5,10,50\). 
The columns show, from left to right, the curvature profile \(\zeta(x)\), the compaction function \(\mathcal{C}(x)\), and the long-wavelength density perturbation \(\tilde{\rho}(x)\), where \(x=r/r_{\rm width}\). 
For each value of \(R_F\), the line style denotes the split amplitude \(s=-0.5,0,0.5\), while the colour scale denotes the Gaussian amplitude \(\mu\).} \label{fig:flat_tophat_unprojected_critical_profiles}
\end{figure}

Figure~\ref{fig:flat_tophat_unprojected_critical_profiles} illustrates the
profiles obtained with the original unprojected split mode for different
values of \(\mu\) and \(s\). This construction provides a controlled
one-parameter deformation of the finite-band profile at fixed central
amplitude. Since the split mode has a nonzero quadratic term in its expansion
around the origin, it also changes the spherical curvature of the central
peak. The threshold variation shown in the figure therefore reflects changes
in both the extended radial structure of the profile and the local curvature
variable. We define $R_F \equiv k_{+}/k_{-}$ and $r_{\rm width}=k^{-1}_0$ for this case. For moderate bandwidths, \(R_F=5\) and \(R_F=10\), the critical profiles show compensated oscillatory tails characteristic of a sharp finite band in Fourier
space. These oscillations affect the compaction function and may
produce several competing local maxima. For the broader case \(R_F=50\), the
profiles become more extended and the compaction function develops a smoother
but radially broad structure. As we can see in the left panel of Fig.~\ref{fig:ham_constraint_example}, these oscillations in the compaction function become smoothed out once the fluctuation re-enters the cosmological horizon.

\begin{figure}[!htbp]
    \centering
    \includegraphics[width=1.0\textwidth]{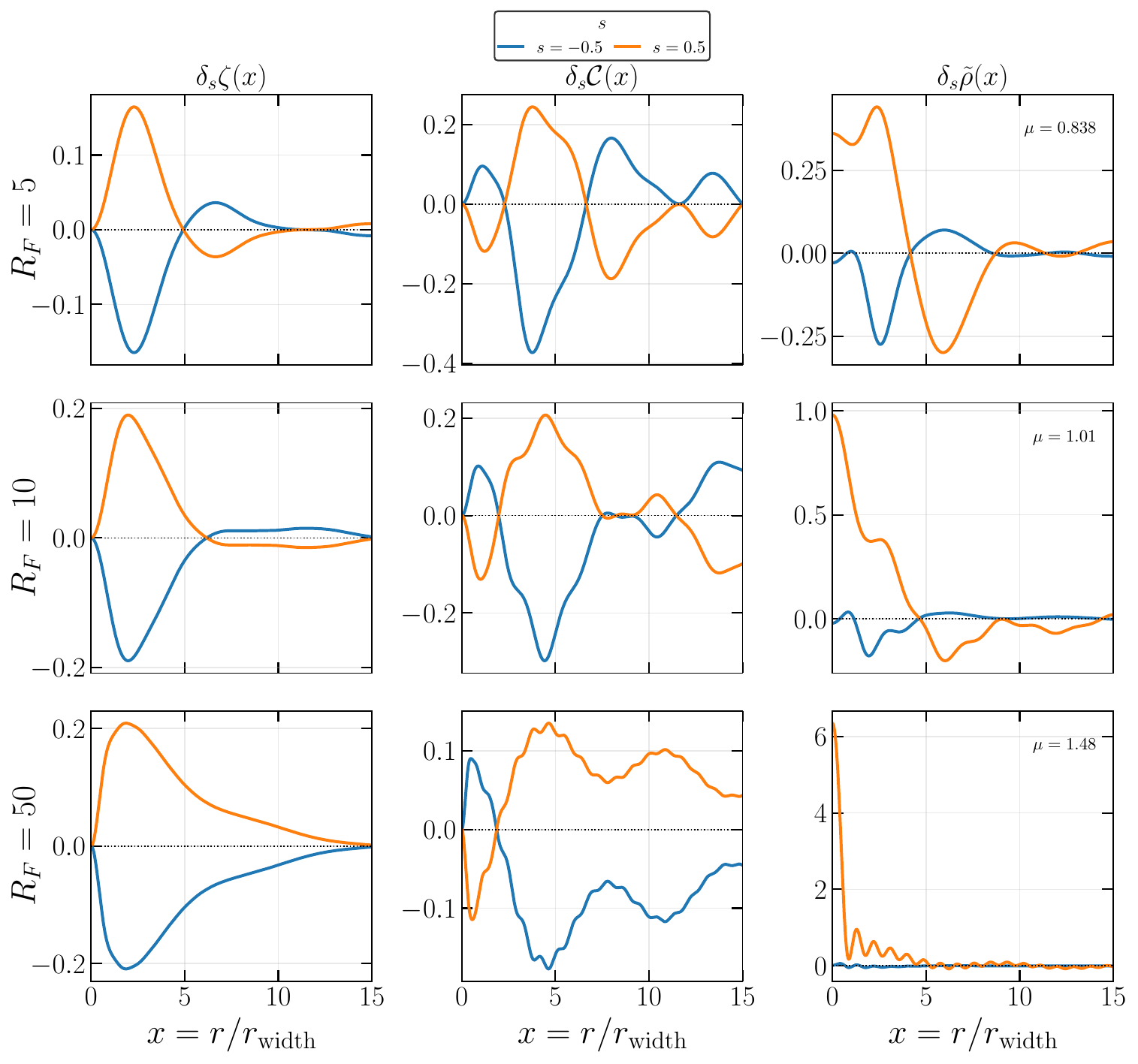}
\caption{Effect of the split deformation on the initial profiles for the finite top-hat spectrum. 
Each row corresponds to a different bandwidth ratio \(R_F\). 
The columns show, from left to right, the residual curvature perturbation
\(\delta_s\zeta(x)\), the residual compaction function
\(\delta_s\mathcal{C}(x)\), and the residual long-wavelength density perturbation
\(\delta_s\tilde{\rho}(x)\), with \(x=r/r_{\rm width}\). The value of \(\mu_c(0)\) used in each row is indicated in the corresponding panels. 
The blue and orange curves correspond to \(s=-0.5\) and \(s=0.5\), respectively.}
\label{fig:flat_tophat_unprojected_profile_differences}
\end{figure}

To make the role of the split deformation more transparent,
Fig.~\ref{fig:flat_tophat_unprojected_profile_differences} shows the difference
between the deformed profiles and the corresponding \(s=0\) profile at fixed
amplitude \(\mu=\mu_c(0)\). For \(R_F=5\) and \(R_F=10\), the deformation produces oscillatory compensated
features in \(\delta_s\zeta\), \(\delta_s{\cal C}\) and $\delta_s \tilde{\rho}$.  The response of the
compaction function is more structured than that of the curvature profile,
because \({\cal C}\) depends on the logarithmic radial derivative of
\(\zeta\), and therefore emphasizes changes in the radial distribution of the
overdensity.  For \(R_F=50\), the deformation becomes broader in real space: the
curvature difference extends over a larger radial interval and the compaction
response develops a slowly varying component with small superimposed
oscillations. Increasing the width of the spectrum enhances the dispersion effect on
\(\delta_s\zeta(x)\).

\begin{figure}[!htbp]
    \centering
    \includegraphics[width=1.0\textwidth]{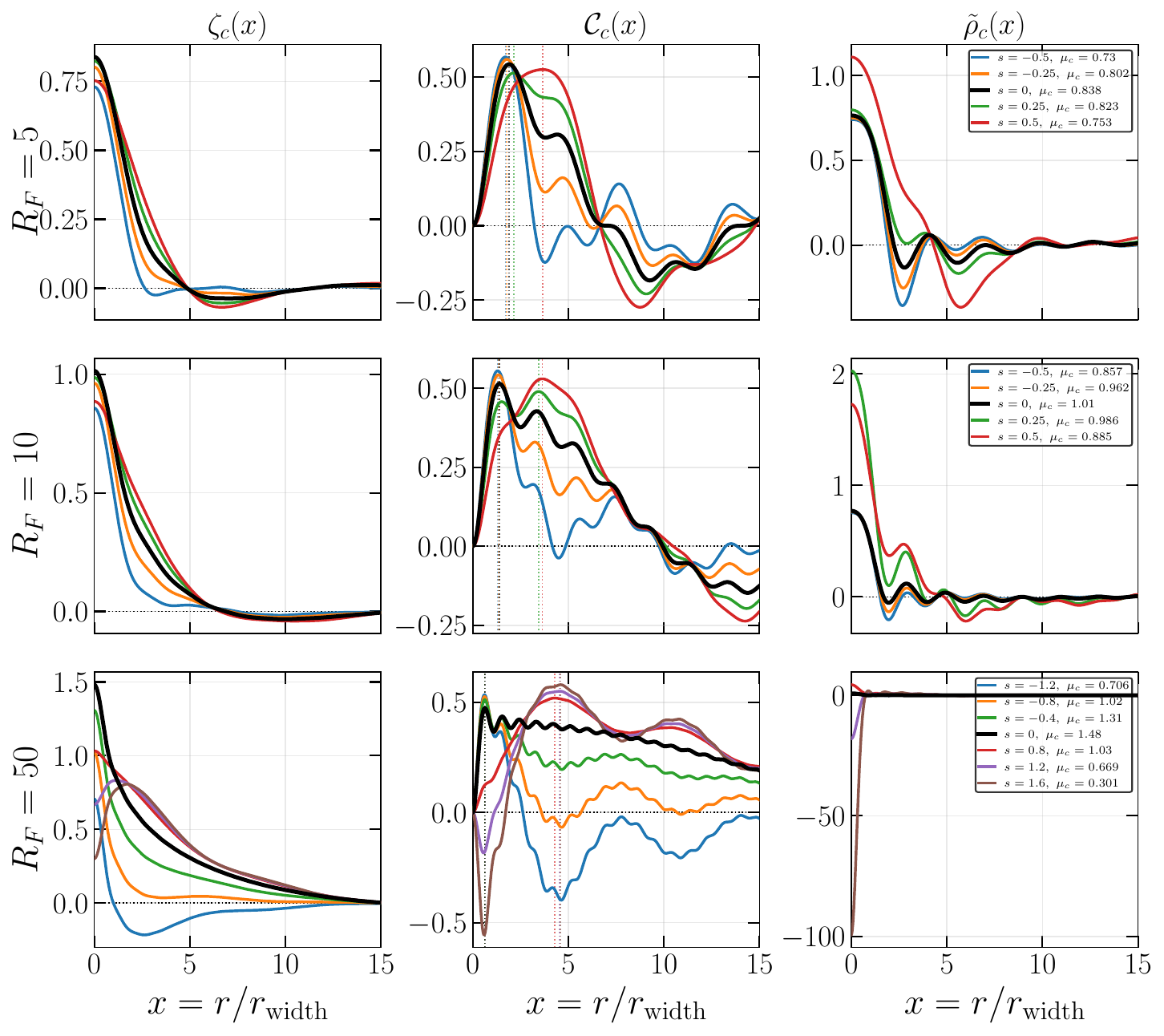}
\caption{Critical initial profiles for the finite top-hat spectrum using the original, unprojected split deformation. Each row corresponds to a different bandwidth ratio \(R_F\), while the columns show, from left to right, the critical curvature perturbation \(\zeta_c(x)\), the compaction function \(\mathcal{C}_c(x)\), and the long-wavelength density perturbation \(\tilde{\rho}_c(x)\), with \(x=r/r_{\rm width}\). 
The coloured curves correspond to different values of the split amplitude \(s\), and the legends indicate the associated threshold amplitudes \(\mu_c(s)\). 
The vertical dotted lines in the compaction panels mark the location of the dominant compaction maximum for each critical profile.} \label{fig:flat_tophat_unprojected_critical_profiles_extended}
\end{figure}

Figure~\ref{fig:flat_tophat_unprojected_critical_profiles_extended} shows the
critical profiles obtained using $\mu_c(s)$. For \(R_F=5\), the deformation mainly changes the width of the central profile and
the relative height of the first compensated oscillation.  The corresponding
compaction profiles show that the dominant compaction maximum can move
significantly with \(s\). For \(R_F=10\), the profiles become more extended and the compaction
develops several competing local maxima associated with the oscillatory
real-space tail. For \(R_F=50\), the real-space profiles are much broader, and the
compaction functions show a radially extended structure.  In this case, large
positive values of \(s\) may displace the dominant curvature maximum away from
the origin. For $R_F = 50$ and $s=1.2,1.6$, we observe the formation of a large underdense region and a negative mass excess near the origin. However, in these cases, PBH formation is sourced by the positive surrounding mass excess.

\begin{figure}[!htbp]
    \centering
    \includegraphics[width=1.0\textwidth]{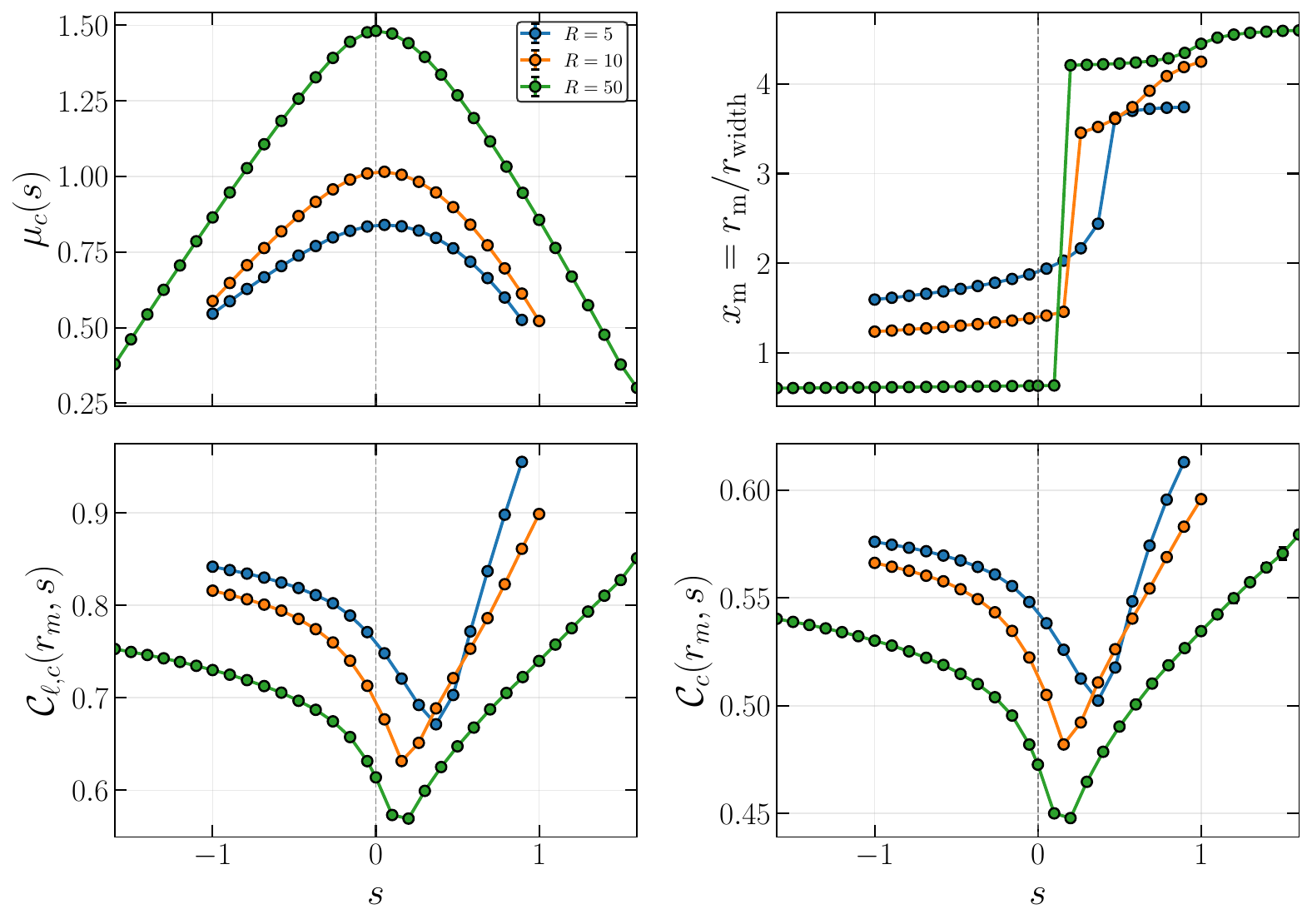}
\caption{
Threshold diagnostics. The different curves correspond to bandwidth
ratios \(R_F=5,10,50\).  The upper-left panel shows the collapse
threshold \(\mu_c(s)\), while the upper-right panel shows the position
\(x_m=r_m/r_{\rm width}\) of the dominant maximum of the linear
compaction function. The lower panels show the corresponding peak values of \({\cal C}_{\ell,c}(r_m,s)\)
and \({\cal C}_{c}(r_m,s)\), respectively. The vertical dashed line marks the reference branch \(s=0\).  }
\label{fig:flat_tophat_unprojected_threshold_diagnostics}
\end{figure}

Figure~\ref{fig:flat_tophat_unprojected_threshold_diagnostics} summarizes the
collapse thresholds obtained for the unprojected finite-band split family.  The
threshold \(\mu_c(s)\) is not monotonic: for all bandwidths shown here it reaches
its largest values close to the reference configuration \(s=0\), while both
negative and sufficiently positive deformations lower the threshold. This indicates that, within this one-parameter family, the reference
configuration requires a comparatively larger central amplitude to collapse,
whereas coherent shape deformations can lower the collapse threshold.

The behaviour of \(x_m\) shows that the effect of the deformation is
not simply a small perturbation of the same compaction maximum. In several cases, and most
clearly for \(R_F=50\), the dominant maximum of \({\cal C}_\ell\) switches from an
inner peak to an outer peak as \(s\) is varied.  This produces the sharp jumps in
\(x_{\rm m}\).  The lower panels show that this branch switching is accompanied
by a pronounced variation of both the linear and nonlinear peak compactions.
The peak value \({\cal C}_{c}(r_m)\) remains less variable than
\({\cal C}_{\ell,c}(r_m)\), because the nonlinear relation partially
compresses the range of large linear-compaction values.

\begin{figure}[!htbp]
    \centering
    \includegraphics[width=1.0\textwidth]{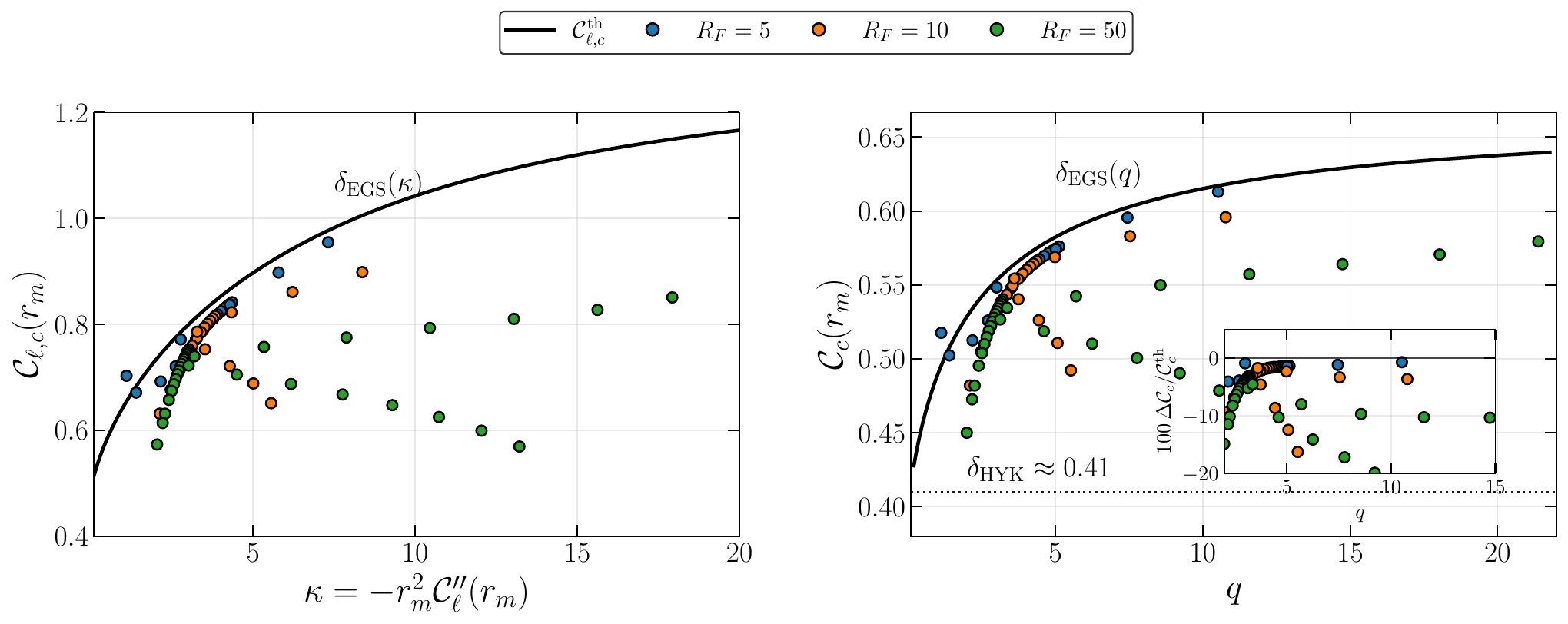}
\caption{
Comparison between the numerical collapse thresholds of the finite top-hat
profiles and the analytic threshold estimates.  The left panel shows the
critical value of the linear compaction peak,
\({\cal C}_{\ell,c}(r_m)\), as a function of the curvature-shape parameter
\(\kappa\), while the right panel shows the corresponding compaction function threshold, \({\cal C}_{c}(r_m)\), as a function of the shape parameter \(q\).
The black curves denote the analytic EGS predictions, and the coloured points
show the numerical thresholds for the finite-band spectra with
\(R_F=5,10,50\). The horizontal dotted line marks the
approximately universal HYK estimate \(\delta_{\rm HYK}\simeq 0.41\). The inset
shows the relative deviation of the numerical compaction function threshold
from the EGS prediction.}
\label{fig:flat_tophat_unprojected_EGS_comparison}
\end{figure}

Figure~\ref{fig:flat_tophat_unprojected_EGS_comparison} compares the numerical thresholds obtained for the finite top-hat family with the analytic EGS \cite{Escriva:2019nsa} and HYK \cite{Harada:2013epa} threshold estimates.  The comparison is shown both in terms of the linear compaction function threshold, \({\cal C}_{\ell,c}\), as a function of \(\kappa\), and in terms of the compaction function threshold, \({\cal C}_{c}\), as a function of \(q\).  For the narrower spectra, \(R_F=5\) and \(R_F=10\), the numerical points approximately follow the analytic trend.  

The agreement becomes poorer for the broadest case, \(R_F=50\).  In this regime the finite-band profiles develop extended compensated tails and several competing maxima of the compaction function.  As a result, the collapse threshold is no longer controlled only by the local shape around a single dominant peak, and the mapping to the EGS one-parameter estimate becomes less accurate.  The relative deviation shown in the inset confirms that the largest differences occur for large \(q\), where the dominant compaction function scale is often associated with an outer radial maximum. This behaviour is consistent with the branch switching seen in
Fig.~\ref{fig:flat_tophat_unprojected_threshold_diagnostics}.  A similar effect
was already found in Ref.~\cite{Escriva:2023qnq}, where numerical simulations
showed that secondary peaks in the compaction function, originating from
overlapping fluctuations (see
also Ref.~\cite{Raatikainen:2023bzk,Raatikainen:2025gpd} in the context of stochastic
inflation), can significantly modify the collapse criterion. In particular, the surrounding region with positive mass excess lowers the formation threshold compared with the analytical estimate $\delta_{\rm EGS}$.

\begin{figure}[!htbp]
    \centering
    \includegraphics[width=1.0\textwidth]{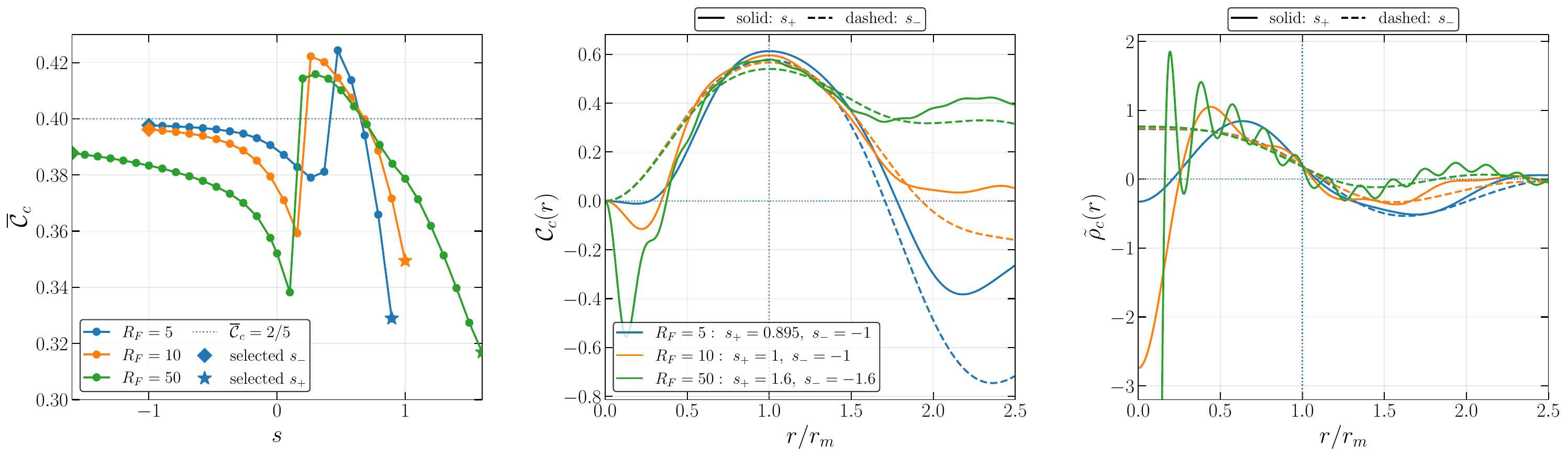}
\caption{Left panel shows the volume-averaged critical compaction
\(\overline{\mathcal C}_c\) as a function of the shape parameter \(s\) for
\(R_F=5,10,50\). The horizontal dotted line marks the reference value
\(\overline{\mathcal C}_c=2/5\), while diamonds and stars indicate the
negative- and positive-\(s\) configurations selected for the profile comparison.
The middle and right panels show the corresponding critical compaction
function \(\mathcal C_c(r)\) and density perturbation
\(\tilde{\rho}_c(r)\), respectively. Solid lines correspond to the selected
positive-\(s\) configurations and dashed lines to the negative-\(s\)
configurations. The radial coordinate is normalized by the corresponding
compaction scale \(r_m\), indicated by the vertical dotted line.}
\label{fig:flat_tophat_three_panel_averaged}
\end{figure}

Figure~\ref{fig:flat_tophat_three_panel_averaged} shows that the averaged critical compaction
\(\overline{\mathcal{C}}_c\) remains close to the reference value
\(\overline{\mathcal{C}}_c \simeq 2/5\) over a significant part of the parameter space,
especially along the negative-\(s\) branches characterized by an overdense region. For sufficiently large positive \(s\),
however, noticeable deviations appear, with abrupt jumps associated with changes in
the dominant compaction peak and with a central underdense region. The middle and right panels show that these deviations
are associated with more structured critical profiles, including pronounced central
underdensities and compensated outer regions.

\begin{figure}[!htbp]
    \centering
    \includegraphics[width=1.0\textwidth]{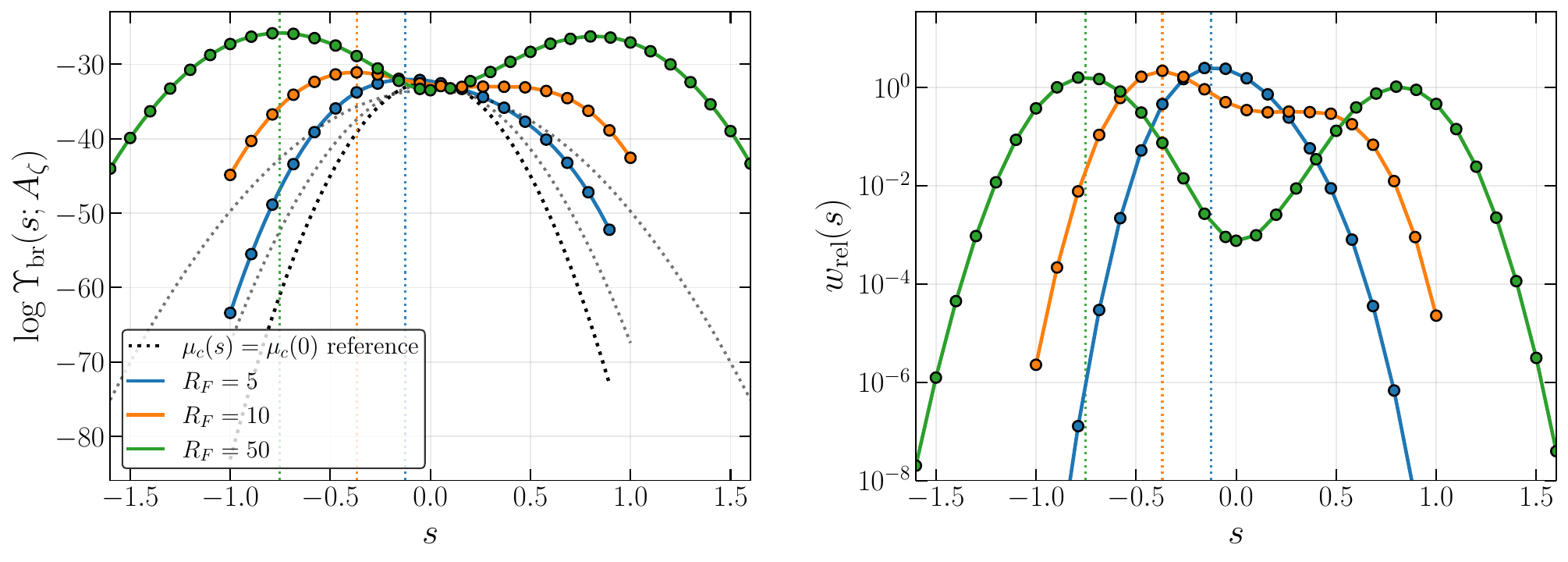}
\caption{Abundance weighting of the finite top-hat split deformations in the original
unprojected construction.  The left panel shows the logarithmic weight
\(\log \Upsilon_{\rm br}(s;A_{\zeta})\) as a function of the split amplitude \(s\), for
bandwidths \(R_F=5,10,50\).  The coloured solid curves include
the numerically determined threshold dependence \(\mu_c(s)\), while the dotted
curves show the corresponding reference weights obtained by keeping the
threshold fixed at its reference value, \(\mu_c(s)=\mu_c(0)\). The vertical
dotted lines mark the value of \(s\) giving the dominant contribution for each
bandwidth. The right panel shows the same information as normalized relative contributions $w_{\rm rel}(s)$, displayed on a logarithmic scale.  }
\label{fig:flat_tophat_unprojected_abundance_weight}
\end{figure}

Figure~\ref{fig:flat_tophat_unprojected_abundance_weight} illustrates how the
shape-dependent threshold modifies the statistical weight of the finite-band
profiles. If the threshold were independent of the split amplitude, the only
effect of changing \(s\) would be the Gaussian action penalty associated with
the coherent deformation. This is represented by the dotted reference curves in
the left panel. Once the numerical dependence \(\mu_c(s)\) is included, the
weight can be significantly enhanced in regions where the collapse threshold is
lower. The dominant contribution therefore need not occur at \(s=0\), even
though \(s=0\) is the most probable shape before imposing the collapse
condition. For \(R_F=5\) and \(R_F=10\), the preferred contribution is shifted moderately
towards negative \(s\), where the threshold is reduced relative to the reference
profile. For \(R_F=50\), the effect is much stronger: the broader real-space
profiles have a more pronounced threshold variation, and the resulting weight
is spread over a wider range of \(s\), with two peaks.  The right panel makes this more explicit
by showing the normalized contribution of each sampled deformation. The broad-band case displays more than one relevant contribution, consistent
with the branch structure and compaction-peak switching observed in the threshold diagnostics. In contrast to Case A, we do not compute the PBH mass function for the
finite-band example. Here our aim is to isolate how the integrated PBH
abundance changes with the available radial-profile dispersion as the
spectral bandwidth is varied. For a broad spectrum, a mass-function
calculation would additionally require tracking the characteristic
formation scale of each configuration and its mapping to the
corresponding horizon mass. This becomes non-trivial when the position
of the compaction maximum varies with the profile or different radial
maxima exchange dominance. We therefore restrict the present analysis
to the abundance ratio $\beta_{\rm disp}/\beta_0$.

Table~\ref{tab:bbks_integrated_peak_abundance_unprojected} quantifies the
effect of integrating over the coherent finite-band shape deformation in the
BBKS peak abundance. The quantity shown is the abundance ratio between the shape-dispersed estimate and the reference no-dispersion result.  The reference
amplitude is fixed by imposing the same peak height
\(\nu_{\rm ref} \approx 8.45\) for the \(s=0\) threshold in each bandwidth. The integration over \(n\) includes both the Gaussian probability cost of the deformation and the change in the cumulative BBKS peak abundance induced by the shape-dependent threshold \(\mu_c(s)\).

The enhancement of the abundance of peaks grows rapidly with the bandwidth.  For \(R_F=5\), the effect is moderate, giving roughly a factor of \(2\) increase relative to the
no-dispersion estimate.  For \(R_F=10\), the enhancement rises to about a factor of \(8\), indicating that the threshold reduction along the preferred deformation is already large enough to compensate a several-sigma Gaussian
shape cost.  The strongest effect occurs for \(R_F=50\), where the integrated
BBKS peak abundance is enhanced by nearly three orders of magnitude,
\(\beta_{\rm disp}/\beta_0\simeq 3.0\times10^3\).  In this broad-band case the dominant contribution comes from a rare coherent deformation,
\(n_\star\simeq -4.5\), but the corresponding reduction of the collapse
threshold produces a much larger BBKS peak abundance. Therefore, as the width of the spectrum increases, we find that the effect of
dispersion becomes progressively larger. This shows that, for very broad spectra,
the dispersion effect can be relevant and may lead to abundance differences of
many orders of magnitude compared with the no-dispersion reference estimate.

These results illustrate the central point of the shape-dispersion calculation:
the statistically dominant contribution is not necessarily the reference profile \(s=0\), nor simply the profile with the lowest threshold.  It is selected by a
competition between the Gaussian cost of realizing the coherent deformation and
the exponential gain associated with lowering the collapse threshold.  Because
the profiles used here come from the original unprojected split family, the
variation with \(s\) includes both a genuine finite-band radial-shape effect and
a change in the BBKS spherical curvature.  A stricter fixed-\((\nu,x_{\rm B})\)
interpretation would require repeating the calculation with the projected
\(q_x\)-orthogonal split mode or integrating explicitly over the curvature
variable, which is what we will briefly show in the next section.

\begin{table}[t]
\centering
\begin{tabular}{c c c c c c c}
\hline\hline
\(R_F\)
& \(\gamma_{\rm BBKS}\)
& \(\mu_c(0)\)
& \(s_\star\)
& \(n_\star\)
& $\nu_c(s_{\star})$
& \(\beta_{\rm disp}/\beta_0\)
\\
\hline
\(5\)  & \(0.757\) & \(0.839\) & \(-0.158\) & \(-1.59\) & $8.27$ & \(2.00\) \\
\(10\) & \(0.652\) & \(1.01\)  & \(-0.369\) & \(-3.07\) & $7.63$ & \(8.06\) \\
\(50\) & \(0.505\) & \(1.48\)  & \(-0.790\) & \(-4.50\) & $5.86$ & \(2.98\times 10^{3}\) \\
\hline\hline
\end{tabular}
\caption{
Integrated BBKS peak-abundance ratio for the finite top-hat spectrum using the
original unprojected split deformation.  For each bandwidth
\(R_F\), the variance amplitude is fixed by the reference
condition
\(\nu_{\rm ref} \approx 8.45\). The abundance ratio is computed following Eq.~\eqref{eq:abundance_peak_ratio}.}
\label{tab:bbks_integrated_peak_abundance_unprojected}
\end{table}

\subsubsection{Residual split mode at fixed BBKS curvature}
\label{sec:subcaseB22}

The split mode introduced above preserves the central height, but it changes the
spherical curvature of the peak. For a stricter peak-conditioned construction we
can instead remove the component of the split mode parallel to the BBKS curvature
direction. For the finite top-hat spectrum we define
\begin{equation}
 S_1\equiv {\sinh(2\Delta)\over 2\Delta},
 \qquad
 S_2\equiv \left[{\sinh(4\Delta)\over 4\Delta}\right]^{1/2},
 \qquad
 \gamma_{\rm BBKS}={S_1\over S_2} .
\end{equation}
The BBKS curvature direction orthogonal to the height direction is
\begin{equation}
 q_x(k)=
 {k^2/\sigma_2-\gamma_{\rm BBKS}/\sigma_0
 \over \sqrt{1-\gamma_{\rm BBKS}^2}} .
\end{equation}
The equal-variance top-hat split mode is given by Eq.~\eqref{eq:qsplit_flat}. Its overlap with the curvature direction is
\begin{equation}
\rho_{\rm split,x}
\equiv
(q_{0,\rm split},q_x)_{\mathcal P}
 =-{\cosh(2\Delta)-1
 \over 2\Delta\,S_2\sqrt{1-\gamma_{\rm BBKS}^2}} .
\end{equation}
The projected residual split mode is therefore
\begin{equation}
q_{0,\rm split}^{\perp}(k)
=
\frac{
q_{0,\rm split}(k)-\rho_{\rm split,x}q_x(k)
}{
\sqrt{1-\rho_{\rm split,x}^2}
}.
\end{equation}
By construction,
\begin{equation}
\inner{q_{0,\rm split}^{\perp}}{q_\nu}=0,
\qquad
\inner{q_{0,\rm split}^{\perp}}{q_x}=0,
\qquad
\inner{q_{0,\rm split}^{\perp}}{q_{0,\rm split}^{\perp}}=1.
\end{equation}
The corresponding real-space profile can be written as
\begin{equation}
 \zeta_G(r,s)=\mu\Psi(u)+s \mathcal{G}_{\rm F}^{\perp}(u),
 \qquad
 u=k_0r,
\end{equation}
where
\begin{equation}
\mathcal G_{\rm F}^{\perp}(u)
=
\frac{
\mathcal G_{\rm F}(u)-\rho_{\rm split,x}X(u)
}{
\sqrt{1-\rho_{\rm split,x}^2}
}.
\end{equation}
Here \(\mathcal{G}_{\rm F}(u)\) is the unprojected top-hat split profile
defined above, and
\begin{equation}
 X(u)= {K_2^{\rm F}(u)/S_2-\gamma_{\rm BBKS}\Psi(u)
 \over \sqrt{1-\gamma_{\rm BBKS}^2}},
 \qquad
 K_2^{\rm F}(u)={\cos(ue^{-\Delta})-\cos(ue^{\Delta})
 \over 2\Delta\,u^2},
\end{equation}
understood with the regular limit \(K_2^{\rm F}(0)=S_1\).  This projected
deformation satisfies
\begin{equation}
 \mathcal{G}_{\rm F}^{\perp}(0)=0,
 \qquad
 \left. {d^2 \mathcal{G}_{\rm F}^{\perp}\over du^2}\right|_{u=0}=0,
\end{equation}
so it leaves both the central height and the spherical BBKS curvature unchanged.
The origin therefore remains a local maximum of \(\zeta_G\) whenever the fixed
BBKS curvature is positive.  Notice, however, that this local peak condition does
not guarantee that the origin is the global maximum of the full radial profile;
large coherent residual modes can still generate a larger outer peak.

\begin{figure}[!htbp]
    \centering
    \includegraphics[width=1.0\textwidth]{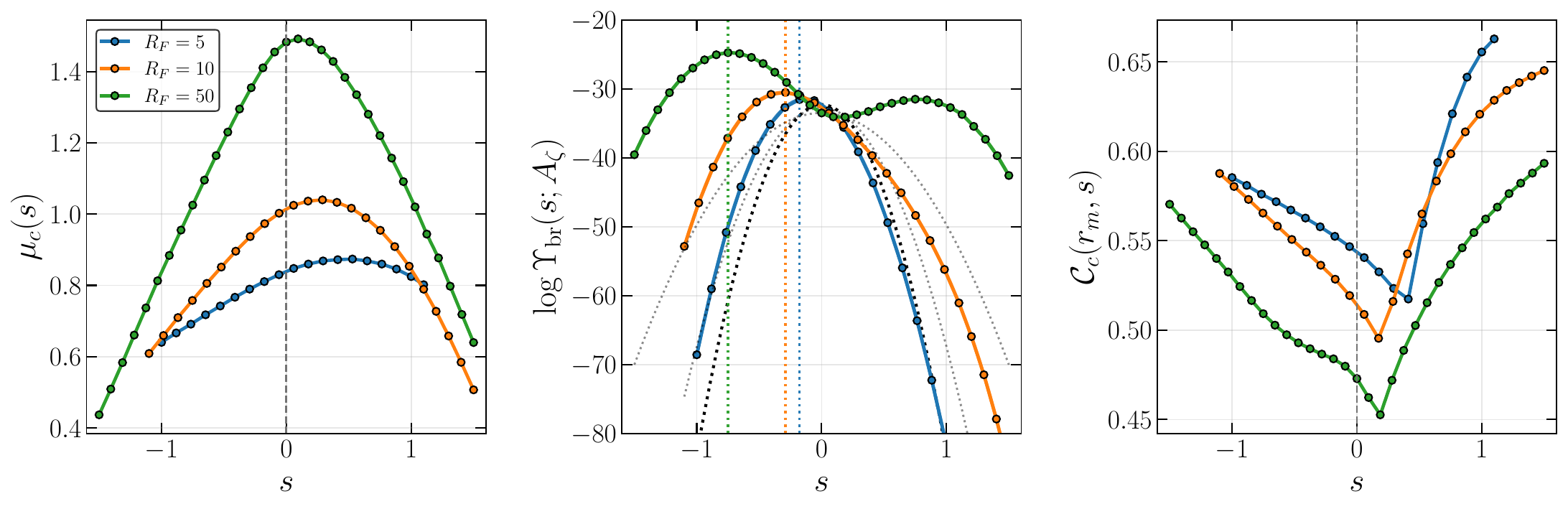}
\caption{Threshold and abundance-weight diagnostics for the projected \(q_x\)-orthogonal finite top-hat split deformation. 
The panels show, from left to right, the threshold amplitude \(\mu_c(s)\), the abundance-weight diagnostic \(\log \Upsilon_{\rm br}(s;A_\zeta)\), and the compaction function peak \(\mathcal{C}_c(r_m,s)\). 
Colours denote the bandwidth ratios \(R_F=5,10,50\). 
The dotted curves in the middle panel correspond to the fixed-threshold reference \(\mu_c(s)=\mu_c(0)\), and the vertical dotted lines mark the dominant branch for each bandwidth.}
\label{fig:flat_tophat_projected_res}
\end{figure}

With this replacement the logarithmic radial derivative entering the compaction
functions becomes
\begin{equation}
 r\zeta_G'(r,s)=\mu D\Psi(u)+sD \mathcal{G}_{\rm F}^{\perp}(u),
 \qquad
 D\equiv u{d\over du},
\end{equation}
and hence
\begin{equation}
 {\cal C}_\ell(r,s)
 =
 -{4\over 3}
 \left[\mu D\Psi(u)+sD \mathcal{G}_{\rm F}^{\perp}(u)\right],
\end{equation}
\begin{equation}
 {\cal C}(r,s)
 =
 -{2\over 3}Y(r,s)\,[2+Y(r,s)],
 \qquad
 Y(r,s)=\mu D\Psi(u)+sD \mathcal{G}_{\rm F}^{\perp}(u).
\end{equation}
Using the same procedure as in the previous case, we perform simulations to determine the threshold. The results are shown in Fig.~\ref{fig:flat_tophat_projected_res} and exhibit qualitatively similar behaviour to that observed in Figs.~\ref{fig:flat_tophat_unprojected_threshold_diagnostics} and  \ref{fig:flat_tophat_unprojected_abundance_weight}.

Table~\ref{tab:bbks_integrated_peak_abundance_projected_qx} summarizes the numerical results. They are consistent with the previous unprojected construction, where the profile was built without imposing the $q_x$-orthogonality condition. After the projection, the same qualitative dependence on the bandwidth is observed: broader spectra lead to a larger effect of profile dispersion. This provides a useful consistency check, since it indicates that the enhancement is not solely associated with changes in the BBKS spherical curvature, but arises from the residual radial-dispersion effect itself.

\begin{table}[t]
\centering
\begin{tabular}{c c c c c c}
\toprule
$R_F$ & $\mu_{c}(0)$ & $s_\star$ & $n_\star$ & $\nu_{c}(s_{\star})$ & $\beta_{\rm disp}/\beta_0$ \\
\midrule
$5$ & $ 0.839$ & $-0.132$ & $-1.33$ & $8.24$ & $2.4$ \\
$10$ & $1.014$ & $-0.307$ & $-2.55$ & $7.75$ & $10.3$ \\
$50$  & $1.483$ & $-0.733$ & $-4.18$ & $5.91$ & $5.2\times 10^{3}$ \\
\bottomrule
\end{tabular}
\caption{
Summary of the numerical results for the projected \(q_x\)-orthogonal finite
top-hat split deformation.  This table is analogous to
Table~\ref{tab:bbks_integrated_peak_abundance_unprojected}, but uses the
projected residual mode, which keeps both the central height and the BBKS
spherical curvature fixed.}
\label{tab:bbks_integrated_peak_abundance_projected_qx}
\end{table}

\section{Conclusions}
\label{sec:conclusions}

In this work we have developed a statistical framework for describing
curvature-profile dispersion in primordial black hole formation.  The central idea is to quantify the Gaussian statistical cost of coherent
departures from a chosen reference profile, and to compare this cost with the possible gain from a reduced PBH formation threshold determined by the nonlinear collapse dynamics. The power spectrum defines a natural Gaussian-action metric on the
space of profiles, and coherent deformations can be normalized with respect to
this metric.  Their amplitudes therefore have a direct statistical
interpretation as standard Gaussian variables, rather than as pointwise
fluctuations of the curvature field.

Within this framework, the usual BBKS peak variables are recovered as special
action-normalized directions.  The peak height, spherical curvature and
quadrupolar Hessian sector correspond to the central value, the trace of the
Hessian and the local ellipsoidal deformation, respectively.  The orthogonal
complement to this BBKS sector describes genuine residual profile dispersion:
coherent radial or angular deformations that can modify the full real-space
profile relevant for collapse while leaving the chosen local peak data fixed.
The monochromatic limit provides a useful consistency check, since height and
curvature are perfectly correlated and independent radial dispersion is absent.
For finite-width spectra this degeneracy is lifted, and additional
finite-action radial degrees of freedom become available.  The same construction
also provides a systematic route to generalize the description of
non-spherical PBH-forming shapes beyond the ellipsoidal sector encoded in the
BBKS Hessian variables~\cite{Bardeen:1985tr}.

We applied the framework to two spherical numerical-collapse examples.  The
first used a sharply peaked finite-width spectrum with a coherent fixed-height
radial split mode, including both Gaussian perturbations and a logarithmic local
non-Gaussian map.  The collapse threshold \(\mu_c(s)\) was found to depend
nontrivially on the shape coordinate \(s\): changing \(s\) modifies the
position of the compaction peak, the compensation tail and the nonlinear
compaction maximum.  The threshold variation is therefore not equivalent to a
simple rescaling of a fixed profile.

Combining the numerical threshold curves with peak statistics shows that the
dominant contribution to the abundance is selected by a competition between the
Gaussian action cost of realizing a coherent deformation and the exponential
gain associated with lowering the collapse threshold. For positive logarithmic
non-Gaussianity this competition leaves the dominant contribution close to the
reference profile associated with the chosen conditioning.  For negative non-Gaussianity, however, rare
coherent shape deformations can substantially lower the threshold and overcome
their Gaussian suppression.  In the examples studied here this produces a large
enhancement of the PBH abundance and can significantly reduce the
power-spectrum amplitude required to obtain a fixed PBH fraction.

This has important observational implications for scenarios in which negative local non-Gaussianity is used to suppress PBH production while retaining a scalar-induced gravitational-wave signal in the PTA band
\cite{NANOGrav:2023gor,EPTA:2023fyk,Reardon:2023gzh,Xu:2023wog,
Franciolini:2023pbf,Wang:2023ost,Choudhury:2023hfm,Inomata:2023zup,Inui:2024fgk}. At fixed scalar power spectrum, our results indicate that coherent profile
dispersion can enhance the PBH abundance relative to estimates that neglect
residual profile dispersion. In this setting, reduced-profile or no-dispersion
calculations may therefore underestimate the PBH abundance, strengthening the
tension between PBH production and PTA-motivated scalar-induced
gravitational-wave scenarios. Conversely, if the target PBH abundance is fixed instead of the scalar power spectrum, the enhanced abundance implies that a smaller scalar power-spectrum amplitude would be required. At leading order, this would reduce the associated scalar-induced gravitational-wave signal, although the precise effect depends on the full joint treatment of the scalar spectrum, non-Gaussian corrections and PBH abundance. A dedicated joint analysis including profile dispersion is therefore needed for accurate estimates.

The second example considered a finite top-hat spectrum.  We found
non-monotonic threshold curves and, especially for broad spectra, branch
switching between competing compaction maxima.  This effect is most visible for
positive values of the dispersion parameter, \(s>0\), where the dominant
compaction maximum can move from the inner peak to an outer radial maximum as
the coherent deformation is varied, producing non-trivial threshold behaviour.
This shows that, for finite-band spectra, the collapse threshold is not
controlled only by the local shape of a single compaction peak, but by the
global radial structure of the profile, in a way similar to the behaviour found
in Ref.~\cite{Escriva:2023qnq}.  After integrating over the coherent shape
variable, the BBKS peak-abundance ratio increases rapidly with the spectral
bandwidth.  We find
\(\beta_{\rm disp}/\beta_0\simeq 2\) for
\(R_F=5\), \(\simeq 8\) for \(R_F=10\), and
\(\simeq 3.0\times10^3\) for \(R_F=50\).  In the broadest case the dominant
differential contribution comes from a rare negative deformation,
\(n_\star\simeq -4\). Thus the profiles responsible for the largest
abundance enhancement are not necessarily those displaying the strongest
positive-\(s\) branch switching, but rather those that optimize the full
statistical weight: the threshold reduction, the BBKS peak abundance and the
Gaussian action cost.

The main conclusion of this work is that the reference profile associated with the chosen conditioning is not generically guaranteed to dominate PBH production. The relevant profile is selected
statistically and dynamically: it is the configuration that optimizes the
balance between the Gaussian cost of the deformation and the threshold reduction
achieved by that deformation. This effect is small when the collapse threshold
is weakly shape-dependent, but it can become large for broad spectra or for
non-Gaussian maps that generate strongly shape-dependent thresholds. In such
cases, the abundance is sensitive both to the nonlinear collapse dynamics and
to the amount of statistically available profile dispersion. It is also worth
emphasizing that the commonly used monochromatic-spectrum approximation is a
restrictive idealization from the perspective of profile dispersion. In the
strict monochromatic limit, the radial structure is fixed once the amplitude is
specified, and independent radial shape dispersion is absent. For finite-width
spectra, however, this degeneracy is lifted: additional statistically allowed
radial shape degrees of freedom become available, and these can affect the
collapse threshold and the resulting PBH abundance. The logarithmic bandwidth
$(L=\ln(k_+/k_-)=\ln R_F)$ provides a useful quantitative measure of the
effective duration of the enhanced feature in e-folds of horizon crossing, up
to slow-roll corrections. In the finite top-hat examples considered here,
$(R_F=5,10,50)$ correspond to approximately $(1.6)$, $(2.3)$, and $(3.9)$
e-folds, respectively. The dispersion effect is already visible for widths of
order a few e-folds and becomes particularly large for the broadest case,
$(L\simeq 3.9)$, where the abundance enhancement reaches several orders of
magnitude. This suggests that profile dispersion can become quantitatively important
for enhanced power-spectrum features extending over several e-folds,
although the precise crossover depends on the shape of the spectrum,
the collapse-threshold dependence on the profile, and the statistical
prescription used to characterize the profile ensemble. Our results indicate that, within
the statistical construction and finite-action deformation families considered
here, abundance estimates based only on the no-dispersion reference branch can
miss an important contribution when the spectrum has finite width. A fully
general assessment of this effect would require extending the same
finite-action treatment to other choices of collapse variables and conditioning
prescriptions. An additional implication of the spherical results is that profile
dispersion can shift the dominant contribution towards lower effective
peak heights, where the usual high-peak near-spherical approximation
is less restrictive. Radial and angular shape dispersion may therefore
become increasingly coupled precisely in the regimes where the spherical
dispersion effect is strongest.

Several extensions should be addressed in future work.  First, the
non-spherical modes identified in the general formalism should be evolved with
relativistic simulations beyond spherical symmetry.  This is especially
important for higher multipoles beyond the BBKS ellipsoid, which would allow a
direct comparison with previous ellipsoidal studies of PBH formation
\cite{Escriva:2024hsv} and with PBH mass-function calculations including
non-spherical effects~\cite{Escriva:2024rsk}.  Second, the present numerical
examples used specific low-dimensional ansaetze to construct the dispersed
profiles.  Other residual shape directions and functional forms should be
explored in order to assess how the abundance depends on the chosen projection
of the full shape space.  A natural extension is therefore to enlarge the
residual shape space in a controlled way: the finite-dimensional examples
considered here represent selected action-normalized directions in the space of
coherent profile deformations, while future work should construct residual
bases conditioned on the BBKS peak variables and study the convergence of the
PBH abundance as additional radial modes and non-spherical multipoles are
included.  In particular, multipoles with \(\ell\geq3\) provide genuine shape
degrees of freedom beyond the local ellipsoidal peak description, and their
collapse thresholds require fully non-spherical relativistic simulations.  This
is especially relevant for broad spectra, where our results indicate that
profile dispersion becomes increasingly important and non-spherical
deformations may give a non-negligible contribution to the PBH abundance. Finally, it would be interesting to consider inflationary models similar to
those of Ref.~\cite{Escriva:2025ftp}, but capable of generating a broader power
spectrum than the one studied there.  In such cases, profile dispersion may
become relevant, and there may exist parameter regions where the PBH abundance
is dominated by the vacuum-bubble channel rather than by the adiabatic channel.

Overall, our results indicate that curvature-profile dispersion should be
treated as part of the statistical description of the primordial field, since it can modify the relative weight of different PBH-forming profiles in the
abundance calculation.  We also emphasize the
importance of relativistic numerical simulations for determining accurately the
critical conditions for PBH formation.

\acknowledgments
This work was supported by JSPS KAKENHI Grant Number 26K17141.

\appendix
\section{Appendix}

\subsection{Spherical harmonics definitions}
\label{appendix:harmonics}

The spherical
Bessel function is defined as
\begin{equation}
j_\ell(z)
=
(-1)^\ell z^\ell
\left(\frac{1}{z}\frac{d}{dz}\right)^\ell
\frac{\sin z}{z}.
\end{equation}
and $Y_{\ell m}$ is a spherical harmonic, which is normalized with the angular-average measure
\(d\Omega/(4\pi)\). In terms of the polar angle \(\theta\) and the azimuthal
angle \(\phi\), we take
\begin{equation}
Y_{\ell m}(\theta,\phi)
=
\left[
(2\ell+1)\frac{(\ell-m)!}{(\ell+m)!}
\right]^{1/2}
P_\ell^m(\cos\theta)\,e^{im\phi},
\qquad
\ell=0,1,2,\ldots,\qquad
0\leq m\leq \ell .
\end{equation}
The negative-\(m\) modes are defined by
\begin{equation}
Y_{\ell,-m}(\theta,\phi)
=
(-1)^mY^*_{\ell m}(\theta,\phi).
\end{equation}
With this complex convention, the spherical harmonics satisfy
\begin{equation}
\int\frac{d\Omega}{4\pi}
Y_{\ell m}(\hat{\bm x})
Y^*_{\ell' m'}(\hat{\bm x})
=
\delta_{\ell\ell'}\delta_{mm'} .
\end{equation}
For a real field, the coefficients of the complex modes obey the corresponding
reality condition. Equivalently, one may work with real harmonics obtained, for
\(m>0\), from the cosine- and sine-type combinations
\begin{equation}
Y_{\ell m}^{(c)}
=
\frac{Y_{\ell m}+(-1)^mY_{\ell,-m}}{\sqrt{2}},
\qquad
Y_{\ell m}^{(s)}
=
\frac{Y_{\ell m}-(-1)^mY_{\ell,-m}}{i\sqrt{2}} .
\end{equation}
Together with \(Y_{\ell0}\), these form an orthonormal real basis,
\begin{equation}
\int\frac{d\Omega}{4\pi}
Y^{a}_{\ell A}(\hat{\bm x})
Y^{b}_{\ell' A'}(\hat{\bm x})
=
\delta_{\ell\ell'}\delta_{AA'}\delta_{ab} .
\end{equation}
This is the convention used in the main text when the independent Gaussian
amplitudes are written as real variables and the quadratic action is written as
a sum of squares.

The associated Legendre polynomials are defined as
\begin{equation}
P_\ell^m(x)
=
(-1)^m(1-x^2)^{m/2}
\frac{d^m}{dx^m}P_\ell(x),
\qquad 0\leq m\leq \ell ,
\end{equation}
where
\begin{equation}
P_\ell(x)
=
\frac{1}{2^\ell \ell!}
\frac{d^\ell}{dx^\ell}(x^2-1)^\ell .
\end{equation}
For negative \(m\), one has
\begin{equation}
P_\ell^{-m}(x)
=
(-1)^m
\frac{(\ell-m)!}{(\ell+m)!}
P_\ell^m(x).
\end{equation}

\subsection{Numerical methodology}
\label{appendix:numerics}
We now briefly describe the numerical methodology of SPriBHoS codes~\cite{Escriva:2019sim,Escriva:2025eqc} used to determine the
collapse threshold for each curvature profile. Since in this work we restrict the
numerical analysis to the monopole sector, the collapse is evolved in spherical
symmetry using the Misner--Sharp formulation in comoving coordinates. We assume a
perfect fluid energy-momentum tensor,
\begin{equation}
    T^{\mu\nu}=(\rho+p)u^\mu u^\nu+p g^{\mu\nu},
\end{equation}
with a linear equation of state
\begin{equation}
    p=w\rho ,
\end{equation}
where in the simulations presented below we take \(w=1/3\), corresponding to the
radiation-dominated epoch.

The line element is written as
\begin{equation}
    ds^2=-A(r,t)^2dt^2+B(r,t)^2dr^2+R(r,t)^2d\Omega^2 ,
    \label{eq:MS_metric}
\end{equation}
where \(A(r,t)\) is the lapse function, \(B(r,t)\) is the radial metric
coefficient and \(R(r,t)\) is the areal radius. We define the proper-time and
proper-radial derivatives as
\begin{equation}
    D_t \equiv \frac{1}{A}\frac{\partial}{\partial t},
    \qquad
    D_r \equiv \frac{1}{B}\frac{\partial}{\partial r}.
\end{equation}
The two basic Misner--Sharp kinematical variables are then
\begin{equation}
    U \equiv D_t R = \frac{\dot R}{A},
    \qquad
    \Gamma \equiv D_r R = \frac{R'}{B}.
\end{equation}
Here and in the following, a dot denotes \(\partial_t\), while a prime denotes
\(\partial_r\). The Misner--Sharp mass \(M(r,t)\) is related to \(U\), \(\Gamma\)
and \(R\) through the constraint
\begin{equation}
    \Gamma^2 = 1+U^2-\frac{2M}{R}.
    \label{eq:MS_constraint}
\end{equation}

For Type-II curvature fluctuations, the areal radius is not monotonic, and there can be points where ($R'=0$). In the standard Misner--Sharp formalism for PBH formation (see Ref.~\cite{Escriva:2019sim} for the set of equations), this leads to terms of the form ($U'/R'$), which are numerically ill-defined at the throat and prevent the simulation of Type-II fluctuations. As shown in Ref.~\cite{Escriva:2025eqc}, introducing the trace of the extrinsic curvature as an auxiliary variable makes it possible to overcome this issue and simulate Type-II fluctuations within the Misner--Sharp formalism (see Ref.~\cite{Uehara:2024yyp} for simulations using the BSSN formalism),

\begin{equation}
    K \equiv -\left(\frac{U'}{R'}+2\frac{U}{R}\right)
      = -\frac{1}{A}\left(\frac{\dot B}{B}+2\frac{\dot R}{R}\right).
    \label{eq:K_definition}
\end{equation}
In the homogeneous FLRW limit this reduces to \(K_b=-3H\). The use of \(K\)
absorbs the potentially singular term \(U'/R'\) and gives a formulation that can
be applied to both type-I and type-II configurations.

For a constant equation of state, the lapse can be obtained from the Euler
equation as
\begin{equation}
    A(r,t)=\left(\frac{\rho_b(t)}{\rho(r,t)}\right)^{\frac{w}{1+w}},
    \label{eq:lapse}
\end{equation}
where \(\rho_b(t)\) is the background FLRW density. The system of evolution
equations used in the simulations is then
\begin{align}
    \dot U
    &=
    -A\left(
        \frac{M}{R^2}+4\pi R w\rho
      \right)
    +\frac{A'\Gamma}{B},
    \label{eq:evol_U}
    \\
    \dot\rho
    &=
    A\rho(1+w)K,
    \label{eq:evol_rho}
    \\
    \dot R
    &=
    AU,
    \label{eq:evol_R}
    \\
    \dot\Gamma
    &=
    \frac{A'U}{B},
    \label{eq:evol_Gamma}
    \\
    \dot B
    &=
    -AB\left(K+2\frac{U}{R}\right),
    \label{eq:evol_B}
    \\
    \dot K
    &=
    A\left[
        \left(K+2\frac{U}{R}\right)^2
        +2\left(\frac{U}{R}\right)^2
        +4\pi\rho(1+3w)
      \right]
      \nonumber \\
    &\hspace{1.0cm}
    -\frac{1}{B^2}
      \left[
        A''
        +A'\left(
            2\frac{R'}{R}
            -\frac{B'}{B}
          \right)
      \right].
    \label{eq:evol_K}
\end{align}
Although the Misner--Sharp mass can also be evolved through
\begin{equation}
    \dot M=-4\pi A w\rho U R^2,
\end{equation}
we reconstruct \(M\) during the evolution using Eq.~\eqref{eq:MS_constraint},
namely
\begin{equation}
    M=\frac{R}{2}\left(1+U^2-\Gamma^2\right).
    \label{eq:M_constraint}
\end{equation}
The Hamiltonian constraint,
\begin{equation}
    \mathcal{H}
    \equiv
    \frac{M'}{B}
    -4\pi R^2\rho\,\Gamma ,
    \label{eq:Hamiltonian_constraint}
\end{equation}
is monitored as a diagnostic of the numerical accuracy.

The initial conditions are imposed when the perturbation is well outside the
cosmological horizon. In this regime we use the gradient expansion \cite{Lyth:2004gb,Tanaka:2007gh},
controlled by
\begin{equation}
    \epsilon(t)\equiv \frac{1}{H(t)L_{\rm ls}(t)} \ll 1 ,
\end{equation}
where \(L_{\rm ls}(t)\) is the physical length scale of the perturbation. At zeroth order
the metric takes the asymptotic form
\begin{equation}
    ds^2=-dt^2+a(t)^2e^{2\zeta(r)}
    \left(dr^2+r^2d\Omega^2\right),
    \label{eq:LW_metric}
\end{equation}
with \(\zeta(r)\) the primordial curvature profile. It is also useful to introduce the areal radial coordinate
$\tilde r$ through
\begin{equation}
    \tilde r \equiv r e^{\zeta(r)},
    \qquad
    \frac{d \tilde{r}}{dr} = e^{\zeta}(1+r \zeta')
    \label{eq:rtilde_relation}
\end{equation}
at leading order in the gradient expansion.

The leading corrections to
the FLRW solution are then written as
\begin{align}
    U &= HR\left(1+\epsilon^2\widetilde U\right),\\
    \rho &= \rho_b\left(1+\epsilon^2\widetilde\rho\right),\\
    R &= a r e^{\zeta(r)}
        \left(1+\epsilon^2\widetilde R\right),\\
    M &= \frac{4\pi}{3}\rho_b R^3
        \left(1+\epsilon^2\widetilde M\right),\\
    B &= a e^{\zeta(r)}
        \left(1+\epsilon^2\widetilde B\right),\\
    A &= 1+\epsilon^2\widetilde A,\\
    K &= -3H\left(1+\epsilon^2\widetilde K\right).
\end{align}
For a given curvature profile \(\zeta(r)\), the first non-vanishing terms are
\begin{align}
    \widetilde U
    &=
    \frac{1}{5+3w}
    e^{-2(\zeta-\zeta_m)}
    \zeta'
    \left(
        \frac{2}{r}+\zeta'
    \right)r_m^2,
    \\
     \label{eq:perturbations}
    \widetilde\rho
    &=
    -\frac{2(1+w)}{5+3w}
    e^{-2(\zeta-\zeta_m)}
    \left[
        \zeta''
        +\zeta'
        \left(
            \frac{2}{r}+\frac{\zeta'}{2}
        \right)
    \right]r_m^2,
    \\
    \widetilde R
    &=
    \frac{1}{1+3w}
    \left(
        -\frac{w}{1+w}\widetilde\rho
        +\widetilde U
    \right),
    \\
    \widetilde M
    &=
    -3(1+w)\widetilde U,
    \\
    \widetilde B
    &=
    -\frac{1}{1+3w}
    \left(\widetilde\rho+2\widetilde U\right),
    \\
    \widetilde A
    &=
    -\frac{w}{1+w}\widetilde\rho,
    \\
    \widetilde K
    &=
    -\frac{\widetilde\rho}{3(1+w)} ,
\end{align}
where \(\zeta_m\equiv \zeta(r_m)\), and \(r_m\) denotes the characteristic scale
of the perturbation. We now introduce the compaction function \cite{Shibata:1999zs} (see \cite{Harada:2023ffo} for a recent discussion), which is defined in the comoving gauge as the ratio of the mass excess to the areal radius,
\begin{equation}
    \mathcal{C}(r,t) = 2 \frac{M(r,t)-M_b(r,t)}{R}
\end{equation}
where $M_b(r,t) = 4 \pi \rho_b(t) R(r,t)^3/3$. We define the length scale of the fluctuation as the location of the maximum of the compaction function $r_m$. At leading order in the gradient expansion \cite{Harada:2015yda} the
compaction function is
\begin{equation}
    \mathcal{C}(r)
    =
    f(w)
    \left[
        1-\left(1+r\zeta'(r)\right)^2
    \right],
    \qquad
    f(w)=\frac{3(1+w)}{5+3w}.
    \label{eq:compaction_zeta}
\end{equation}

The radial derivatives in Eqs.~\eqref{eq:evol_U}--\eqref{eq:evol_K} are computed
using a pseudo-spectral Chebyshev collocation method. 
Black-hole formation is identified through the null expansions
\begin{equation}
    \Theta_\pm=\frac{2}{R}\left(U\pm\Gamma\right).
    \label{eq:null_expansions}
\end{equation}
An apparent horizon forms when
\begin{equation}
    \Theta_+(r_\ast,t)=0,
    \qquad
    \Theta_-(r_\ast,t)<0,
\end{equation}
or equivalently when \(2M/R=1\). We track the peak value of the compaction function to determine whether a fluctuation collapses or disperses \cite{Escriva:2019sim}. For each profile shape, we vary the amplitude of the curvature perturbation and determine the critical value separating dispersal from apparent-horizon formation. We assess the accuracy of our simulations by monitoring the Hamiltonian constraint equation Eq.~\eqref{eq:Hamiltonian_constraint}, with similar behaviour as in \cite{Escriva:2025eqc}.

\begin{figure}[t]
    \centering
    \includegraphics[width=1.0\textwidth]{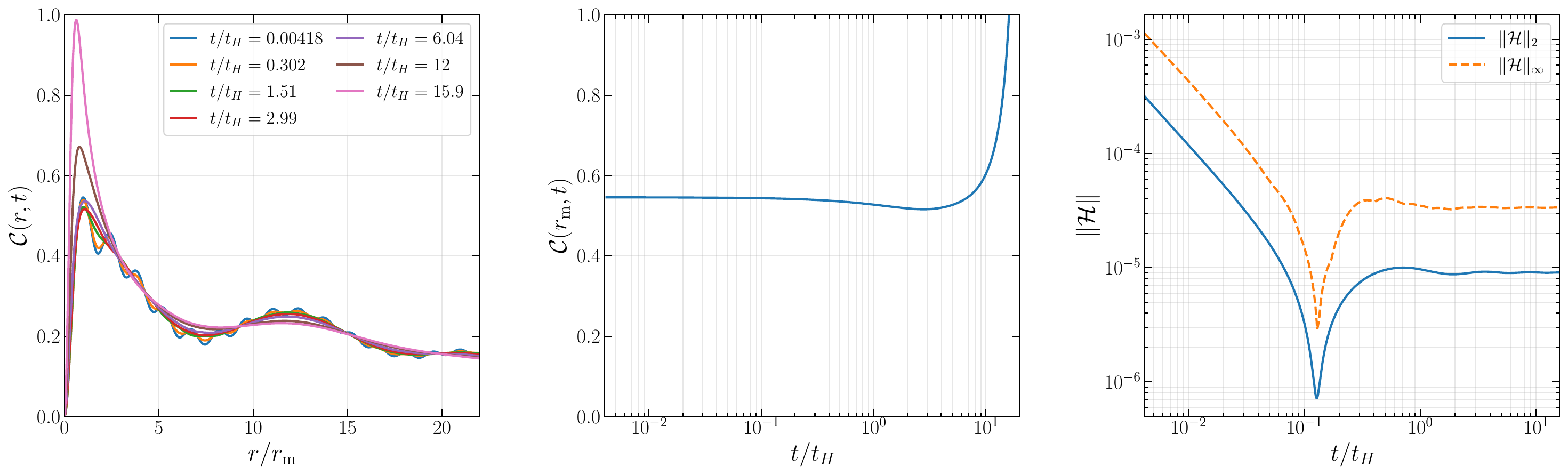}
    \caption{Left-panel: evolution of the compaction function \(\mathcal{C}(r,t)\) at selected times, with the radial coordinate normalized
by the initial compaction scale \(r_m\).
Middle-panel: evolution of the maximum compaction, $\mathcal{C}(r_{\rm m} , t)$, as a function of
\(t/t_H\). Right-panel: time evolution of the \(L_2\) and
\(L_\infty\) norms of the Hamiltonian constraint.}
    \label{fig:ham_constraint_example}
\end{figure}

An example of the numerical evolution is shown in Fig.~\ref{fig:ham_constraint_example}.
The figure illustrates the dynamical evolution of a representative collapsing configuration.
At early times, the compaction profiles remain nearly unchanged, while the most significant
evolution occurs at later times, when the inner peak becomes progressively enhanced.
This behaviour is also visible in the middle panel: the maximum compaction initially
decreases slightly and subsequently grows rapidly, eventually reaching
\(\mathcal{C}(r_m,t) \approx 1\), which we use as the criterion signalling PBH formation,
following Ref.~\cite{Escriva:2019sim}.
The right panel shows the corresponding evolution of the Hamiltonian-constraint violation.
Both the \(L_2\) and \(L_\infty\) norms decrease during the early stages of the evolution
and remain well below unity throughout the time interval used to determine whether
black-hole formation occurs.
The other simulation cases considered in this work exhibit similarly robust behaviour of the
Hamiltonian constraint.

\bibliographystyle{JHEP}
\bibliography{references}

\end{document}